\documentclass[a4paper,12pt,english,oneside]{book}

\usepackage{amsmath,amssymb}
\usepackage[english]{babel}
\usepackage{tikz}
\usepackage[normalem]{ulem}
\usepackage{indentfirst}
\usepackage[T1]{fontenc}
\usepackage[utf8]{inputenc}
\usepackage{graphicx}
\usepackage{graphics}
\usepackage{xcolor}
\usepackage{epigraph, varwidth}
\usepackage{pdfpages}
\usepackage{setspace}
\usepackage[font=small,labelfont=bf]{caption}
\usepackage{pgfplots}
\usepackage{epstopdf}
\usepackage{subfig}
\usepackage{afterpage}
\usepackage{array}
\usepackage{emptypage}
\usepackage[retainorgcmds]{IEEEtrantools}
\usepackage[acronym,nonumberlist,sort=use]{glossaries}
\usepackage{enumitem}
\usepackage[colorlinks=true, allcolors=black]{hyperref}
\usepackage{caption}
\usepackage[super]{nth}
\usepackage{cite}

\newcommand{\be}{\begin{equation}}
\newcommand{\ee}{\end{equation}}
\newcommand{\bal}{\begin{aligned}}
\newcommand{\eal}{\end{aligned}}
\newcommand{\beq}{\begin{eqnarray}}
\newcommand{\eeq}{\end{eqnarray}}
\newcommand{\ba}{\begin{array}}
\newcommand{\ea}{\end{array}}
\newcommand{\bs}{\boldsymbol}
\newcommand{\bi}{\begin{itemize}}
\newcommand{\ei}{\end{itemize}}
\newtheorem{theorem}{Theorem}
\newcommand{\bt}{\begin{theorem}}
\newcommand{\et}{\end{theorem}}
\newcommand{\nn}{\nonumber}

\usepackage[bottom=2cm,top=3cm,left=3cm,right=2cm]{geometry}

\newglossary{abbrev}{abs}{abo}{List of Abbreviations}
\newglossaryentry{RTBP}{
    name        = RTBP ,
    description = Restricted three-body problem ,
    type        = abbrev
}
\newglossaryentry{CRTBP}{
    name        = CRTBP ,
    description = Circular restricted three-body problem ,
    type        = abbrev
}
\newglossaryentry{PCRTBP}{
    name        = PCRTBP ,
    description = Planar circular restricted three-body problem ,
    type        = abbrev
}
\newglossaryentry{EMS}{
    name        = EMS ,
    description = Earth-Moon system ,
    type        = abbrev
}
\newglossaryentry{Li}{
    name        = $\text{L}_i$,
    description = $i$-th Lagrangian point ,
    type        = abbrev
}
\newglossaryentry{RK}{
    name        = RK,
    description = Runge-Kutta ,
    type        = abbrev
}
\newglossaryentry{ZVC}{
    name        = ZVC,
    description = Zero-velocity curve ,
    type        = abbrev
}
\newglossaryentry{UPO}{
    name        = UPO,
    description = Unstable periodic orbit ,
    type        = abbrev
}
\newglossaryentry{ODE}{
    name        = ODE,
    description = Ordinary differential equation ,
    type        = abbrev
}
\makenoidxglossaries

\begin{document}

\pagenumbering{gobble}
%
%\includepdf[pages=-]{folhas_de_rosto/folha_de_rosto_pt.pdf}
%
%\includepdf[pages=-]{folhas_de_rosto/ficha_catalografica.pdf}

\includepdf[pages=-]{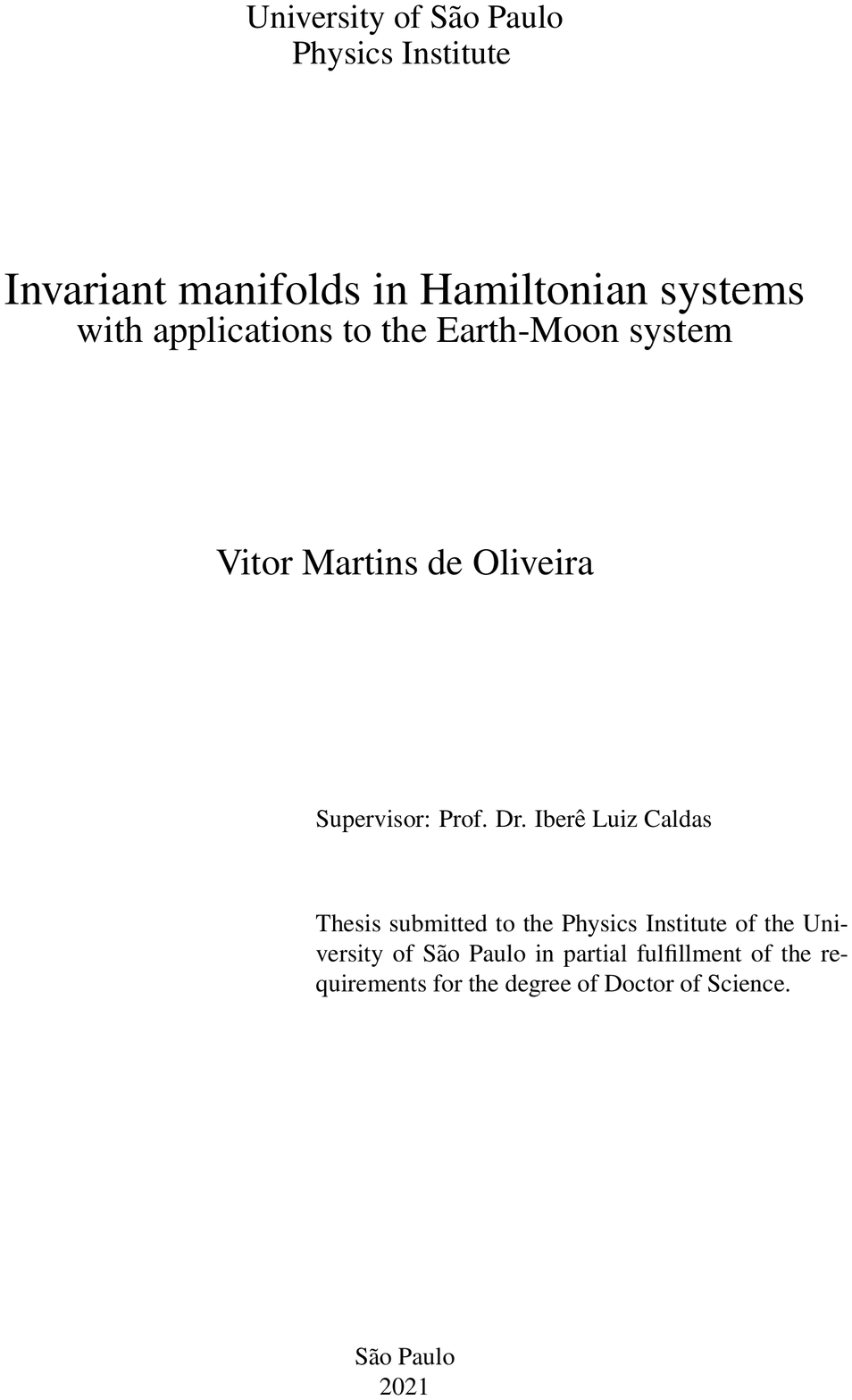}

%\blankpage

\begin{center}
\section*{Acknowledgments}
\end{center}

\begin{center}

The last four years were a time of growth, both personally and professionally. And it wouldn't have been the same without a large number of people involved in my life.

First of all, I am thankful to my awesome wife, that walked this journey with me and always stood by my side.

I am thankful to my amazing family, that always supported my dreams and taught me all the values I carry dear.

I am thankful to Prof. Iberê, an excellent advisor and a wonderful person, whose example I will always try my best to follow.

I am thankful to the Oscillations Control Group, which is formed by an incredible group of people and that has provided me with lifetime friends.

I am thankful to the staff of the Institute of Physics, especially to the secretaries of the Applied Physics Department, whose incredible efficiency allowed me to work with tranquility.

I am especially thankful to Prof. Priscilla, whose guidance was of paramount importance for this work.

Lastly, I am thankful to everyone who directly or indirectly helped me achieve my goals.

\vspace*{0.5cm}

This study was financed in part by the Coordenação de
Aperfeiçoamento de Pessoal de Nível Superior - Brasil (CAPES) -
Finance Code 001, the São Paulo Research Foundation (FAPESP) - Grant No. 2018/03211-6, and the National Council for Scientific and Technological Development - Brazil (CNPq) - Grant Nos. 302665/2017-0 and 407299/2018-1. 

\end{center}

%\begin{center}
%\section*{Agradecimentos}
%\end{center}
%
%\begin{center}
%
%Os últimos quatro anos foram um período de crescimento, tanto pessoal quanto profissional. E este período não teria sido o mesmo sem um grande número de pessoas envolvidas em minha vida.
%
%Primeiramente, sou grato à minha esposa, a qual trilhou esta jornada comigo e sempre esteve ao meu lado.
%
%Sou muito grato à minha família, que sempre apoiou os meus sonhos e me ensinou todos os valores os quais eu levo comigo.
%
%Eu sou grato ao prof. Iberê, um ótimo orientador e uma ótima pessoa, cujo exemplo profissional eu sempre tentarei seguir.
%
%Sou grato ao grupo Controle de Oscilações, o qual é formado por um excelente grupo de pessoas e que me proporcionou amigos para toda a vida.
%
%Sou grato aos funcionários do Instituto de Física da USP, especialmente às secretárias do Departamento de Física Aplicada, cuja eficiência me permitiu trabalhar com tranquilidade.
%
%Sou especialmente grato à prof. Priscilla, cujos conselhos foram primordiais para a realização deste trabalho.
%
%Por último, sou grato a todos que direta ou indiretamente me ajudaram a alcançar os meus objetivos.
%
%\vspace*{0.5cm}
%
%O presente trabalho foi realizado com apoio da Coordenação de Aperfeiçoamento de Pessoal de Nível Superior - Brasil (CAPES) - Código de Financiamento 001, da Fundação de Amparo à Pesquisa do Estado de São Paulo (FAPESP), processo nº 2018/03211-6, e do Conselho Nacional de Desenvolvimento Científico  e Tecnológico - Brasil (CNPq), processos nº 302665/2017-0 e nº 407299/2018-1. 
%
%\end{center}

%\newpage
%\vspace*{\fill}

\newpage
\vspace*{\fill}
\epigraph{Kirk ``Spock, these cadets of yours, how good are they? How will they respond under real pressure?'' \\ Spock ``As with all living things, each according to his gifts".}{Star Trek II: The Wrath of Kahn}

% \newpage
% \vspace*{\fill}
% \epigraph{The most difficult in chess is to see moves with knight back}{GM Vasyl Ivanchuk}

%\newpage
%\vspace*{\fill}
%\epigraph{Epigraph}{Author}

%\blankpage
\newpage

\chapter*{Abstract}

\thispagestyle{empty}

Invariant manifolds are the skeleton of the chaotic dynamics in Hamiltonian systems. In Celestial Mechanics, these geometrical structures are applied to a multitude of physical and practical problems, such as to the description of the natural transport of asteroids, and to the construction of trajectories for artificial satellites. In this work, we focus our investigation on the motion of a body with negligible mass that moves due to the gravitational attraction of both the Earth and the Moon. As a model, we adopt the planar circular restricted three-body problem, a near-integrable Hamiltonian system with two degrees of freedom, and consider a situation where all orbits inside the Earth's or the Moon's realm are free to move between these regions but are bounded within the system. We derive the equations of motion for the problem and explain in detail all the numerical procedures that are carried out, from the determination of periodic orbits to the calculation of two-dimensional invariant manifolds. By varying the Jacobi constant of motion, we observe that the system undergoes a transition from a mixed phase space with a far-reaching stickiness effect, to a global chaos scenario, and back to a mixed phase space, although now with localized stickiness. During this process, the Lyapunov orbit manifolds spread throughout the phase space, displaying a close relationship with the shape and location of regular regions, and also with the transport of orbits between the realms, while the invariant manifolds associated with certain unstable periodic orbits, formed by the destruction of the last KAM torus of the regular regions, are related to the behavior of stickiness and, consequently, to dynamically trapping transit orbits. Our results provide a visual description of the influence of invariant manifolds in the dynamical properties of the Earth-Moon system and could contribute to the understanding of the connection between dynamics and geometry in Hamiltonian systems.

\vspace{.5cm}
\textbf{Keywords}: Invariant manifolds, Hamiltonian systems, Three-body problem, Chaos

{\normalsize\tableofcontents}
\thispagestyle{empty}

{\normalsize\listoffigures}
\thispagestyle{empty}

{\printnoidxglossary[type=abbrev]}
\thispagestyle{empty}

\clearpage
\pagestyle{headings}

\mainmatter

\setcounter{page}{8}

% \doublespacing

\chapter{Introduction}
\label{ch:introduction}

% {\color{red} Possible changes to the manuscript
% 
% \begin{itemize}
% \item include some historical notes in the introduction
% \item make graphs comparing non-regularized with regularized system
% \item make graphs comparing different RK methods (has to be appendix)
% \item make just one or two graphs showing error
% \item make graph and a paragraph on the horseshoe dynamics associated with homoclinic intersections
% \end{itemize}
% 
% }

Hamiltonian systems constitute an important class of dynamical systems, and are represented by a myriad of physical problems in fields such as Celestial Mechanics, Plasma Physics, and Quantum Mechanics \cite{Szebehely1967,Lichtenberg1992,Ozorio1990book}. Near-integrable Hamiltonian systems, in special, display a rich dynamical structure, with the coexistence of both regular and chaotic motion (\emph{mixed} phase space). The structure of the system's phase space, i.e., the location and shape of the regular solutions and chaotic sea, depends on the system's parameters and constants of motion.

% {\color{red} The history of Hamiltonian systems...}

Periodic solutions are of special significance, since they locally determine the behavior of other solutions in the phase space. In particular, \emph{unstable periodic orbits} immersed in the chaotic sea possess associated \emph{stable} and \emph{unstable invariant manifolds}, which are composed by all the solutions that converge to them in a given time direction. The importance of these geometrical structures was already perceived by Poincaré when he studied the three-body problem and concluded that the intersection between invariant manifolds lead to complicated motion in the system \cite{Poincare1890,Alligood2012}. Later, it was showed that such intersections are indeed the cornerstone of Hamiltonian chaos \cite{Smale1967,Ott2002}.

%%%%

Much of the interesting phenomena related to invariant manifolds can be observed from studying these curves in area-preserving maps, which are easier to investigate than continuous-time systems and, in certain circumstances, can be derived directly from a Hamiltonian flow \cite{Mackay1987,Meiss1992}. Notably, invariant manifolds are deeply connected with transport properties in dynamical systems, acting as transport barriers, partial transport barriers and transport channels \cite{Conley1968,Meiss2015,Ciro2016plasma}.

% {\color{red}
% Scaterring? Or something related to my research maybe
% }

%%%%

In Celestial Mechanics, invariant manifolds were widely investigated, being associated to resonant transitions in the orbits of Jupiter comets \cite{Koon2000}, to the jumping mechanism of Trojan asteroids \cite{Oshima2015}, to stickiness effect in barred spiral galaxies \cite{Contopoulos2010}, and more recently, to the pathways of comets in the outer Solar System \cite{Todorovic2020}. In Orbital Dynamics, these geometrical structures were also employed in a variety of applications, such as the construction of a transfer trajectory to and from a periodic orbit named Halo orbit, which was first exploited by the Genesis mission \cite{Gomez2001,Lo2001}, the determination of low energy transfers to the Moon \cite{Koon2001}, and also a trajectory design for a mission to Titan \cite{Gawlik2009}.

%%%%

In this work, we make use of efficient numerical methods to \emph{visually illustrate} the influence of invariant manifolds in the dynamical properties of Hamiltonian systems. We focus our investigation on the motion of a small body that is under the gravitational influence of both the Earth and the Moon, the \emph{Earth-Moon system}. As a model, we adopt the \emph{Planar Circular Restricted Three-Body Problem}, a hallmark of Hamiltonian mechanics. We select a range of values for the \emph{Jacobi constant} in which all orbits are \emph{bounded} within the system, and we carry out our analyses on a Poincaré map which is calculated on a suitable surface of section in the neighborhood of the Moon. Our interest lies on how the dynamical and geometrical objects present in the system's phase space evolve and interact with each other as we vary the constant of motion.

% {\color{red} The interest on the three-body problem began...(maybe change three-body problem with poincare back there). The system can be reduced...}

Periodic solutions of the \emph{Circular Restricted Three-Body Problem} with Earth and Moon's masses have been extensively studied for a long time \cite{Broucke1968,Henon1974}. One of the main families of periodic solutions, that also exists in the planar case, is composed by the \emph{Lyapunov orbits}, which are \emph{unstable periodic orbits} around the \emph{Lagrangian equilibrium points} \cite{Koon2008}. Here, we are interested in the Lyapunov orbits around $L_1$, which are strategic localized in the \emph{neck} region between the Earth and the Moon's realms. Another class of solutions that we investigate is formed by the \emph{direct periodic orbits} around the Moon, specifically the \emph{Low Prograde Orbits} and the \emph{Distant Prograde Orbits} \cite{Szebehely1967,Restrepo2018}.

Due to the presence of invariant manifolds, there are natural connections between the aforementioned unstable periodic orbits, and also between these and other periodic orbits in the system, which, for example, translates as low-cost transfers for space mission design \cite{Mingotti2012,Cox2020}. Given that these solutions can change stability or appear/disappear as the Jacobi constant varies (\emph{periodic orbit bifurcation}), it is important to determine which solutions coexist for the same system's configuration in order to define which connections are possible \cite{Folta2015}.

It is worth mentioning that calculating periodic orbits is generally a difficult task, specially in more complicated versions of the three-body problem which considers, for example, the influence of the Sun, the eccentricity of the lunar orbit, or the mass of the third particle \cite{Szebehely1967}. With this, another relevance of studying periodic solutions in the circular restricted three-body problem, whose determination already relies on numerical or ad hoc mathematical methods \cite{Gomez2001}, is that these solutions can be used as a basis for finding orbits in such complicated models \cite{Hadjidemetriou1975,Leiva2008}.

%%%%

Invariant manifolds in the planar circular restricted three-body problem are two-dimensional geometrical structures which can be represented as disconnected curves in the surface of section, hence behaving differently from one-dimensional manifolds in planar maps \cite{Koon2008,Ozorio1990}. A clear depiction of how the manifolds are organized in the phase space is important, and gives information on the distribution of homo/heteroclinic intersections \cite{Koon2000,Gidea2007}, and on the destruction of KAM tori \cite{Simo2000}. Here, we illustrate how these structures occupy the phase space alongside dynamical objects, such as invariant tori and chaotic orbits, and how they affect the transport properties of the system. We also study the stability properties of the periodic orbits associated with the regular regions, in order to fully describe the dynamical changes in the system.

% Our results contribute to the understanding of the fundamental relashionship between dynamics and geometry in Hamiltonian systems.

%%%%

Our investigation is supported by numerical procedures, from the calculation of the Lagrangian point location, passing through the tracing of unstable periodic orbits, and up to the determination of the intersection between invariant manifolds and the defined surface of section, all of which are explained in details in the text. In order to obtain a reliable result, then, the precision of these procedures must always be of concern, as should be the case for any computational work. During our analyses, we keep track of the numerical errors in every step of the way, ensuring that they are low enough ($\lesssim10^{-10}$), specially regarding the conservation of the constant of motion. 

%%%%

This manuscript is organized as follows. In Chapter~\ref{ch:hamiltonian_systems}, we introduce the main concepts of Hamiltonian systems that are relevant to our work, along with the practical aspects of computationally treating dynamical systems. In Chapter~\ref{ch:invariant_manifolds}, we define one- and two-dimensional invariant manifolds, and talk about the numerical methods we use for calculating these geometrical structures. In Chapter~\ref{ch:earth_moon_system}, the Earth-Moon system is presented and analyzed, as modeled by the planar circular restricted three-body problem. In Chapter~\ref{ch:conclusion}, we give our conclusion regarding our investigation, and discuss perspectives for future research. Finally, in Chapter~\ref{ch:intellectual}, we comment on the scientific papers and the open source code that resulted from this work.

\chapter{Hamiltonian systems}
\label{ch:hamiltonian_systems}

The Earth-Moon system, as modeled by the planar circular restricted three-body problem, is a near-integrable Hamiltonian system with two degrees of freedom. In this chapter, we give a brief presentation on some concepts in Hamiltonian systems as viewed from a Dynamical Systems perspective. We discuss both the theoretical and the computational aspects regarding the dynamical description of these systems.

\section{Dynamical systems concepts}
\label{sec:dynamical_systems}

A system whose state evolves with time is called a \emph{dynamical system} \cite{Arrowsmith1990}. Depending on the nature of the time variable, it can be separated into two classes. When time is a \emph{continuous} variable, the system is usually described by a set of first-order ordinary differential equations

\be
\dfrac{d\bs{x}}{dt}=\bs{F}(\bs{x},t),
\label{eq:ODE}
\ee

\noindent with $t\in\mathbb{R}$. On the other hand, when the time variable is \emph{discrete}, the system is described by a mapping function

\be
\bs{x}_{n+1}=\bs{M}(\bs{x}_n),
\label{eq:map}
\ee

\noindent with $n\in\mathbb{N}$. For us, $\bs{F}$ is a \emph{vector field}, while $\bs{M}$ is a bijection, with both $\bs{M}$ and $\bs{M}^{-1}$ differentiable.

%We denote the solution of Eq.~\eqref{eq:ODE} with initial condition $\bs{x}_0$ and at time $t$ by $\bs{\varphi}(\bs{x}_0,t)$, whereas an orbit 

The \emph{state} of a dynamical system is given by the variable $\bs{x}\in\mathbb{R}^N$, which takes values on the $N$-dimensional \emph{phase space}. One distinctive property of Hamiltonian systems is that they preserve volumes in phase space under a time evolution, i.e., phase space volumes are incompressible. This result is known as \emph{Liouville's theorem} \cite{Ott2002}.

\subsubsection{Computational implementation}

Numerical computation of orbits in discrete-time systems is straightforward, with errors associated only to the computational precision used to represent numerical values, and to the calculation of the map itself. For continuous-time systems, on the other hand, numerical methods for integrating ordinary differential equations (\gls{ODE}s) are necessary, making these systems somewhat more difficult to study.
 
There are various methods for finding the solution of ODEs \cite{Press2007}. We are interested in the Runge-Kutta (\gls{RK}) methods, which are easy to implement and versatile. These methods assume a solution for Eq.~\eqref{eq:ODE} of the form

\be
\bs{x}_{n+1}=\bs{x}_n+h\sum_{i=1}^sb_i\bs{k}_i,
\label{eq:RK}
\ee

\noindent with coefficients $k_i$ given by

\be
\bs{k}_i=\bs{F}\left(x_n+h\sum_{j=1}^{s}a_{ij}\bs{k}_j,t_n+c_ih\right),
\label{eq:k_RK}
\ee

\noindent where $c_i$,  $b_i$  and $s$ are the nodes, weights, and stage of the method, respectively. The difference between the various RK methods are the values of these parameters \cite{Butcher2003}. In particular, the step $h$ gives the time increment for the operation: $t_{n+1}=t_n+h$. A lower step-size means a higher precision in Eq.~\eqref{eq:RK}, but also a higher computational cost.

Runge-Kutta methods advance a solution of an ODE one time step $h$, and utilizes the vector field to obtain information on the state's derivative in $s$ different points in this interval. Equation~\eqref{eq:RK} is a truncated series approximation, and the error term is used to attribute a parameter called \emph{order} to the method. A $n^{th}$ order method has truncation error proportional to $h^{n+1}$.

An embedded RK method solves Eq.~\eqref{eq:RK} by using two schemes of different orders at the same time. The difference between the calculated solutions give a more reliable accuracy measure to the procedure. We can then adapt the size of the step for each iteration according to this accuracy, reducing the total computational time. In practice, we keep the step-size between $10^{-3}$ and $10^{-11}$, and require an accuracy of $10^{-13}$.

The errors associated with the numerical integrators, along with the errors introduced by other methods needed for our investigation, are carried throughout the simulation. Therefore, higher-order RK methods are necessary in order for us to obtain a reliable result. We extensively tested and compared various methods, including, but not limited to, the $4^{th}$ and $5^{th}$-order RK Cash-Karp \cite{Cash1990}, the $4^{th}$ and $5^{th}$-order RK Dormand-Prince \cite{Dormand1980}, and the $7^{th}$ and $8^{th}$-order RK Dormand-Prince \cite{Prince1981}. The most efficient one for our analysis was found to be the $8^{th}$ and $9^{th}$-order RK Prince-Dormand \cite{Galassi2001}.

We also tested symplectic integrators in some situations, for which we obtained similar results. These methods were developed with the preservation of symplectic areas in mind, and work very well especially for long time simulations \cite{Stuchi2002}. For our investigation, however, given that we use a higher-order RK, do not integrate the equations of motion for long periods of times, and use a low-dimensional model, such methods, which are harder to implement, presented no discernible improvement.

\subsubsection{Linear stability}

When we analyze the system's \emph{phase space}, two types of solutions for Eqs.~\eqref{eq:ODE} and \eqref{eq:map} are of special interest: \textbf{(i)} solutions where the state does not change over time and \textbf{(ii)} solutions where the state of the system repeats itself after a period of time.

A solution of type \textbf{(i)} is called an \emph{equilibrium point} $\bs{x}_e$ for a continuous system, and a \emph{fixed point} $\tilde{\bs{x}}_f$ for a discrete system. In this case, we have

%\be
%\begin{aligned}
%\dfrac{d\bs{x}_e}{dt}&= 0, \ \ \ \ \ \ \forall t\in\mathbb{R}, \\
%\bs{M}(\tilde{\bs{x}}_f) &= \tilde{\bs{x}}_f, \ \ \ \ \forall n\in\mathbb{N}.
%\label{eq:equilibrium_fixed_points}
%\end{aligned}
%\ee

\be
\begin{aligned}
\bs{F}(\bs{x}_e,t)&= 0, \ \ \ \ \ \ \forall t\in\mathbb{R}, \\
\bs{M}^n(\tilde{\bs{x}}_f) &= \tilde{\bs{x}}_f, \ \ \ \ \forall n\in\mathbb{N},
\label{eq:equilibrium_fixed_points}
\end{aligned}
\ee

\noindent where $\bs{M}^n$ stands for the function $\bs{M}$ being applied $n$ times.

%\begin{subequations}
%\label{eq:equilibrium_fixed_points}
%\begin{align}
%\dfrac{d\bs{x_e}}{dt}&= 0, \ \ \ \ \ \ \forall t\in\mathbb{R}, \label{1} \\
%\bs{M}(\tilde{\bs{x}}_f) &= \tilde{\bs{x}}_f, \ \ \ \ \forall n\in\mathbb{N}. \label{2}
%\end{align}
%\end{subequations}

A solution of type \textbf{(ii)} is called \emph{periodic}, and the \emph{trajectory} formed by this solution in space is called a \emph{periodic orbit}. Denoting by $\bs{\varphi}(\bs{x}_0,t)$ the  solution of Eq.~\eqref{eq:ODE} with initial condition $\bs{x}_0$ and at time $t$, we have

\be
\begin{aligned}
\bs{\varphi}(\bs{x}_p,t) &= \bs{\varphi}(\bs{x}_p,t+T), \ \ \ \ \ \ T\in\mathbb{Q}^{+}, \\
\bs{M}^k(\tilde{\bs{x}}_p) &= \tilde{\bs{x}}_p, \ \ \ \ \ \ \ \  \ \ \ \ \ \ \ \ \ \ \ k\in\mathbb{N},
\label{eq:periodic_orbit}
\end{aligned}
\ee

\noindent where $T$ and $k$ are the \emph{periods} of the orbits. The periodic orbits are then given by $\bs{\varphi}(\bs{x}_p,t)$ for $t\in[0,T]$ in the continuous system and by $\{\tilde{\bs{x}}_p,\bs{M}(\tilde{\bs{x}}_p),\dots,\bs{M}^{k-1}(\tilde{\bs{x}}_p)\}\equiv\{\tilde{\bs{x}}_0,\tilde{\bs{x}}_{1},\dots,\tilde{\bs{x}}_{k-1}\}$ in the discrete system.

Both these types of orbits are important because they determine the dynamical properties on their neighborhood \cite{Alligood2012}. Orbits near a \emph{stable equilibrium} or a \emph{stable periodic orbit} stay near these for all time, while orbits near an \emph{unstable equilibrium} or an \emph{unstable periodic orbit} will \emph{typically} move away from it. To determine one's stability is therefore necessary and, in order to do so, we carry out a local linear analysis.

For discrete systems, we analyze the Jacobian of the map $\bs{DM}$ for fixed points, and the Jacobian of the iterated map $\bs{D}\bs{M}^k$ for a periodic orbit of period $k$. By calculating the \emph{eigenvalues} and \emph{eigenvectors} of these matrices, we can define the stable and unstable subspaces around these points \cite{Alligood2012}.

For autonomous continuous systems, where $\bs{F}$ does not depend on time, the Jacobian of the vector field $\bs{DF}$ is used to analyze the stability properties of an equilibrium point, since it gives a good first approximation for the vector field expanded in a Taylor series around this point \cite{Ozorio1990book}. For a periodic orbit in such systems, we have to analyze the \emph{monodromy matrix} associated with this orbit \cite{Meyer2008}.

Let us consider a solution $\bs{\varphi}(\bs{x}_p+\delta\bs{x}_p,t)$ that starts in the neighborhood of a periodic solution $\bs{\varphi}(\bs{x}_p,t)$. The displacement between these solutions over time is described, to first order, by the \emph{state transition matrix} $\Phi(t)$ \cite{Mireles2006}, which is given by

\be
\Phi(t)=
\left(
\ba{ccc}
\dfrac{\partial \varphi_1}{\partial x_1} & \hdots & \dfrac{\partial \varphi_1}{\partial x_N}\\
\vdots & \ddots & \vdots\\
\dfrac{\partial \varphi_N}{\partial x_1} & \hdots & \dfrac{\partial \varphi_N}{\partial x_N}\\
\ea
\right),
\label{eq:transition_matrix}
\ee

\noindent with $\bs{\varphi}=(\varphi_1,\dots,\varphi_N)$ and $\bs{x}=(x_1,\dots,x_N)$. The monodromy matrix $\mathcal{M}$ is then given by the state transition matrix evaluated after one period $T$ of the periodic orbit, 

\be
\mathcal{M}=\Phi(T),
\label{eq:monodromy_matrix}
\ee

\noindent and it defines a local mapping that characterizes the behavior of small perturbations near the periodic orbit  \cite{Koon2008,Mailybaev2019}

\be
\delta\bs{x}_{n+1}=\mathcal{M}\delta\bs{x}_n.
\label{eq:monodromy_matrix_map}
\ee

If we differentiate Eq.~\eqref{eq:transition_matrix} with respect to time and consider the autonomous form of Eq.~\eqref{eq:ODE}, we can show that the time evolution of the transition matrix $\Phi$ is given by

\be
\bal
\dfrac{d\Phi}{dt}&=J\Phi,\\
\Phi(t_0)&=\mathbb{I},\\
\eal
\label{eq:transtition_matrix_evolution}
\ee

\noindent where $J\equiv\bs{DF}$ and $\mathbb{I}$ stands for the identity matrix.

Therefore, after determining $\bs{\varphi}(\bs{x}_p,t)$, we can obtain $\mathcal{M}$ by integrating Eqs.~\eqref{eq:transtition_matrix_evolution} for one period $T$. However, it is important to note that the Jacobian of the field $J$ depends on the state of the system, making it necessary for us to calculate the evolution of the periodic orbit along with the evolution of the transition matrix. In practice, we reduce the computational cost of this calculation by writing $\Phi$ in a vector form, appending it to the state vector $\bs{x}$, and integrating both using the same numerical routine.

The stability of equilibria and periodic orbits, which we have discussed so far, is analyzed considering the system with a given set of parameters. Another type of stability, that is going to be important for us, is called \emph{structural stability} and it concerns what happens to the system when we slightly change said parameters. In this context, there can exist \emph{bifurcations}, where periodic orbits appear, disappear or change stability, for example \cite{Contopoulos2004}.

\subsubsection{Periodic orbit determination}

In dynamical systems with dissipative terms, a periodic orbit may be the final state of a set of orbits in the system either forward or backward in time. If this is the case, the determination of such periodic orbits involve only calculating the time evolution of a trajectory in this set. In Hamiltonian systems, where volumes are preserved in the phase space, this situation is not possible and numerical methods are usually necessary for the determination of periodic orbits.

For discrete systems, computation of periodic orbits generally involve solving Eq.~\eqref{eq:periodic_orbit} using some variation of the \emph{Newton's method} \cite{Miller2000}. For continuous-time systems, there are methods which also involve minimization procedures such as the \emph{Levenberg-Marquardt} \cite{Marquardt1963, Raphaldini2020}, or a discretized version of the monodromy matrix \cite{Baranger1988,Simonovic1999}. All these methods are efficient and need a good initial guess to work.

For the circular restricted three-body problem, which is used as a model for the Earth-Moon system, there are more sophisticated methods such as expanding the system's pseudo-potential in \emph{Legendre polynomials} and finding a formal series solution for certain families of periodic orbits, or using the Hamiltonian formalism of the problem to describe it using \emph{normal forms} with higher-order terms \cite{Gomez2001,Koon2008}. These methods, however, are harder to implement and demand higher computational time. 

In this work, we use the method described in Ref.~\cite{Mireles2006}, which explores a symmetry in the solutions of the circular restricted three-body problem in order to find both stable and unstable periodic orbits in a suitable surface of section. To facilitate the comprehension of the method, we will explain it in details in Sec.~\ref{sec:phase_space}, only after the Earth-Moon system is introduced.

\section{Poincaré map}
\label{sec:poincare_map}

\begin{figure}[h!]
\centering
\includegraphics[scale=0.6]{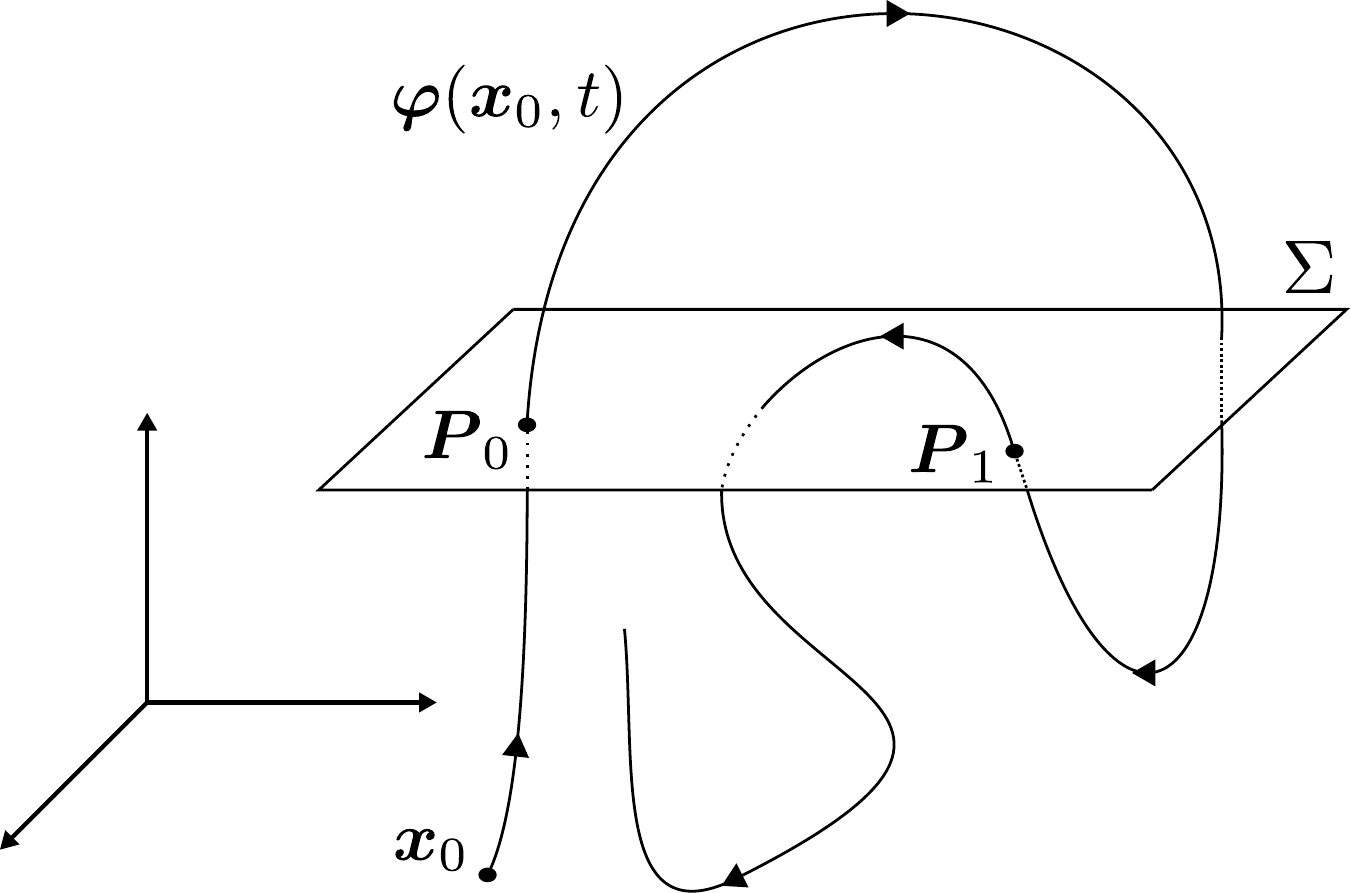}
\caption[Poincaré map construction]{Construction of a Poincaré map for a three-dimensional flow. All the upward piercings of the continuous-time trajectory $\bs{\varphi}(\bs{x}_0,t)$ with the two-dimensional surface of section $\Sigma$ are recorded, and the map $\bs{M}$ is implicitly defined by $\bs{M}(\bs{P}_i)=\bs{P}_{i+1}$, for $i=0,1,\dots$}
\label{fig:poincare_map}
\end{figure}

The dynamical structure of a $N$-dimensional continuous-time system can be captured by a $(N-1)$-dimensional mapping. This can be achieved by considering the intersections of a continuous-time trajectory, in a given direction, with a predefined \emph{surface of section}, as is shown in Fig.~\ref{fig:poincare_map}. A map can then be implicitly defined as a function that images one intersection to the next. The discrete system resulting from this procedure is called a \emph{Poincaré map}.

The construction of a Poincaré map is a powerful tool for analyzing dynamical systems as the generated mapping reflects the dynamical structure of the continuous system by preserving the qualitative properties of the flow, such as periodic and chaotic behavior. A periodic solution, in particular, is represented by a fixed point or a periodic orbit on the map, and the stability of the discrete trajectory is the same as that of its continuous counterpart \cite{Koon2008}.

Poincaré maps derived from Hamiltonian flows maintain the volume-preserving character of the corresponding continuous-time system \cite{Ozorio1990book}. For example, three-dimensional Hamiltonian flows, or Hamiltonian flows that are restricted to a three-dimensional space, generate a two-dimensional area-preserving Poincaré map. This will be the case for the Earth-Moon system.

\subsubsection{Numerical considerations}

In order to obtain one point on a Poincaré map, we need to evolve the system until the solution crosses our chosen surface of section. This means that the error propagated in each step of the numerical integration scheme yields an error on the map, which is specially concerning for orbits integrated for a long period of time. Since our investigation is carried out by analyzing the system on a surface of section, this issue reinforces the importance of selecting a high-precision integration method.

The procedure we use for determining the intersection between a trajectory and the surface of section goes as follows. While integrating the system, we verify if two consecutive points of the trajectory fall on different sides of the surface of section $\Sigma$. If this happens, we select the backward point $\bs{x}_{n}$ and calculate a new point $\bs{x}_{n+1}$ using Eq.~\eqref{eq:RK} but with a new step-size which is half of the previous one, $h/2$. We then repeat this process until the difference between $\bs{x}_{n+1}$ and $\Sigma$ is lower than $10^{-10}$.

The method described above is simple to implement and sufficiently precise. We have to be careful, though, and keep the step-size higher than $10^{-15}$ in order to avoid floating-point error. Furthermore, after calculating the intersection, we have to continue the numerical integration of the trajectory using the original backward point, so as not to insert the method error into the orbit calculation.
% 
% Finally, we have also tested computing Poincaré maps by changing the dependent variable according to the definition of $\Sigma$, and rewritting the vector field accordingly \cite{Henon1982}. Even though this method enables a intersection estimation with just one new step, it was observed to be not efficient in terms of preserving .

\vfill

\section{Integrability and KAM theory}
\label{sec:KAM}

A \emph{Hamiltonian system} is one that can be completely described by a \emph{Hamiltonian function} $\mathcal{H}(\bs{q},\bs{p},t)$, which is expressed in terms of the \emph{canonically conjugate variables} $(\bs{q},\bs{p})\in\mathbb{R}^{2D}$. The dimensionality $D$ of these variables is the number of degrees of freedom in the system \cite{Ott2002}.

\subsubsection{Integrable Hamiltonian systems}

We call \emph{constant of motion} a function $f$ of the system's dynamical variables that stays constant as the system evolves in time. The existence of a constant of motion implies that the system's dynamics is restricted to a ($2D-1$)-dimensional surface defined by $f=k$, where $k$ is constant. For example, in an autonomous Hamiltonian system, the Hamiltonian function $\mathcal{H}(\bs{q},\bs{p})$ is a constant of motion that restricts the system to a surface $\mathcal{H}=k$, for any $(\bs{q},\bs{p})$ in the phase space. 

In practice, a constant of motion also serves as a parameter from which we can estimate a precision for the numerical integration of the equations of motion. This is very useful since it is an error estimation which is based on the system's dynamics and not on the method itself.

An autonomous Hamiltonian system is said to be \emph{integrable} if it has $D$ independent constants of motion $f_i$. In this case, the dynamics of the system occurs in a $D$-dimensional surface defined by $f_i=k_i$, $i=1,\dots,D$. It can be shown that the topology of such a surface is that of a $D$-dimensional torus \cite{Ozorio1990book}. It is then possible to analytically describe the motion of any orbit in this type of system.

For a two-degrees-of-freedom case, the motion is given in a two-dimensional torus, which can be parametrized by two frequencies, one for each degree of freedom. If the ratio between the frequencies is rational, the orbit is periodic. Conversely, if the frequencies are incommensurable, the trajectory fills the entire toroidal surface \cite{Lichtenberg1992}.

\subsubsection{Near-integrable Hamiltonian systems}

The \emph{KAM theory} states that, if we perturb the Hamiltonian of an integrable system with a small volume-preserving perturbation, there still exists a finite fraction of \emph{regular} trajectories, those which are described by a KAM torus. The other fraction of trajectories describe \emph{chaotic} motion, i.e., they are not periodic and have sensitive dependence on initial conditions. When analyzed in a surface of section, the regular trajectories form invariant circles, while the chaotic ones densely fill a finite region of this surface \cite{Lichtenberg1992,Meiss1992}. This is the dynamical scenario of a \emph{near-integrable} Hamiltonian system, which is composed of regions of regular motion, or stability, along with chaotic areas in the phase space.

Given a small volume-preserving perturbation, some invariant curves are more easily destroyed than others, and bifurcate into a group of elliptic (stable) and hyperbolic (unstable) periodic orbits, following the \emph{Poincaré-Birkhoff} theorem \cite{Ott2002}. When this happens to the last KAM curve of the invariant family surrounding an elliptic point, a resonant island chain is created around the regular region and in direct contact with the chaotic sea. This is an important mechanism because the unstable periodic orbits created in this bifurcation, and their associated invariant manifolds, influence the chaotic motion around the regular region.

Orbits in the Hamiltonian chaotic sea may linger and concentrate in certain regions of the phase space, specially around stability regions. Such phenomenon is called \emph{stickiness}, and it can be associated with invariant manifolds in the system \cite{Contopoulos2004,Contopoulos2010}. The general behavior of the stickiness effect will be an important component describing the dynamical properties of the Earth-Moon system.

\subsubsection{Monodromy matrix properties}

The monodromy matrix of a periodic orbit in an autonomous Hamiltonian flow has two unity eigenvalues. The first one is associated with the flow direction, which is tangent to the orbit, and the second one is related to the fact that the Hamiltonian function is a constant of motion \cite{Koon2008}. The other eigenvalues may determine the stability of the periodic orbit in the following manner.

A stable periodic orbit has pairs of eigenvalues that are complex conjugate to one another. Conversely, an unstable periodic orbit has at least one pair of real eigenvalues, where one is the inverse of the other \cite{Meyer2008}. While the eigenvalue $\lambda>1$ defines an unstable direction for the linearized system around the orbit, the eigenvalue $1/\lambda$ defines the stable direction. These eigenvectors will later be used for calculating the invariant manifolds associated with an unstable periodic orbit.

Let us now restrict ourselves to two degrees of freedom. In this case, we can analyze the stability of a periodic orbit by looking at the \emph{trace} of the monodromy matrix $\mathcal{M}$. If $0<Tr(\mathcal{M})<4$, the orbit is stable, while $Tr(\mathcal{H})<0$ or $Tr(\mathcal{H})>4$ implies an unstable orbit \cite{Aguiar1987}. Therefore, the sum of the monodromy matrix eigenvalues may indicate a \emph{periodic orbit bifurcation} if it approaches either $0$ or $4$ as we change the value of the system's constant of motion.

\subsubsection{Newtonian and Lagrangian formalisms}

Before closing this chapter, it is important that we make an observation on the coordinate system that we use during our work.

We derive the equations of motion for a small body in the Earth-Moon system from a Newtonian approach, which describes the system in terms of the variables position and velocity $(\bs{x},\dot{\bs{x}})$. This system is Hamiltonian in the sense that there exists a coordinate transformation from which we can obtain a set of canonically conjugate variables $(\bs{q},\bs{p})$ and a corresponding Hamiltonian function. However, we keep the former set of variables throughout our investigation, which is natural for problems that may involve real-life applications.

The resulting equations of motion are a set of ordinary differential equations with an incompressible vector field ($\bs{\nabla \cdot F}=0$). Therefore, the system is still volume-preserving in this coordinate system and the scenario described for the phase space of Hamiltonian flows still holds in our situation \cite{Salamon1989}. We could also have arrived at the same equations of motion by defining a \emph{Lagrangian function}, which would be associated to the Hamiltonian function through a \emph{Legendre transformation} \cite{Koon2008}.

%\section{Closed and open systems}
%\label{sec:open}
%
%\begin{itemize}
%\item Transitivity - this will be important for the area-preserving Hénon map and when we talk about escape in Sec. X
%\end{itemize}

\chapter{Invariant manifolds}
\label{ch:invariant_manifolds}

%\be\nn
%H(x, p, t) = \dfrac{p^2}{2} + \sum_{n=-\infty}^{\infty} A \sin{(x - 2\pi n t + \phi_n)}
%\ee

%In dynamical systems theory, an invariant manifold is a set of trajectories that asymptotically converge to an unstable fixed point, an unstable equilibrium point, or an unstable periodic orbit (\gls{UPO}). When the convergence happens forward in time, we call such set \emph{stable}. Conversely, if the orbits converge backward in time, i.e., if the orbits originate from the point they are associated with, we call such set \emph{unstable}.

In dynamical systems theory, the set of states whose temporal evolution asymptotically converges to an unstable fixed point, an unstable equilibrium point, or an unstable periodic orbit (\gls{UPO}), is called the \emph{invariant manifold of said point or orbit}. When the convergence happens forward in time, we call such set \emph{stable manifold}. Conversely, if the trajectories converge backward in time, i.e., if they originate from the point, or the orbit, they are associated with, we call such set \emph{unstable manifold}.

The stable and unstable manifolds are geometrical structures related to the stable and unstable subspaces of an unstable point, or unstable periodic orbit, respectively. These sets are \emph{manifolds} since they are locally homeomorphic to the Euclidean space and they are \emph{invariant} because the trajectory formed by anyone of their elements, which is obtained by solving Eq.~\eqref{eq:ODE} or Eq.~\eqref{eq:map}, also belongs to the corresponding set by definition.

In this chapter, we define these geometrical structures, show how to numerically calculate them, and discuss their relationship to the system's dynamical properties. We first describe one-dimensional invariant manifolds in two-dimensional area-preserving maps, and we analyze a suitable version of the Hénon map as an example. Later, we consider two-dimensional manifolds in autonomous two-degrees-of-freedom Hamiltonian systems, which will be the case for the Earth-Moon system.

\section{One-dimensional manifolds}
\label{sec:1d_manifolds}

Let $\bs{M}$ be a two-dimensional bijection, with both $\bs{M}$ and $\bs{M}^{-1}$ differentiable, and let $\bs{p}$ be a \emph{saddle}, i.e., an unstable fixed point, in the mapping $\bs{M}$. We define the \emph{stable manifold} $W^s$ and the \emph{unstable manifold} $W^u$ associated with $\bs{p}$ as \cite{Alligood2012}

\be
\begin{aligned}
W^s(\bs{p})&=\{\bs{x}\in U\subset \mathbb{R}^2 \ | \ \bs{M}^{n}(\bs{x})\to\bs{p} \ \text{as} \ n\to\infty\},\\
W^u(\bs{p})&=\{\bs{x}\in U\subset \mathbb{R}^2 \ | \ \bs{M}^{-n}(\bs{x})\to\bs{p} \ \text{as} \ n\to\infty\}.\\
\end{aligned}
\label{eq:manifold_1d_definition}
\ee

\begin{figure}[h!]
\centering
\subfloat[Integrable]{\includegraphics[scale=0.55]{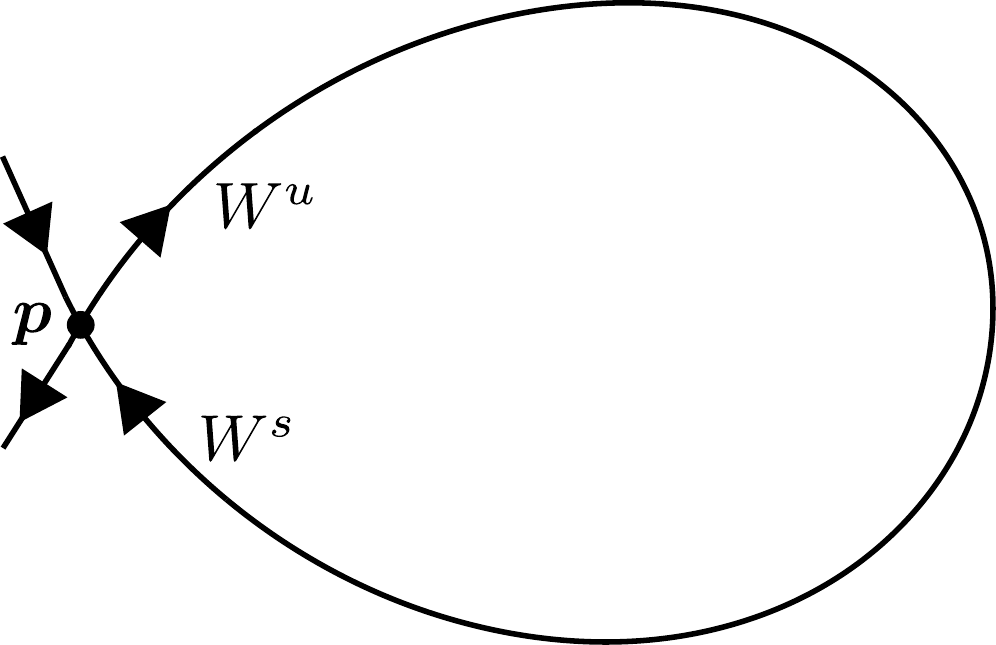}
\label{subfig:separatrix}} \ \ \ \ \ 
\subfloat[Nonintegrable]{\includegraphics[scale=0.55]{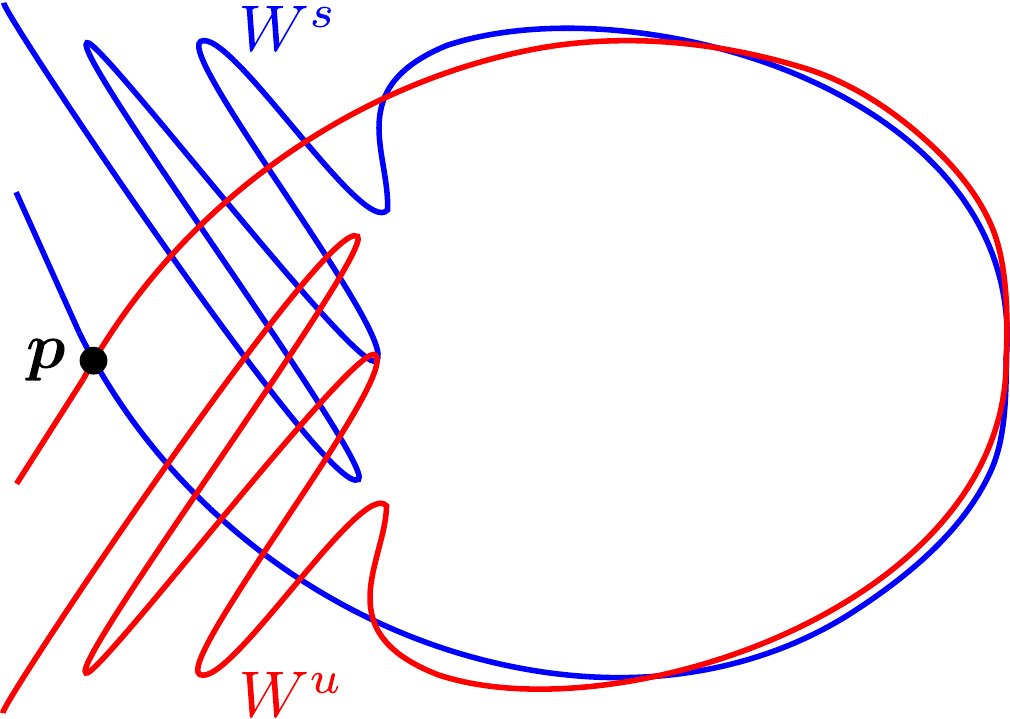}
\label{subfig:manifolds}}
\caption[One-dimensional invariant manifolds]{Invariant manifolds $W^u$ and $W^s$ associated with a saddle $\bs{p}$ for an \protect\subref{subfig:separatrix} integrable and a \protect\subref{subfig:manifolds} non-integrable case.}
\label{fig:1d_manifolds_sketch}
\end{figure}

It can be shown that the invariant manifolds $W^s(\bs{p})$ and $W^u(\bs{p})$ are, in this case, unidimensional curves. Furthermore, these geometrical structures locally follow the direction of the eigenvectors of the Jacobian matrix evaluated at the saddle $\bs{DM}(\bs{p})$.\footnote{These results are given by the \emph{Stable Manifold Theorem}, see Ref.~\cite{Alligood2012}.} The stable manifold is tangent to the eigenvector in the stable subspace, while the unstable manifold is tangent to the unstable eigenvector, with each of these having two branches. In Fig.~\ref{fig:1d_manifolds_sketch}, we present the invariant manifolds associated with a saddle $\bs{p}$ for an integrable and a non-integrable case.

In the integrable case, Fig.~\ref{subfig:separatrix}, both manifolds emerge from the saddle and smoothly reconnects to the unstable point. These curves work as a separatrix, which divides the space into regions with different dynamical properties. In the non-integrable case, Fig.~\ref{subfig:manifolds}, the manifolds oscillate as they return to the initial point, forming \emph{lobes}. The closer they are to $\bs{p}$ the longer are the lobes. This situation leads to crossings between the manifolds, which compose an invariant set of chaotic orbits \cite{Smale1967,Ott2002}. The trajectories near such crossings then follow their dynamics and also behave chaotically, being this the main element for chaos in the system \cite{Reichl2004}.

The crossings in Fig.~\ref{subfig:manifolds} are called \emph{homoclinic} for these happen between manifolds associated with the same saddle, contrary to the \emph{heteroclinic} case where the manifolds of different unstable points intersect each other. Due to the described scenario, the invariant manifolds are sometimes regarded as the skeleton of the system's underlying dynamics. Therefore, it is of paramount importance to determine how these geometrical structures are spatially organized.

\begin{figure}[h!]
\centering
\includegraphics[scale=0.55]{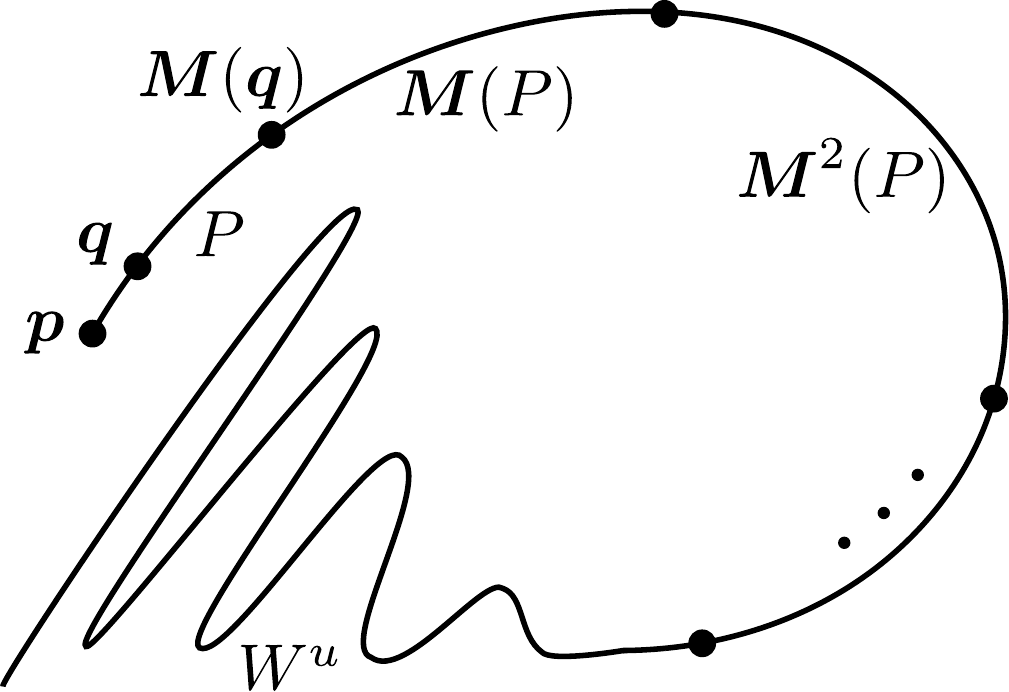}
\caption[The primary segment method]{The primary segment method for tracing invariant manifolds. After calculating a primary segment $P=[\bs{q},\bs{M}(\bs{q})]\subset W^u$, we can map it to determine the rest of the manifold $W^u$ associated with the saddle $\bs{p}$.}
\label{fig:primary_segment_method}
\end{figure}

Numerical tools for determining invariant manifolds are necessary since, generally, we are not able to obtain these curves analytically \cite{Hobson1993}. One of the main ideas consists in choosing a piece of the invariant manifold, called a \emph{primary segment}, and using it to trace the rest of the manifold. In Fig.~\ref{fig:primary_segment_method}, we sketch this procedure for a branch of the unstable manifold. The primary segment $P$ is formed by the points on the manifold between ${\bs{q}}$ and its image under the map ${\bs{M}}({\bs{q}})$. By construction, the images and preimages of P under ${\bs{M}}$ compose the invariant manifold, i.e., $W^u({\bs{p}})=\cup_{i=-\infty}^{\infty}{\bs{M}}^i(P)$.

In practice, we calculate $\bs{DM}(\bs{p})$ and its eigenvectors, either analytically or numerically, and we choose a primary segment very close to the saddle and in the direction of the eigenvector of interest. Next, we select a few points on the segment and iterate them with the map and its inverse. The orbits generated will then trace the invariant manifold to a certain precision. By increasing the number of initial conditions, we increase the curve precision, but also the computational cost.

The definition of invariant manifolds, Eqs.~\eqref{eq:manifold_1d_definition}, can be readily extended to unstable periodic orbits. Let $\gamma=\{\bs{p}_1, \bs{p}_2,\dots, \bs{p}_m\}$ be an UPO of period $m$ of the map $\bs{M}$. The invariant manifolds associated with $\gamma$ are given by \cite{Alligood2012}

\be
\begin{aligned}
W^s(\gamma)&=\{\bs{x}\in U\subset \mathbb{R}^2 \ | \ \bs{M}^{n}(\bs{x})\to\bs{M}^{n}(\bs{p}_i) \ \text{as} \ n\to\infty, \ i=1,\dots, m \},\\
W^u(\gamma)&=\{\bs{x}\in U\subset \mathbb{R}^2 \ | \ \bs{M}^{-n}(\bs{x})\to\bs{M}^{-n}(\bs{p}_i) \ \text{as} \ n\to\infty, \ i=1,\dots, m \}.\\
\end{aligned}
\label{eq:manifold_1d_upo}
\ee

The manifolds locally follow the direction of the eigenvectors of $\bs{DM}^m(\bs{p}_i)$ in this case, which is natural since every $\bs{p}_i$ is a fixed point of the map $\bs{M}^m$. In practice, we calculate a primary segment for each member of $\gamma$ and follow the procedure described before to trace these curves.

\subsubsection{The area-preserving Hénon map}

In order to illustrate some of the main dynamical aspects of one-dimensional invariant manifolds, we analyze the \emph{Hénon map} \cite{Henon1976}. This map was initially presented as a simple dynamical system which contains a set of solutions that converge to a \emph{chaotic attractor}. Since then, it has become an archetype of two-dimensional discrete systems. Here, we study an area-preserving version of the map which is given by

\be
\bs{H}:
\left\{
\begin{aligned}
x_{n+1} &= a-x_n^2+y_n,\\
y_{n+1} &= x_n,
\end{aligned}\right.
\label{eq:henon}
\ee

\noindent and whose inverse is given by

\be
\bs{H}^{-1}:
\left\{
\begin{aligned}
x_{n-1} &= y_n,\\
y_{n-1} &= -a+y_n^2+x_n.
\end{aligned}\right.
\label{eq:henon_inverse}
\ee

\begin{figure}[h!]
   \centering
   \includegraphics[width=0.94\textwidth]{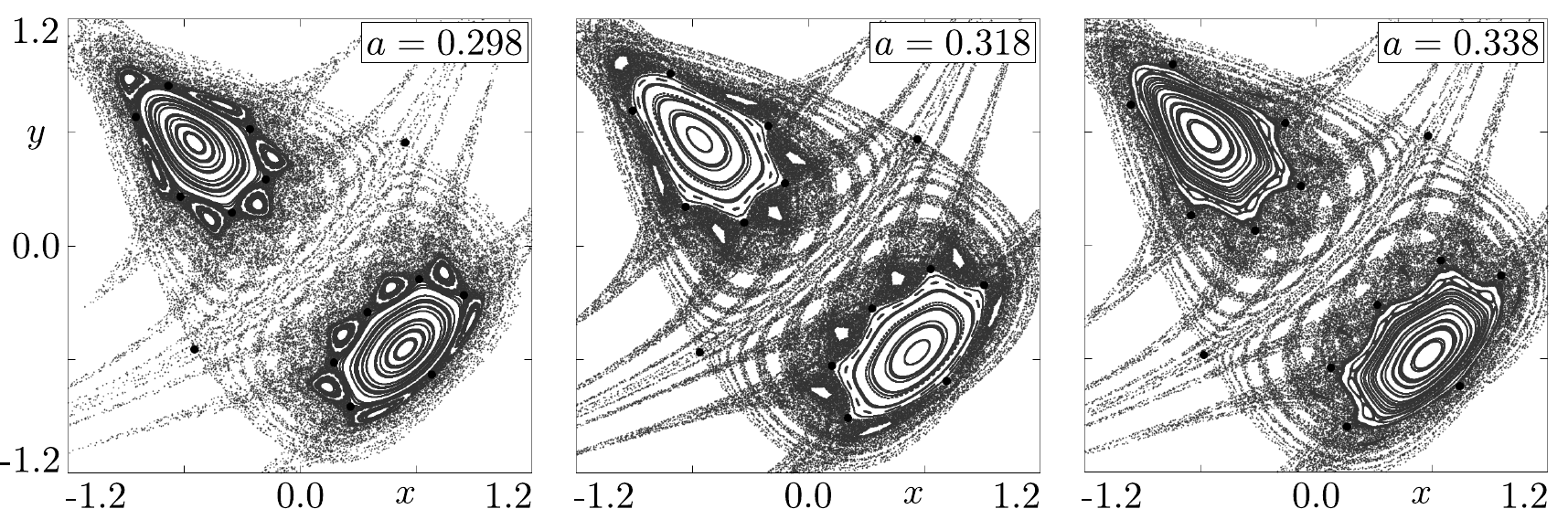}
   \caption[Phase space of the area-preserving H\'enon map]{Phase space $x$-$y$ of the area-preserving H\'enon map $\bs{H}$. The small black circles mark two fixed saddles at $x=y=\pm\sqrt{a}$, and a periodic saddle of period 12 around the invariant curves.}
   \label{fig:henon_phase_space}
\end{figure}

The Hénon mapping is an open system. Most of the initial conditions leads to regular solutions that converge to $x=-\infty$ when iterated under $\bs{H}$, and to $y=+\infty$ when iterated with its inverse, $\bs{H}^{-1}$. There is, however, a region of the system that is composed by invariant stable solutions, near which the unbounded orbits experiment a chaotic transient. In Fig.~\ref{fig:henon_phase_space}, we present the system's phase space $x$-$y$ for three different values of the parameter $a$.

%The Hénon mapping is an open system. Most of the initial conditions leads to regular solutions that converge to $(-\infty,-\infty)$ when iterated under $\bs{H}$, and to $(+\infty,+\infty)$ when iterated with its inverse, $\bs{H}^{-1}$. There is, however, a region of the system composed by invariant stable solutions, near which the unbounded orbits experiment a chaotic transient. In Fig.~\ref{fig:henon_phase_space}, we present the system's phase space $x$-$y$ for three different values of the parameter $a$.

Equations.~\eqref{eq:henon} have two fixed points, which are located at $x=y=\pm\sqrt{a}$, both of them being saddles. For the chosen range of parameter values, $0.298\leq a \leq 0.338$, there are also a period-2 center, i.e., a stable periodic orbit of period 2, which is surrounded by invariant curves, and a period-12 UPO that is located around the regular region. The UPO is related to a resonant island chain which detaches from the regular region and is destroyed as $a$ increases.

There are two main features we observe from Fig.~\ref{fig:henon_phase_space}. The first one is a higher concentration of orbits, \emph{stickiness}, around the resonant island chain. The second one is the shape of the chaotic sea, which is mostly confined to a definite region of the phase space, and also shows an intricate pattern that is specially visible between the regular regions. We now proceed to calculate the invariant manifolds associated with both fixed saddles and the period-12 UPO in order to show that the aforementioned phase space features have a close relationship with the spatial distribution of these geometrical structures.

The primary segment method can trace invariant manifolds with a small error, but there are some improvements that can be made to enhance its computational efficiency. Since these curves have complex structures with elongated and bended sections, there are some parts of the manifolds that may require more or less points for a good resolution.
We can then start with a small number of initial conditions on the primary segment, and later use some criteria, such as the manifold's local curvature, to add or remove points as needed \cite{Hobson1993}. We can also interpolate the images of the primary segment as we calculate them \cite{Ciro2018}.

\begin{figure}[h!]
   \centering
   \includegraphics[width=0.94\textwidth]{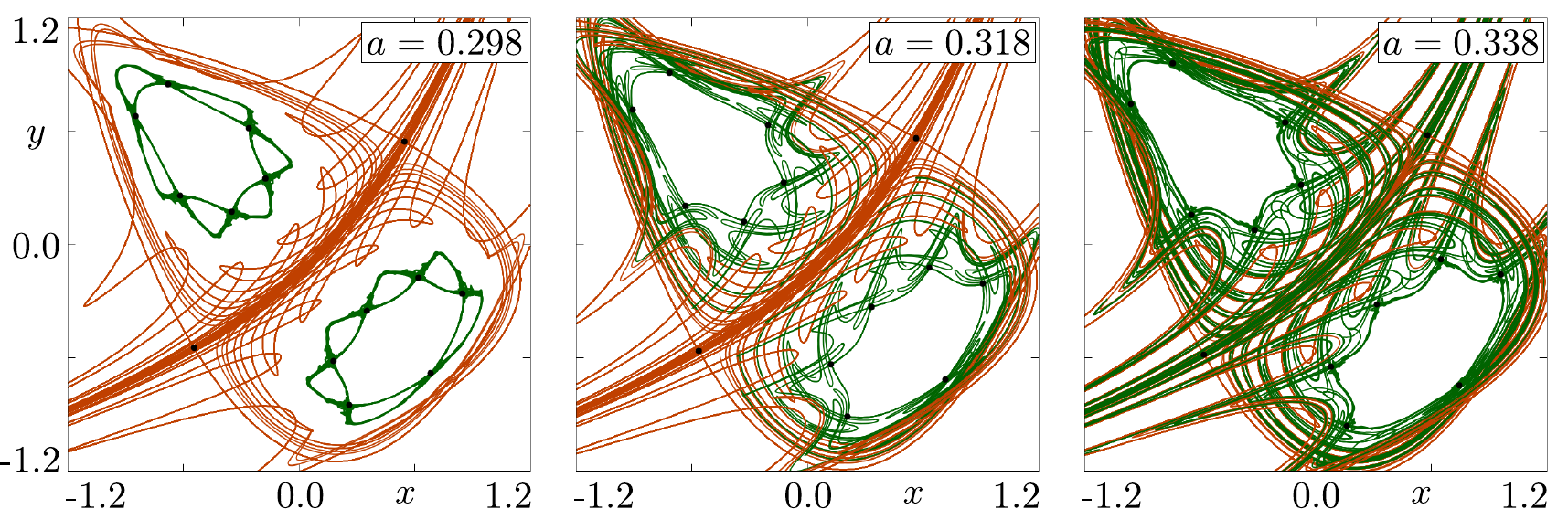}
   \caption[Invariant manifolds in the area-preserving H\'enon map]{Invariant manifolds of the fixed points (orange) and the period-12 orbit (green), traced in the phase space $x$-$y$ for different values of $a$. As the control parameter increases, these curves become more entangled.}
   \label{fig:henon_manifolds}
\end{figure}

The results are presented in Fig.~\ref{fig:henon_manifolds}, where the invariant manifolds associated with the fixed saddles, both stable and unstable, are colored orange, and the ones associated with the period-12 UPO are in green. These curves were traced using the open-source software Automan\footnote{Available at \url{http://yorke.if.usp.br/OscilControlData/AUTOMAN/}.}, which follows the primary segment method as described in Ref.~\cite{Ciro2018}. By comparison to Fig.~\ref{fig:henon_phase_space}, we can readily observe the role of each structure in the phase space configuration. While the fixed saddles' manifolds delimit the region where the interesting dynamics takes place, the manifolds associated with the UPO are related to the stickiness effect.

Another aspect observable from Fig.~\ref{fig:henon_manifolds} is the interaction between both groups of manifolds. Initially, for $a=0.298$, the orange and green curves are essentially isolated from each other. They do eventually cross, given that they are immersed in the same chaotic sea, but only after a large number of lobes were formed by both manifolds. Later, for $a=0.318$, the crossings start sooner and the manifolds become entangled as they spread across the phase space. Finally, for $a=0.338$, both curves seem similarly organized, as the manifolds with the same type of stability cannot cross each other and, therefore, these must evolve parallel to each other as they compete for space.

To understand the influence of these geometrical structures in the transport properties of the system, we calculate the \emph{transit time} profile for the different values of the control parameter. In order to do so, we first define a large box that contains our depicted domain. Then, we count how many iterations it takes for an initial condition to reach this box both under $\bs{H}$, denoted by $n_f$, and under $\bs{H}^{-1}$, denoted by $n_b$. Lastly, we calculate the \emph{transit time} associated with each initial condition, which is given by $n_f+n_b$.

\begin{figure}[h!]
   \centering
   \includegraphics[width=0.98\textwidth]{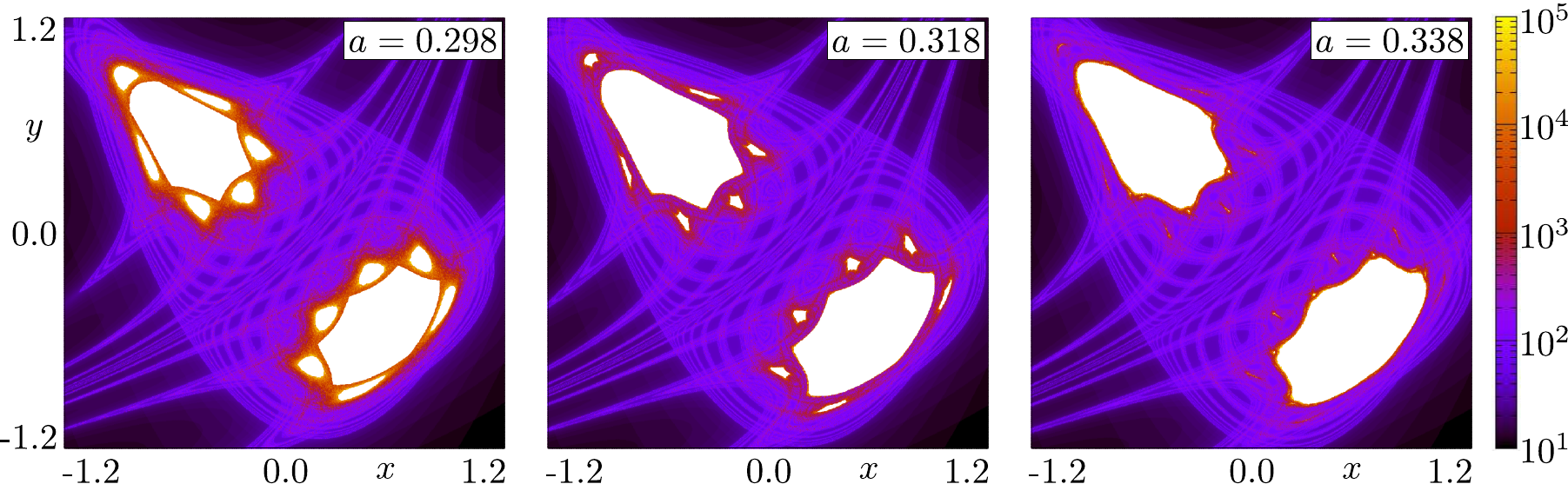}
   \caption[Transit time profile for the area-preserving H\'enon map]{Transit time profile for the area-preserving H\'enon map $\bs{H}$ in a logarithmic scale, measured in the phase space $x$-$y$ for different values of $a$. The white areas contain bounded orbits, which are not considered for the analysis.}
   \label{fig:henon_transit}
\end{figure}

The transit time is a measure of how long it takes for an unbounded orbit to enter and exit a region of interest. In Fig.~\ref{fig:henon_transit}, we show the transit time profiles for a grid of $1000\times1000$ initial conditions in $[-1.2,1.2]\times[-1.2,1.2]$, whereas the outer hit box chosen was $[-1000,1000]\times[-1000,1000]$. We observe that the orbits with larger transit times are concentrated around the period-12 island chain, as expected. As the parameter $a$ is increased, the islands get smaller and the number of orbits with longer transit times decrease. The self-similar pattern formed in the chaotic sea is also highlighted, indicating the presence of an underlying geometrical structure which divides the phase space into long and short transit time orbits.

By comparing Figs.~\ref{fig:henon_manifolds}~and~\ref{fig:henon_transit}, we are able to conclude that the fixed saddle manifolds are responsible for the transport of orbits into and out of the chaotic region of the phase space. Meanwhile, the invariant manifolds associated with the period-12 UPO are responsible for temporarily trapping the orbits near the regular regions. As we vary the control parameter, these geometrical structures become entangled, leading to a more uniform transit time profile.

As we mentioned, intersections between invariant manifolds are related to the emergence of chaos in conservative systems, and, as we have seen, to the chaotic transport between different regions of the phase space. By analyzing the homoclinic intersections between the manifolds of the period-12 UPO, and the heteroclinic intersections between the UPO manifolds and the ones associated with the fixed saddles, we can also show that the phase space distribution of these two sets present similar properties as the control parameter is increased \cite{deOliveira2020dynamical}.

The results observed here should hold qualitatively for any range of $a$ in which there is an island chain being detached from the regular region and, consequently, a suitable UPO can be chosen for the analysis. The fixed saddles, in particular, are not created through a bifurcation, and the influence of their manifolds on the system's dynamical properties can be verified for any value of the control parameter.

\begin{figure}[h!]
   \centering
   \includegraphics[width=0.75\textwidth]{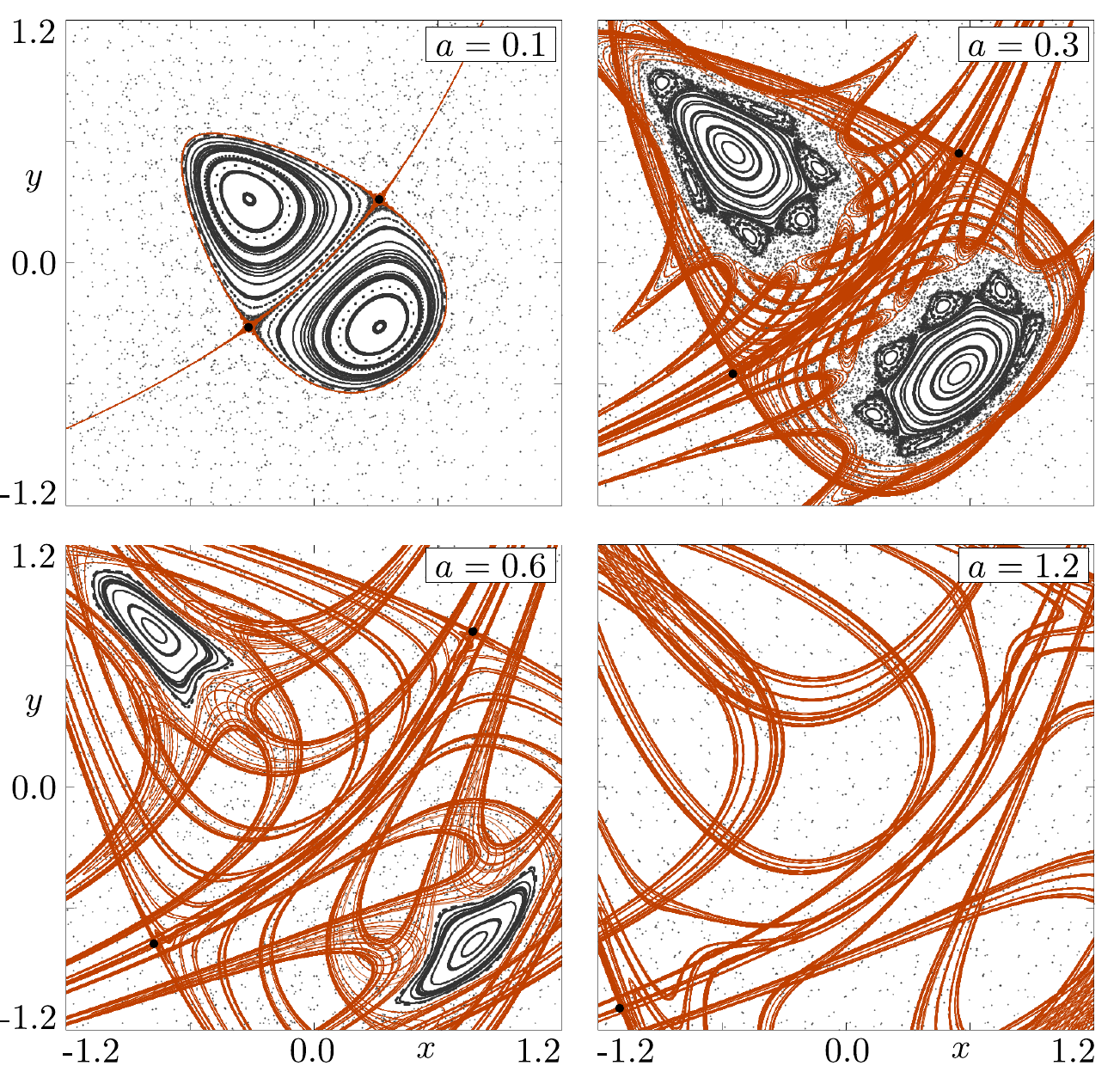}
   \caption[Destruction of regular solutions in the area-preserving H\'enon map]{Destruction of regular solutions in the area-preserving H\'enon map $\bs{H}$. As the control parameter $a$ is increased, the fixed saddles manifolds invade the stable regions in the phase space $x$-$y$, and transport these orbits to infinity.
   }
   \label{fig:henon_destruction}
\end{figure}

In Fig.~\ref{fig:henon_destruction}, we present the phase space $x$-$y$ of the Hénon map $\bs{H}$, along with the fixed saddles manifolds, for different values of the control parameter in the range $0.1<a<1.2$. Initially, the manifolds are very localized around the saddle points. As the parameter increases, they advance to a larger area of the phase space, invading the regions that contain regular solutions. After a solution is destroyed, the new orbit follows the dynamics of the manifolds towards infinity. Eventually, both regular regions disappear and there is no more transient chaos in the system.

A remark here is in order. The map $\bs{H}$ is area-preserving but, since $\det(\bs{DH})=-1$, it is not orientation-preserving, which means that it cannot be derived from a Hamiltonian flow \cite{Meiss1992}. Even though there is an orientation-preserving version of the Hénon map \cite{Devaney1984}, we chose $\bs{H}$ because it is a convenient system to clearly depict the relationship between system's dynamical properties and invariant manifolds spatial distribution. The concepts that were illustrated in this section, such as the idea that these geometrical structures can play different dynamical roles in the system, will also be observed in the Earth-Moon system.

\vfill

\section{Two-dimensional manifolds}
\label{sec:2d_manifolds}

Let us now consider an autonomous Hamiltonian system with two degrees of freedom, and let $\alpha$ be an unstable periodic orbit in this continuous-time system. As was discussed, the stability of $\alpha$ is determined by the eigenvalues of the monodromy matrix, which is calculated at a point $\tilde{\bs{p}}$ of the UPO.

The eigenvalues of the monodromy matrix are related to eigenvectors which locally determine an unstable and a stable direction. This means that there exist two sets of orbits, one that converges to $\tilde{\bs{p}}$ forward in time, and another that does it backward in time. By extending these local solutions to the entire space, and by following Eqs.~\eqref{eq:manifold_1d_definition}, we can define the invariant manifolds associated with $\tilde{\bs{p}}$ as

\be
\begin{aligned}
W^s(\tilde{\bs{p}})&=\{\bs{x}\in V\subset \mathbb{R}^4 \ | \ \bs{\varphi}(\bs{x},t)\to\tilde{\bs{p}} \ \text{as} \ t\to\infty\},\\
W^u(\tilde{\bs{p}})&=\{\bs{x}\in V\subset \mathbb{R}^4 \ | \ \bs{\varphi}(\bs{x},t)\to\tilde{\bs{p}} \ \text{as} \ t\to-\infty\},\\
\end{aligned}
\label{eq:manifold_2d_definition_local}
\ee

\noindent where $\bs{\varphi}(\bs{x},t)$ is a solution of the system's equations of motion with initial condition $\bs{x}$ and at time $t$. From now on, we will suppress the superscripts $s$ and $u$ for simplicity when there is no confusion.

The invariant manifolds associated with the unstable periodic orbit $\alpha$ is then given by the union of the invariant manifolds of each point of $\alpha$. Formally, we write

\be
W(\alpha)=\bigcup_{\tilde{\bs{p}}\in\alpha}W(\tilde{\bs{p}}).
\label{eq:manifold_2d_definition}
\ee

These geometrical structures are two-dimensional surfaces that are locally homeomorphic to cylinders \cite{Ozorio1990}. Due to the presence of a constant of motion in the system, its dynamics is restricted to a three-dimensional surface. Therefore, it is also useful to define the intersection between $W(\alpha)$ and a two-dimensional surface of section $\Sigma$. We denote such intersections by $\Gamma(\alpha)$, and we can naturally order them by following the tracing of the invariant manifolds and counting the crossings. We have

\be
\Gamma(\alpha)=W(\alpha)\cap\Sigma=\bigcup_{i=1}^{\infty}\Gamma_i(\alpha).
\label{eq:manifold_crossing}
\ee

\begin{figure}[h!]
\centering
\includegraphics[scale=0.65]{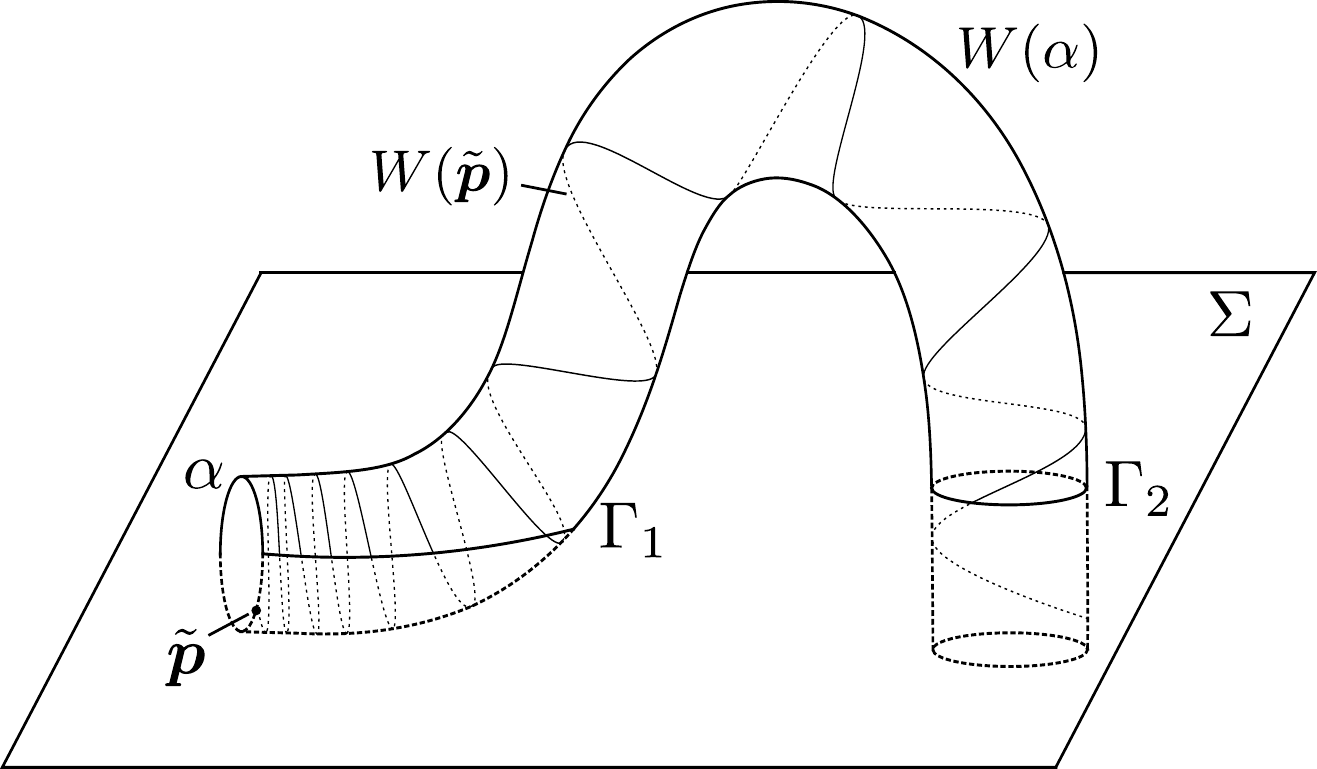}
\caption[Two-dimensional invariant manifold]{Invariant manifold $W(\alpha)$ of the UPO $\alpha$, along with a surface of section $\Sigma$. While the first intersection $\Gamma_1$ is a line, the second one $\Gamma_2$ is a closed curve. An one-dimensional manifold $W(\tilde{\bs{p}})$ associated with an arbitrary $\tilde{\bs{p}}\in\alpha$ is also shown. $W(\alpha)$ is the union of all $W(\tilde{\bs{p}})$.}
\label{fig:2d_manifolds_sketch}
\end{figure}

In Fig.~\ref{fig:2d_manifolds_sketch}, we sketch a branch of an invariant manifold associated with the UPO $\alpha$, along with a surface of section $\Sigma$. Here, $\alpha$ has an ellipse-like shape in the full space and forms a fixed saddle in the Poincaré map. We choose an arbitrary point $\tilde{\bs{p}}$ in the UPO, and depict one of the manifolds $W(\tilde{\bs{p}})$ that compose $W(\alpha)$. Close to the UPO, $W(\alpha)$ is transverse to $\Sigma$, resulting in a line $\Gamma_1$ on the map which is similar to the manifolds seen in two-dimensional maps. However, $W(\alpha)$ eventually disconnects from $\Sigma$, and re-intersects it transversely, resulting in a closed curve $\Gamma_2$ that is homeomorphic to a circle. In this case, the invariant manifolds form a discontinuous structure in the Poincaré map.

Hence, the topology of two-dimensional invariant manifolds in the Poincaré map depends on the chosen surface of section. If $\Sigma$ is always transverse to $W$, the intersection $\Gamma$ does have similar geometrical properties to the manifolds in planar maps. We will observe both situations, continuous and discontinuous manifolds, in our analysis for the planar Earth-Moon system.

We now describe a method to numerically compute $W(\alpha)$ \cite{Koon2008,Mireles2006}. First, we choose a point $\tilde{\bs{p}}_0$ on the unstable periodic orbit, and we calculate the monodromy matrix by integrating the equations of motion along with the transition matrix. Then, we numerically determine the stable and unstable eigenvectors, whose normalized versions we are going to denote by $\bs{e}_0^s$ and $\bs{e}_0^u$, respectively, and we choose a point $\bs{q}_0$ in the direction of such eigenvectors:

\be
\bs{q}_0 = \tilde{\bs{p}}_0 \pm \delta\bs{e}_0^{s,u},
\label{eq:ic_upo_man_2d}
\ee

\noindent where $\delta$ is a small enough displacement so we are still in the linear region of the UPO, but is large enough so our solution can leave the orbit's neighborhood. Usually, we set $\delta=10^{-6}$.

The initial conditions \eqref{eq:ic_upo_man_2d} lead to the manifold $W({\tilde{\bs{p}}}_0)$ when integrated in the appropriate time direction, where the plus or minus sign refers to the branches of the manifold. In order to trace the rest of $W(\alpha)$, we first discretize the UPO in $\tilde{N}$ points and, instead of calculating the monodromy matrix for each one of these points $\tilde{\bs{p}}_k$, we determine the respective eigenvectors $\bs{e}_k$ by propagating the initial eigenvectors $\bs{e}_0$ using the transition matrix $\Phi$ in the following manner

\be
\bs{e}_k^{s,u} = \dfrac{\Phi(t_k)\bs{e}_0^{s,u}}{||\Phi(t_k)\bs{e}_0^{s,u}||},
\label{eq:ic_upo_man_2d_propagation}
\ee

\noindent where $k=1,\dots,\tilde{N}-1$. The time $t_k<T$ corresponds to the time it takes for going from $\bs{q}_0$ to $\bs{q}_k$, with $T$ being the UPO's period.

Determining the eigenvectors $\bs{e}_k$ using Eq.~\eqref{eq:ic_upo_man_2d_propagation} instead of calculating and using the respective monodromy matrices requires less computational cost since we need to make just one loop around the UPO after obtaining the first monodromy matrix. Furthermore, we avoid the additional errors that would be introduced by calculating $\tilde{N}-1$ monodromy matrices.

The two-dimensional invariant manifold $W(\alpha)$ is traced by choosing an initial condition $\bs{q}_k$ in the direction of each eigenvector $\bs{e}_k$, and then integrating it in the appropriate time direction. Finally, we have

\be
W(\alpha)=\bigcup_{k=0}^{\tilde{N}-1}W(\bs{q}_k),
\label{eq:manifold_2d_numerical}
\ee

\noindent where $\bs{q}_k = \tilde{\bs{p}}_k \pm \delta\bs{e}_k^{s,u}$.

Equation \eqref{eq:manifold_2d_numerical} is a numerical approximation of the manifold which depends on two parameters: the number $\tilde{N}$ of points selected in the UPO, and the integration time for each solution $\bs{\varphi}(\bs{q}_k,t)$. These parameters are adjusted depending on the type of investigation that is being carried out. Also, there is always the trade-off between manifold resolution and total computation time.

The determination of $\Gamma(\alpha)$ directly follows by using the standard methods for calculating a Poincaré map on each of the manifolds $W(\bs{q}_k)$. The parameter $\tilde{N}$ is specially important here, since a high number of points on the Poincaré map is usually necessary in order to obtain a good resolution for the intersection between the invariant manifold and the surface of section.

In the next chapter, we will carry out the aforementioned procedure, and calculate two-dimensional invariant manifolds in the Earth-Moon system for different families of unstable periodic orbits.

\chapter{The Earth-Moon system}
\label{ch:earth_moon_system}

In this chapter, we investigate the dynamical and geometrical properties of the Earth-Moon system as modeled by the \emph{planar circular restricted three-body problem} (\gls{PCRTBP}). This model provides a good first approximation for studying the motion of a small body, such as an artificial satellite or an asteroid, under the gravitational influence of a two-body system, such as the Earth-Moon or the Sun-Jupiter system \cite{Murray1999}. In this context, we assume that the system's motion is limited to a plane (\emph{planar}), the primaries describe circular orbits around their common center of mass (\emph{circular}), and the third particle's mass is negligible (\emph{restricted}).

Initially, we derive the equations of motion for the general three-body problem in a rotational coordinate system. Then, we reduce the system to a simpler form, the planar circular restricted case, and present the main aspects of the model. Later, by considering a mass parameter which corresponds to the Earth-Moon system, we calculate the phase space configuration on a suitable surface of section, and we analyze the bifurcations that occur in the system. We then trace the invariant manifolds associated with certain unstable periodic orbits and we study these structures as represented in the surface of section, relating them to the phase space configuration. Finally, we define the \emph{multiplicative transit time} to illustrate the influence of the invariant manifolds in the system's transport properties, and the \emph{mean escape measure} in order to examine the features of the transient dynamics when escape is considered in the system.

\section{Physical model}
\label{sec:model}

The study of the physical system formed by three particles orbiting each other under the influence of their mutual gravitational attraction is named \emph{the three-body problem}. According to \emph{Newton's law}, the force acting on each particle with mass $m_i$ is given by

\be
m_i\ddot{{\bs{r}}}_i=-\sum_{\substack{j=1\\j\neq i}}^{3}\frac{m_im_jG}{{r}_{ij}^{3}}{\bs{r}}_{ij},
\ee

\noindent where ${\bs{r}}_{i}$ is the position vector of the $i$-th particle, ${\bs{r}}_{ij}={\bs{r}}_i-{\bs{r}}_j$ is the relative position vector of the $i$-th particle with respect to the $j$-th particle, and the superscript dot notation stands for the time derivative.

\begin{figure}[h!]
\centering
\includegraphics[scale=0.6]{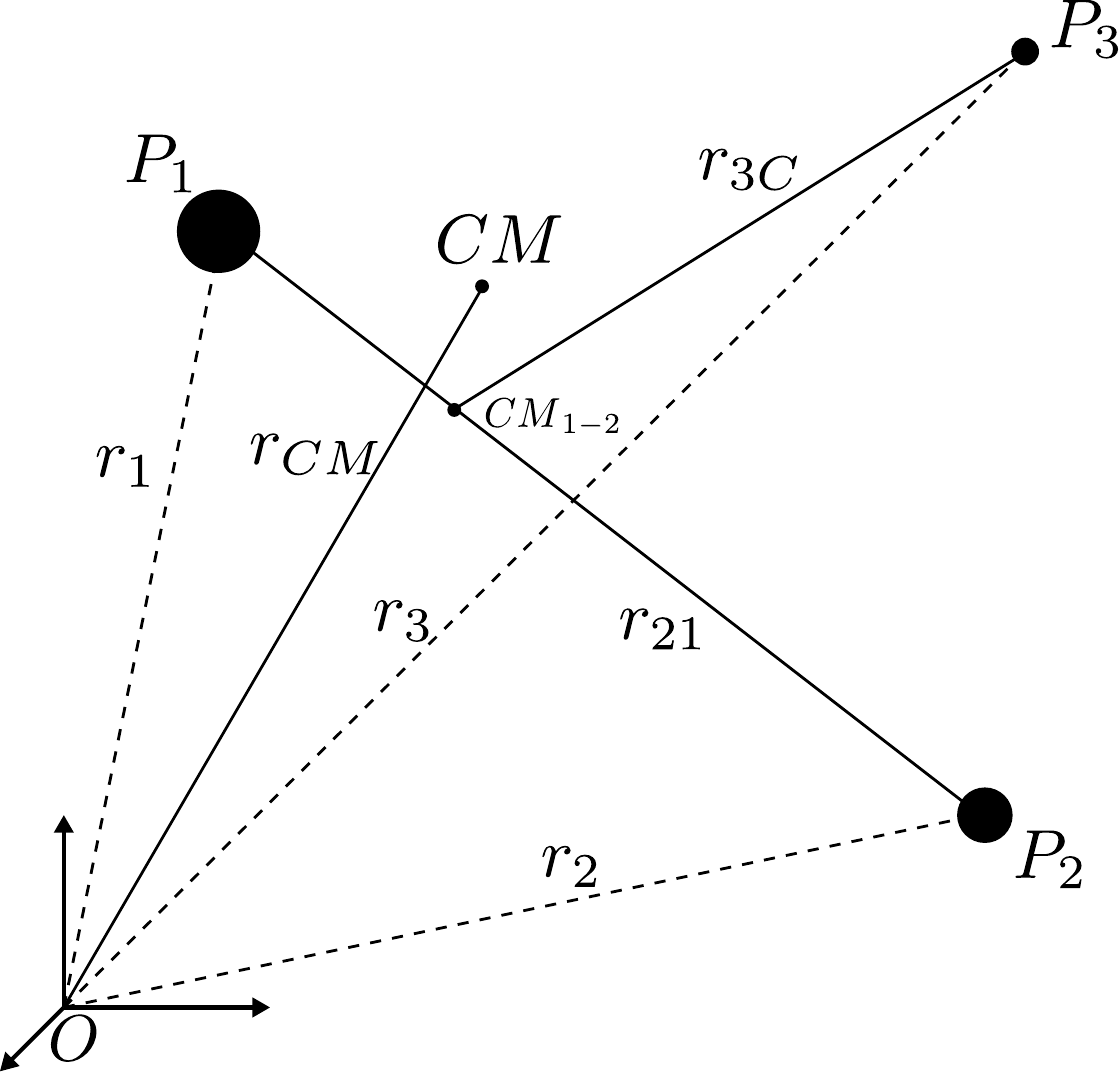}
\caption[Cartesian and Jacobi coordinates]{Cartesian coordinates ${\bs{r}}_{1}$, ${\bs{r}}_{2}$, ${\bs{r}}_{3}$, and Jacobi coordinates ${\bs{r}}_{21}$, ${\bs{r}}_{3C}$, ${\bs{r}}_{CM}$, for the three-body problem, where $P_1$ and $P_2$ are the primaries. $CM$ stands for the system's center of mass and $CM_{1-2}$ for the center of mass of the subsystem $P_1$-$P_2$.}
\label{fig:jacobi_coordinates}
\end{figure}

Considering that the particles have different masses, and naming the two particles with bigger masses the \emph{primaries}, we can write the system in a coordinate frame defined by the distance between the primaries ${{r}}_{21}$, the distance between the third particle and the primaries' center of mass ${{r}}_{3C}$, and the position of the system's center of mass ${{r}}_{CM}$, as shown in Fig.~\ref{fig:jacobi_coordinates}. These are called the \emph{Jacobi coordinates} and are given by \cite{Belbruno2004}

\be
\begin{aligned}
{\bs{r}}_{21} & = {\bs{r}}_{2} - {\bs{r}}_{1}, \\
{\bs{r}}_{3C} & = {\bs{r}}_{3} - \frac{1}{m_{12}}(m_1{\bs{r}}_{1}+m_2{\bs{r}}_{2}), \\
{\bs{r}}_{CM} & = \frac{1}{M}(m_1{\bs{r}}_{1}+m_2{\bs{r}}_{2}+m_3{\bs{r}}_{3}),
\end{aligned}
\ee

\noindent where $m_{12}=m_1+m_2$ is the sum of the primaries' masses, and $M=m_1+m_2+m_3$ is the total mass of the system. 

With some algebraic manipulation, we end up with the equations of motion for the three-body problem in the Jacobi coordinates:

\be
\begin{aligned}
\ddot{\bs{r}}_{21} &= - \left[\frac{m_{12}G}{r_{21}^3} + \frac{m_3G}{m_{12}}\left(\frac{m_2}{{r}_{31}^3}+\frac{m_1}{{r}_{32}^3}\right)\right]{\bs{r}}_{21} - m_3G\left(\frac{1}{r_{31}^3}-\frac{1}{r_{32}^3}\right){\bs{r}}_{3C}, \\
\ddot{\bs{r}}_{3C} &= -\frac{m_1m_2MG}{m_{12}^2}\left(\frac{1}{r_{31}^3}-\frac{1}{r_{32}^3}\right){\bs{r}}_{21}-\frac{MG}{m_{12}}\left(\frac{m_1}{r_{31}^3}+\frac{m_2}{r_{32}^3}\right){\bs{r}}_{3C},\\
\ddot{\bs{r}}_{CM}&=0,
\label{eq:motion_jac}
\end{aligned}
\ee

%\noindent where $r_{31}=\left|{\bs{r}}_{3C}+\frac{m_2}{m_{12}}{\bs{r}}_{21}\right|$ and $r_{32}=\left|{\bs{r}}_{3C}-\frac{m_1}{m_{12}}{\bs{r}}_{21}\right|$.

\noindent where $r_{31}=|{\bs{r}}_{3C}+\frac{m_2}{m_{12}}{\bs{r}}_{21}|$ and $r_{32}=|{\bs{r}}_{3C}-\frac{m_1}{m_{12}}{\bs{r}}_{21}|$.

%\noindent where $r_{31}=|{\bs{r}}_{3C}+({m_2}/{m_{12}}){\bs{r}}_{21}|$ and $r_{32}=|{\bs{r}}_{3C}-({m_1}/{m_{12}}){\bs{r}}_{21}|$.

Equations~\eqref{eq:motion_jac} describe the problem as two coupled second order differential equations: one for the motion of the primaries relative to each other, and one for the motion of the third particle relative to the center of mass of the primaries' subsystem. From $\ddot{\bs{r}}_{CM}=0$, which is due to the conservation of linear momentum, we can move the origin of the coordinate system to the center of mass.

\subsubsection{The planar circular restricted three-body problem}

Let us now assume that the third particle $P_3$ is a test particle. By setting $m_3=0$ in Eqs.~\eqref{eq:motion_jac}, we have

\be
\begin{aligned}
\ddot{\bs{r}}_{21} &= - \frac{m_{12}G}{r_{21}^3}{\bs{r}}_{21}, \\
\ddot{\bs{r}}_{3C} &= -\frac{m_1m_2G}{m_{12}}\left(\frac{1}{r_{31}^3}-\frac{1}{r_{32}^3}\right){\bs{r}}_{21}-G\left(\frac{m_1}{r_{31}^3}+\frac{m_2}{r_{32}^3}\right){\bs{r}}_{3C}.
\label{eq:motion_jac_restricted}
\end{aligned}
\ee

The first equation in \eqref{eq:motion_jac_restricted} is the standard differential equation for the two-body pro\-blem, that describes the motion of the particle $P_2$ around $P_1$, and can be analytically solved. We are interested in the second equation, which describes the motion of $P_3$ under the gravitational field generated by the other two particles.

\begin{figure}[h!]
\centering
\includegraphics[scale=0.6]{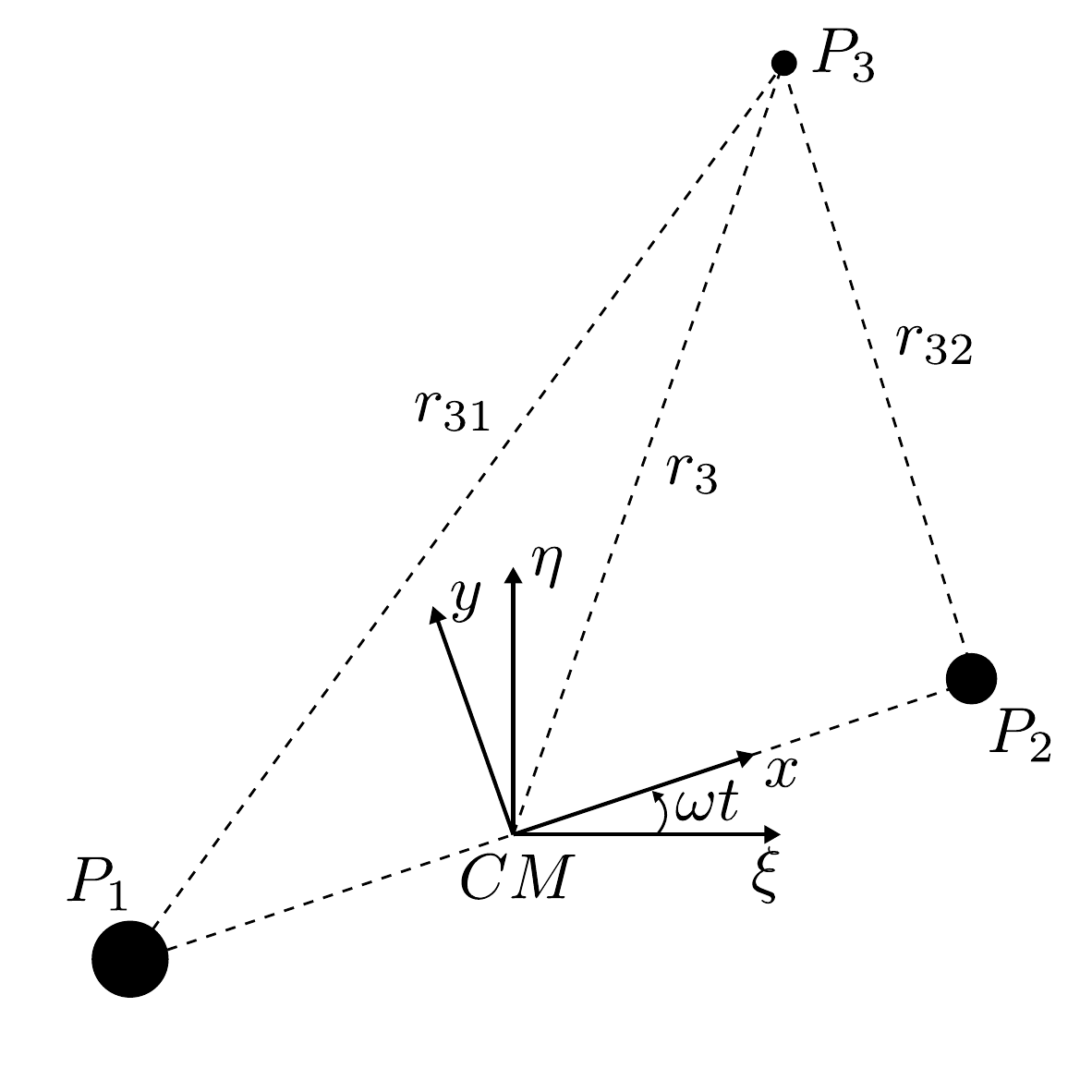}
\caption[Inertial and rotational reference frames]{Inertial (sideral) and rotational (synodic) reference frames. The system's center of mass is the origin of both frames in the restricted case, and the rotational frame follows the motion of the primaries. For the dimensionless problem, we have $\omega=1$.}
\label{fig:frames}
\end{figure}

We will rewrite the equation for $\ddot{\bs{r}}_{3C}$ using two different coordinate systems. One is called \emph{sideral system}, denoted by the variables $(\xi,\eta)$, and the other is called \emph{synodic system} and is going to be denoted by the variables $(x,y)$, as shown in Fig.~\ref{fig:frames}. The former is an \emph{inertial} reference system while the latter is a \emph{rotational} reference system that follows the primaries' motion \cite{Murray1999}.

We assume here that the primaries describe circular orbits around their common center of mass. To render the problem dimensionless, we scale the length unit and the mass unit so that $r_{21}=1$ and $m_{12}G=1$, respectively. It then follows from \emph{Kepler's third law} that the common mean motion of the primaries is $\omega=1$ and that their period of revolution corresponds to $2\pi$ units of time. Lastly, we define the \emph{mass parameter}

\be
\mu=\frac{m_2}{m_1+m_2},
\label{eq:mass_parameter}
\ee

\noindent where $m_1>m_2$. For the Earth-Moon system, $m_1$ stands for the Earth's mass, while $m_2$ for the Moon's mass, and we have $\mu=1.2150\times10^{-2}$.

Since we are dealing with a negligible third particle's mass, the center of mass of the subsystem formed by the primaries is the origin of the coordinate system and, therefore, $r_{3C}=r_3$. Furthermore, we have ${\bs{r}}_{21}(t)=(\cos t, \sin t)$, ${\bs{r}}_{1}(t)=-\mu{\bs{r}}_{21}(t)$, and ${\bs{r}}_{2}(t)=(1-\mu){\bs{r}}_{21}(t)$. Due to our scalings, we also have  $m_1G=1-\mu$ and $m_2G=\mu$.

The last assumption that we make is that the motion of the third particle is restricted to the plane that contains the primaries. This is an useful restriction given that the geometrical structures we investigate in this work are naturally represented in a two-dimensional surface of section.

With these considerations, we can finally write down the dimensionless equations of motion of the PCRTBP for the sideral coordinate system as

\be
\begin{aligned}
\ddot{\xi} &= \dfrac{\partial\Psi}{\partial\xi},\\
\ddot{\eta} &= \dfrac{\partial\Psi}{\partial\eta},\\
\end{aligned}
\label{eq:motion_inertial}
\ee

\noindent where the potential $\Psi$ is given by

\be
\Psi = \frac{1-\mu}{r_{31}}+\frac{\mu}{r_{32}},
\label{eq:tilde_omega}
\ee

\noindent with

\be
\begin{aligned}
r_{31}&=\sqrt{(\xi+\mu\cos t)^2+(\eta+\mu\sin t)^2},\\
r_{32}&=\sqrt{(\xi-(1-\mu)\cos t)^2+(\eta-(1-\mu)\sin t)^2}.\\
\end{aligned}
\label{eq:r_1_2_inertial}
\ee

The sideral and synodic coordinate systems are related by a rotation matrix ${\bs{R}}(t)$:

\be
\left(
\begin{array}{c}
x \\
y 
\end{array} \right)
={\bs{R}}(t)
\left(
\begin{array}{c}
\xi \\
\eta
\end{array} \right),
~\
~\
\left(
\begin{array}{c}
\xi \\
\eta
\end{array} \right)
={\bs{R}}^{-1}(t)
\left(
\begin{array}{c}
x \\
y
\end{array} \right),
\label{eq:rotation}
\ee

\noindent where

\be
\nonumber
{\bs{R}}(t)=\left(
\begin{array}{cc}
\cos{t} & \sin{t} \\
-\sin{t} & \cos{t}
\end{array} \right)
~\
\text{and}
~\
{\bs{R}}^{-1}(t)=\left(
\begin{array}{cc}
\cos{t} & -\sin{t} \\
\sin{t} & \cos{t}
\end{array} \right).
\ee

The equations of motion of the PCRTBP for the synodic coordinate system are then determined by applying the transformation \eqref{eq:rotation} to Eqs.~\eqref{eq:motion_inertial} and are given by

\be
\begin{aligned}
\ddot{x}-2\dot{y} &= \dfrac{\partial\Omega}{\partial x},\\
\ddot{y}+2\dot{x} &= \dfrac{\partial\Omega}{\partial y},\\
\end{aligned}
\label{eq:motion_rotational}
\ee

\noindent where the pseudo-potential $\Omega$ and the distance between the third mass and the primaries are, respectively, 

\be
\Omega = \frac{1}{2}(x^2+y^2)+\frac{1-\mu}{r_{31}}+\frac{\mu}{r_{32}},
\label{eq:omega}
\ee

\noindent and

\be
\begin{aligned}
r_{31}&=\sqrt{(x+\mu)^2+y^2},\\
r_{32}&=\sqrt{(x-(1-\mu))^2+y^2}.\\
\end{aligned}
\label{eq:r_1_2_rotational}
\ee

On the one hand, the rotating coordinate system induces two new terms in the equations of motion: one that corresponds to the Coriolis force (proportional to $\dot{x}$ and $\dot{y}$), and another that represents the inertial force (proportional to $x$ and $y$) and is incorporated in the definition of $\Omega$. On the other hand, Eqs.~\eqref{eq:motion_rotational} are a set of \emph{autonomous} equations, while Eqs.~\eqref{eq:motion_inertial} explicitly depend on time. In this work, we will always consider the PCRTBP in the synodic coordinate system.

Equations \eqref{eq:motion_rotational} are a set of second-order differential equations. In order to numerically integrate these, we write them as a four-dimensional first-order differential equation:

\be
\dfrac{d}{dt}\left(x,y,\dot{x},\dot{y}\right)=\left(\dot{x},\dot{y},\dfrac{\partial\Omega}{\partial x}+2\dot{y},\dfrac{\partial\Omega}{\partial y}-2\dot{x}\right),
\label{eq:motion_rotational_first_order}
\ee

\noindent which has the same format of Eq.~\eqref{eq:ODE}.

% \renewcommand{\arraystretch}{1.6}
% \be
% \dfrac{d}{dt}\left(
% \begin{array}{c}
% x\\
% y\\
% \dot{x}\\
% \dot{y}
% \end{array}
% \right)=
% \left(
% \begin{array}{c}
% \dot{x}\\
% \dot{y}\\
% \dfrac{\partial\Omega}{\partial x}+2\dot{y}\\
% \dfrac{\partial\Omega}{\partial y}-2\dot{x}
% \end{array}
% \right)
% \label{eq:motion_F}
% \ee

The PCRTBP can also be transformed to a Hamiltonian form (see App. \ref{app:PCRTBP_hamiltonian}). For small $\mu$, this model is a near-integrable Hamiltonian system. For $\mu=0$, the system is reduced to the integrable case of a massless particle moving under the gravitational field generated by $P_1$.

\subsubsection{Lagrangian equilibrium points}

\begin{figure}[h!]
\centering
\includegraphics[scale=0.6]{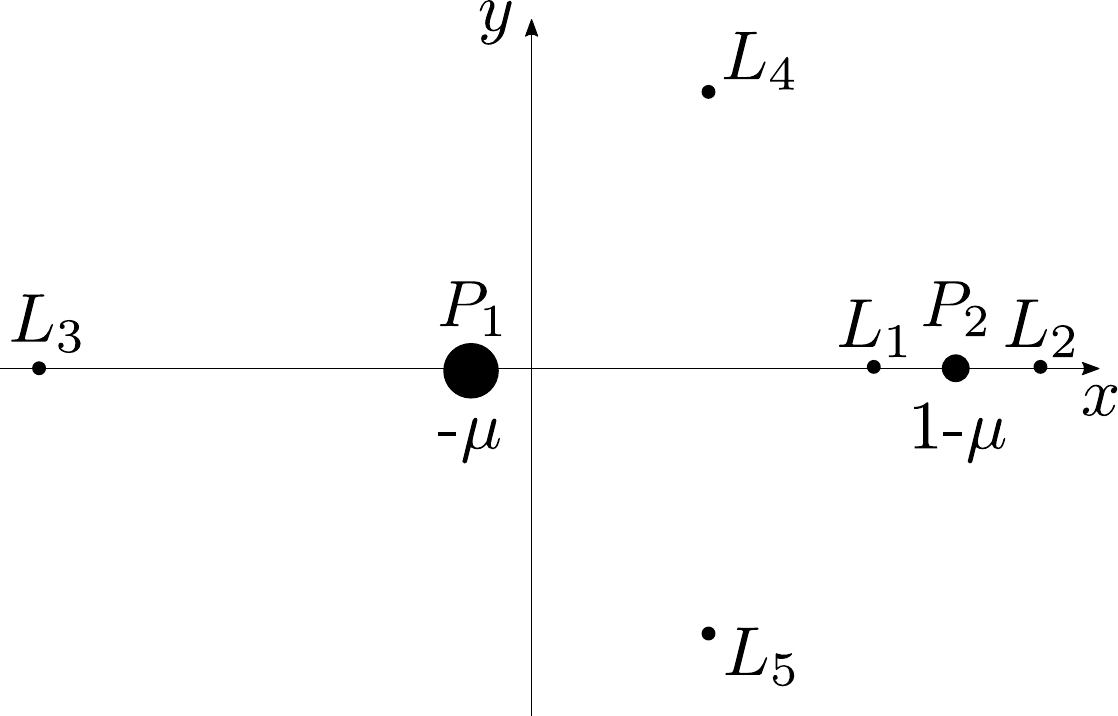}
\caption[Lagrangian equilibrium points]{Location of the Lagrangian equilibrium points. The collinear points $L_1$, $L_2$, and $L_3$ are unstable, while the triangular points $L_4$ and $L_5$ are stable for the Earth-Moon system.}
\label{fig:lagrangian_points}
\end{figure}

There are \emph{five} equilibrium points in the PCRTBP, which correspond to the critical points of the pseudo-potential $\Omega$ in  Eqs.~\eqref{eq:motion_rotational}.  We call these the \emph{Lagrangian points} and denote them by \gls{Li}, with $i=1,\dots,5$. Figure \ref{fig:lagrangian_points} shows a schematic location of these points along with the primaries.

The Lagrangian points are divided in two groups. The first group is formed by the three points that are collinear to the primaries, and are called \emph{collinear equilibrium points}. The first one, $L_1$, is situated between the primaries; the second one, $L_2$, is right of $P_2$; and the third one, $L_3$, is left of $P_1$. 
The second group is composed by the two points that form an equilateral triangle with the primaries, called the \emph{triangular equilibrium points}. The \emph{leading} point, with $y>0$, is denoted by $L_4$, while the \emph{trailing point}, with $y<0$, is denoted by $L_5$. It is important to note that the rotational system moves in the \emph{counterclockwise} direction.

The location of the Lagrangian points depend only on the mass parameter. Furthermore, by carrying out a linear approximation analysis, we can show that the collinear equilibrium points are unstable, i.e., typical orbits in the vicinity of these points will move away from them, while the triangular equilibrium points are stable under certain conditions, which are satisfied for the Earth-Moon system. A detailed calculation of the Lagrangian points location and of their stability properties is presented in App.~\ref{app:stability}.

\subsubsection{Hill stability}

The PCRTBP has \emph{one} constant of motion, named \emph{Jacobi constant}, which is given by

\be
C = 2\Omega(x,y)-\dot{x}^2-\dot{y}^2.
\label{eq:constant}
\ee

\noindent Thus, even though the PCRTBP is a four-dimensional system, the solutions of Eqs.~\eqref{eq:motion_rotational} lie on a three-dimensional surface defined by $C$. A derivation of the Jacobi constant is given in App.~\ref{app:jacobi}.

\begin{figure}[h!]
\centering
\includegraphics[scale=0.95]{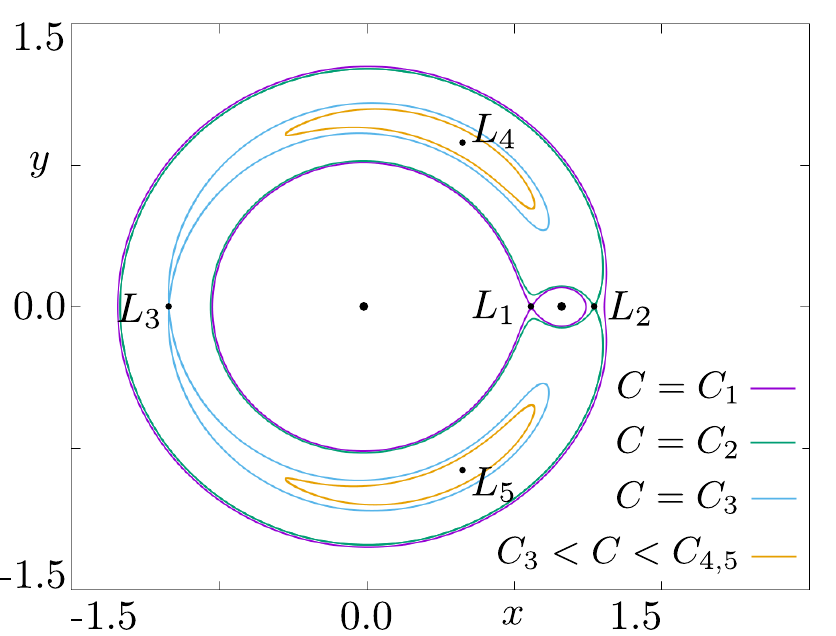}
\caption[Zero-velocity curves]{Zero-velocity curves in the Earth-Moon system for selected $C$, where $C_i=C(L_i)$. The possibility of motion between the realms of the primaries or in and out of the outer region depends on the value of the Jacobi constant.}
\label{fig:zvc_colored}
\end{figure}

There is a special set of level curves of Eq.~\eqref{eq:constant} called the \emph{zero-velocity curves} (\gls{ZVC}s), which are defined by $\dot{x}^2=\dot{y}^2=0$. In Fig.~\ref{fig:zvc_colored}, we present the ZVCs for four different values of $C$ in the Earth-Moon system. We use the notation $C_i=C(L_i)$ for the Jacobi constant calculated at the $i$-th Lagrangian point, where $C_1>C_2>C_3>C_4=C_5$. 

The ZVCs are important because they delimit the area on the coordinate space $x$-$y$ that is accessible to the third body for a given $C$, which is called the \emph{Hill region} $\mathcal{H}$. Since $\dot{x}^2+\dot{y}^2\geq0$ in Eq.~\eqref{eq:constant}, we have

\be
\mathcal{H}=\{(x,y)\in\mathbb{R}^2~|~2\Omega-C\geq0\}.
\label{eq:hill_region}
\ee

\noindent Conversely, the interior of a ZVC defines the \emph{forbidden region}, where the motion of the third body is not permitted.

From Fig.~\ref{fig:zvc_colored}, we can see that the topological properties of the Hill region depend on the Jacobi constant. In this work, we are interested in the situation where $\mathcal{H}$ is composed by two disconnected areas: an inner one $\mathcal{H}_I$, which includes the primaries, and an outer one $\mathcal{H}_O$. This is achieved by choosing a Jacobi constant between $C_1$ and $C_2$, i.e., $C_1 > C > C_2$.

\begin{figure}[h!]
\centering
\includegraphics[scale=0.85]{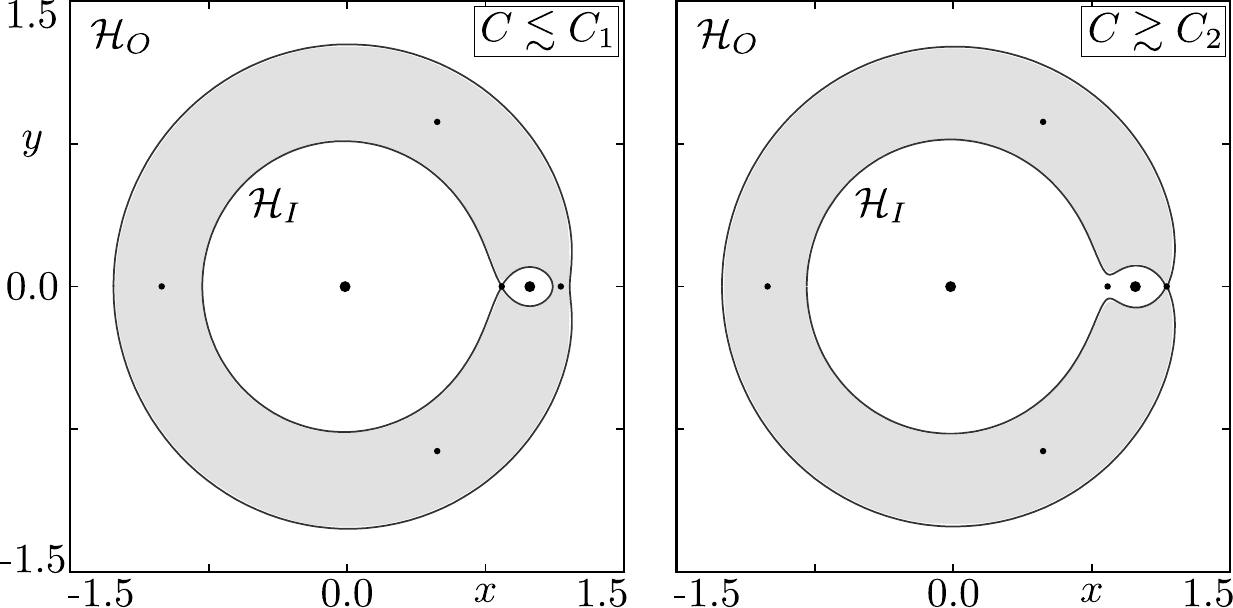}
\caption[The bounded Earth-Moon system]{The bounded Earth-Moon system. For $C_1 > C > C_2$, the Hill region (white) is composed of two disconnected areas $\mathcal{H}_I$ and $\mathcal{H}_O$. Due to the forbidden region (gray), if the third body starts its motion inside the inner area, it will always be confined to it.}
\label{fig:ems_hill}
\end{figure}

In Fig.~\ref{fig:ems_hill}, we show the Hill region (in white), and the forbidden region (in gray), for the Earth-Moon system considering $C\lesssim C_1$ and $C\gtrsim C_2$, where $C_1 \approx 3.1883$ and $C_2 \approx 3.1722$. The inner region is formed by the Earth's realm, to the left of $L_1$, along with the Moon's realm, to the right of $L_1$. We can observe that, for our chosen range of $C$, orbits that start in the vicinity of any primary can transit between the Earth and the Moon's realms but cannot leave and are, therefore, \emph{bounded} within the system. We call this phenomenon \emph{Hill stability}.

%\begin{figure}[h!]
%\centering
%\includegraphics[scale=0.85]{ems_hill_region_smaller.pdf}
%\caption[The bounded Earth-Moon system]{The bounded Earth-Moon system. For $C_1 > C > C_2$, the Hill region (white) is composed of two disconnected areas $\mathcal{H}_I$ and $\mathcal{H}_O$. Due to the forbidden region (gray), if the third body starts its motion inside the inner area, it will always be confined to it.}
%\label{fig:ems_hill}
%\end{figure}

\subsubsection{Family of Lyapunov orbits}

The last aspect of the PCRTBP that we are going to mention is the existence of an uniparametric family of unstable periodic orbits around each of the collinear Lagrangian points. These are called \emph{Lyapunov orbits} and they are going to be important when we investigate the geometrical properties of the system.

\begin{figure}[h!]
\centering
\includegraphics[scale=0.85]{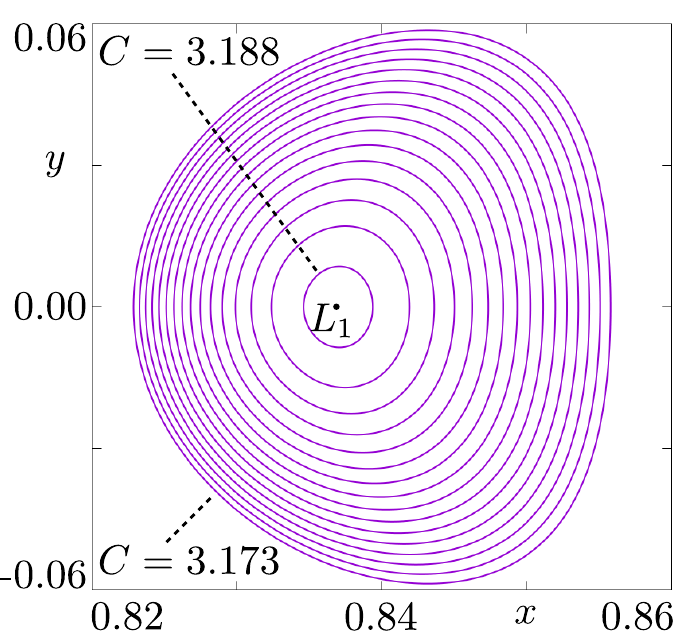}
\caption[Family of Lyapunov orbits around $L_1$]{Family of Lyapunov orbits around $L_1$ in the coordinate space $x$-$y$ for selected $C$ in our chosen range. These orbits are situated in the neck region between the Moon's and the Earth's realm.}
\label{fig:ems_lyapunov_family}
\end{figure}

The only family of Lyapunov orbits that exists for our chosen range of Jacobi constant is the one around $L_1$, since this is the only Lagrangian point inside the Hill region. In Fig.~\ref{fig:ems_lyapunov_family}, we present some members of such family from $C=3.188$ to $C=3.173$. The first members are small ellipses formed in the linear region of $L_1$ (note the difference in scale in the figure's axes). By lowering the value of $C$, the members become larger in length, while the \emph{neck} region between the Moon and the Earth's realms also becomes larger, as shown in Fig.~\ref{fig:ems_hill}.

The numerical computation of a Lyapunov orbit for a given C involves using a method for calculating UPOs in the system, along with the determination of a suitable initial condition in the linear region, and a continuation procedure that uses previously computed members of the family as reference for tracing new members. These techniques are described in the next section.

\subsubsection{The full picture}

\begin{figure}[h!]
\centering
\includegraphics[scale=0.85]{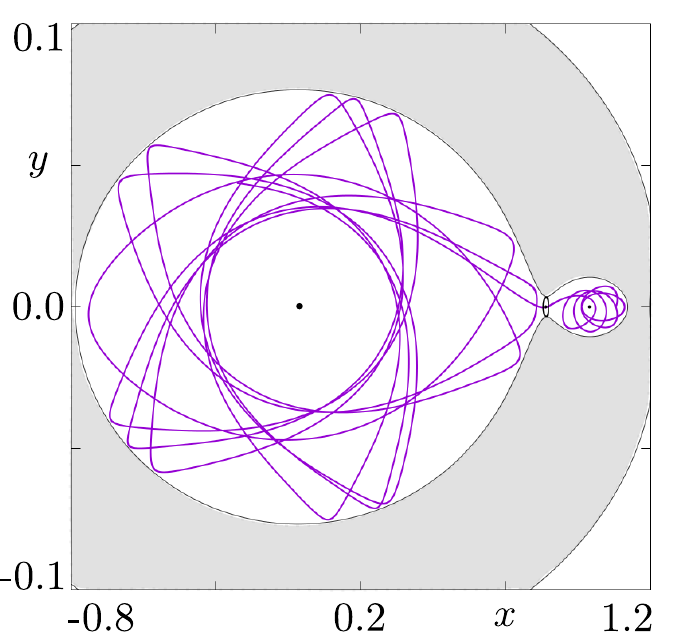}
\caption[Example of an orbit in the Earth-Moon system]{Example of an orbit in the Earth-Moon system for $C=3.183$. The initial condition is given inside the Moon's realm and the trajectory eventually transfers to the Earth's realm through the neck region.}
\label{fig:ems_orbit_example}
\end{figure}

Figure~\ref{fig:ems_orbit_example} shows an example of an orbit in the Earth-Moon system for $C=3.183$. The initial condition is given by  $(x_0,y_0,\dot{x}_0,\dot{y}_0)=(1.08,0.00,0.08,0.22)$, and the trajectory is integrated for $t=50$ units of time. The gray area corresponds to the forbidden region and the Lyapunov orbit for this Jacobi constant is shown in black. It is interesting to note here that, in order for the trajectory to move from the Moon's to the Earth's realm, it has to go through the Lyapunov orbit.

\section{Periodic orbits and regularization}
\label{sec:periodic_orbits_regularization}

In this section, we explain the numerical method that we use for determining periodic orbits in the planar circular restricted three-body problem, and how to find a suitable initial condition for tracing a member of the Lyapunov family. We also present the Levi-Civita transformation and derive the regularized equations of motion, which are necessary to preserve the integration's precision near the primaries.

\subsubsection{Determination of periodic orbits}

It can be shown that the PCRTBP has the following symmetry \cite{Mireles2006}: if $\bs{\varphi}({\bs{x}}_0,t)$ is a solution of Eqs.~\eqref{eq:motion_rotational} with initial condition ${\boldsymbol{x}}_0=(x_0,y_0,\dot{x}_0,\dot{y}_0)$ and $0\leq t\leq \tilde{T}$, then $\hat{\bs{\varphi}}(A{\bs{x}}_{\tilde{T}},t)=A\bs{\varphi}({\bs{x}}_{\tilde{T}},-t)$ is also a solution of the system for $0\leq t\leq \tilde{T}$ and ${\bs{x}}_{\tilde{T}}=\bs{\varphi}({\bs{x}}_0,\tilde{T})$. The symmetry matrix $A$ is given by

\be
A=
\left(
\ba{cccc}
1 & 0 & 0 & 0 \\
0 & -1 & 0 & 0 \\
0 & 0 & -1 & 0 \\
0 & 0 & 0 & 1
\ea
\right)
\ee
\noindent in the basis of the rotational reference frame $(x,y,\dot{x},\dot{y})$.

Now, suppose that we take an initial condition $\bs{x}_0=(x_0,0,0,\dot{y}_0)$ with $\dot{y}_0>0$ and, after integrating the system for $\tilde{T}$ units of time, it reaches the configuration $\bs{x}_{\tilde{T}}=(x_{\tilde{T}},0,0,\dot{y}_{\tilde{T}})$, where $\dot{y}_{\tilde{T}}<0$. By applying the above mentioned symmetry, we find that,  in the coordinate space $x$-$y$, the curve formed by $\bs{\varphi}(A{\bs{x}}_{\tilde{T}},t)$ is a reflection over the $x$-axis of the curve formed by $\bs{\varphi}({\bs{x}}_0,t)$ for $0\leq t \leq \tilde{T}$. Given that $A{\bs{x}}_{\tilde{T}}={\bs{x}}_{\tilde{T}}$, we have that $\bs{\varphi}({\bs{x}}_0,t)$ is actually a periodic orbit with period $T=2\tilde{T}$.

Therefore, the problem of finding periodic orbits in the PCRTBP that are perpendicular to the line $y=0$, which is the case for the Lyapunov orbits, is reduced to the problem of determining an initial condition $\bs{x}_0=(x_0,0,0,\dot{y}_0>0)$, and a half-period $\tilde{T}$, so that $\bs{\varphi}({\bs{x}}_0,\tilde{T})=(x_{\tilde{T}},0,0,\dot{y}_{\tilde{T}}<0)$ for a suitable initial guess. This can be done iteratively by using \emph{Newton's method} in the following manner.

First, we choose an initial guess for the initial condition, ${\bs{x}}^{(0)}_0=(x_0,0,0,\dot{y}^{(0)}_0)$, and one for the half-period, $\tilde{T}^{(0)}$. We then fix the $x_0$ coordinate and we look for $\dot{y}_0$ and $\tilde{T}$ so that $\dot{x}(\tilde{T})={y}(\tilde{T})=0$. In order to do that, we define the function $\bs{h}(\dot{y}_0,\tilde{T})=(\varphi_3(\dot{y}_0,\tilde{T}),\varphi_2(\dot{y}_0,\tilde{T}))$, where $\varphi_2$ and $\varphi_3$ are components of the solution $\bs{\varphi}=(\varphi_1,\varphi_2,\varphi_3,\varphi_4)$, and we calculate its root using Newton's method:

\be
\left(
\ba{c}
\dot{y}_0^{(n+1)} \\
\tilde{T}^{(n+1)}
\ea
\right)=
\left(
\ba{c}
\dot{y}_0^{(n)} \\
\tilde{T}^{(n)}
\ea
\right)-
{[\bs{D}\bs{h}(\dot{y}_0^{(n)},\tilde{T}^{(n)})]}^{-1}
\left(
\ba{c}
\varphi_3^{(n)}(\dot{y}_0^{(n)},\tilde{T}^{(n)}) \\
\varphi_2^{(n)}(\dot{y}_0^{(n)},\tilde{T}^{(n)})
\ea
\right),
\label{eq:newton_method}
\ee

\noindent with $\bs{D}\bs{h}(\dot{y},t)$ given by

\be
\nn
\bs{D}\bs{h}(\dot{y},t)=
\left(
\ba{cc}
\dfrac{\partial \varphi_3}{\partial \dot{y}} & \dfrac{\partial \varphi_3}{\partial t} \\
\dfrac{\partial \varphi_2}{\partial \dot{y}} & \dfrac{\partial \varphi_2}{\partial t}
\ea
\right)=
\left(
\ba{cc}
\Phi_{3,4} & f_3 \\
\Phi_{2,4} & f_2
\ea
\right),
\ee

\noindent where terms $\Phi_{3,4}$ and $\Phi_{2,4}$ belong to the state transition matrix, while $f_3$ and $f_2$ are components of the vector field $\bs{F}=(f_1,f_2,f_3,f_4)$ .

Hence, instead of looking for a periodic orbit in the five-dimensional extended space $(x,y,\dot{x},\dot{y},t)$, we just have to solve the two-dimensional Newton's method, where the matrix $\bs{D}\bs{h}$ can be analytically inverted before performing the method. The fact that we can use the system's symmetry and look for the half-period of the orbit, rather than the full period, makes it easier for the method to converge, especially regarding unstable periodic orbits where the sensibility of the initial guess is higher.

In practice, we iterate Eq.~\eqref{eq:newton_method} a fixed, but sufficient, number of times or until a predetermined precision for the periodic orbit is reached. Due to the low-dimensionality of the method, errors below $10^{-10}$ can easily be attained. It is important to note, however, that this procedure does not impose any kind of conservation for constants of motion and, even though it is very robust, the precision of the Jacobi constant must be checked at all times.

\subsubsection{Tracing Lyapunov orbits}

In order to trace the Lyapunov orbit for a given Jacobi constant, we need to determine an adequate initial guess and use the aforementioned method to calculate the correct initial condition and period of the orbit. This is achieved in two steps: first, we find the periodic solutions of the linearized system centered at the Lagrangian point; and, second, we use a continuation procedure to extend these solutions closer to the predefined value of the Jacobi constant $C^{*}$. We will focus here in the Lyapunov family around the Lagrangian point $L_1$.

Let us use the following transformation in our system and move the origin of the reference frame to $L_1$

\be
\begin{aligned}
\tilde{x}&=\frac{x-x_{L_{1}}}{\gamma}=\frac{x-(1-\mu)+\gamma}{\gamma},\\
\tilde{y}&=\frac{y-y_{L_{1}}}{\gamma}=\frac{y}{\gamma},
\end{aligned}
\label{eq:center_transformation}
\ee

\noindent where $\gamma$ is the distance between the Moon and $L_1$, and the new normalization for the length unit.

Using the new set of variables defined by Eqs.~\eqref{eq:center_transformation} to rewrite the equations of motion, Eqs.~\eqref{eq:motion_rotational}, and also expanding the pseudo-potential $\Omega$, which is given by Eq.~\eqref{eq:omega}, we arrive at \cite{Gomez2001}

\be
\begin{aligned}
\ddot{\tilde{x}}-2\dot{\tilde{y}}-(1+2c_2)\tilde{x}&=\frac{\partial}{\partial \tilde{x}}\sum_{n=3}^{\infty}c_n{\tilde{r}}^nP_n\left(\frac{\tilde{x}}{\tilde{r}}\right), \\
\ddot{\tilde{y}}+2\dot{\tilde{x}}+(c_2-1)\tilde{y}&=\frac{\partial}{\partial \tilde{y}}\sum_{n=3}^{\infty}c_n{\tilde{r}}^nP_n\left(\frac{\tilde{x}}{\tilde{r}}\right),
\end{aligned}
\label{eq:motion_centered_lagrangian_point}
\ee

\noindent where $P_n$ are the Legendre polynomials of order $n$,  $\tilde{r}^2=\tilde{x}^2+\tilde{y}^2$, and $c_n$ are defined as

\be\nn
c_n=\frac{1}{\gamma^3}\left[\mu+{(-1)}^n(1-\mu){\left(\frac{\gamma}{1-\gamma}\right)}^{n+1}\right].
\ee

If we remove the terms for $n\geq2$ from the right-hand side of Eqs.~\eqref{eq:motion_centered_lagrangian_point} and also restrict the analysis to the central subspace, the solution of the system is given by

\be
\begin{aligned}
\tilde{x}&=\alpha cos{(\omega t)}, \\
\tilde{y}&=-k \alpha sen{(\omega t)}, \\
\end{aligned}
\label{eq:linear_solution}
\ee

\noindent where $k=(\omega^2+2c_2+1)/(2\omega)$, $\omega={((2-c_2+{(9c_2^2-8c_2)}^{1/2})/2)}^{1/2}$, and $\alpha$ is an arbitrary amplitude that can be calculated from Eq.~\eqref{eq:constant}.

The continuation procedure for finding a Lyapunov orbit with a predefined Jacobi constant $C^*$ is as follows. First, we use Eqs.~\eqref{eq:linear_solution} as an educated guess to find \emph{two} members of the Lyapunov family in the linear region, which are separated by a value $\delta<<1$ in the $x$-axis. The calculated initial condition for these orbits are ${\bs{x}}^1_0=(x_0,0,0,\dot{y}^1_0)$ and ${\bs{x}}^2_0=(x_0+\delta,0,0,\dot{y}^2_0)$, with half-periods given by $\tilde{T}^1$ and $\tilde{T}^2$. We then use the difference between these solutions, $\Delta^2_1={\bs{x}}^2_0-{\bs{x}}^1_0$, and the difference between their half-periods, $\Delta \tilde{T}^2_1 = \tilde{T}^2 - \tilde{T}^1$, as a reference for a new initial guess, ${{\bs{x}}^3_0}^{(0)}={\bs{x}}^2_0+\Delta^2_1$ and ${\tilde{T}}^{3^{(0)}}={\tilde{T}}^2+\Delta {\tilde{T}}^2_1$, which will return the initial condition of a third Lyapunov orbit after using Newton's method. Next, we define a precision $\varepsilon$, and carry out the procedure explained above until we have $|C({\bs{x}}^n_0)-C^*|<\epsilon$, where ${\bs{x}}^n_0={\bs{x}}^{n-1}_0+\Delta^n_{n-1}$. Lastly, we use a bisection procedure around this solution in order to arbitrarily increase our precision. 

\subsubsection{Levi-Civita transformation}

The singularities in Eqs.~\eqref{eq:motion_rotational_first_order}, which are due to the primaries, introduce a technical issue in our analysis. In order to ensure our desired precision, the time step of the numerical integration must decrease when the orbit passes near the Earth or the Moon, which is not always possible since both the field and the velocity goes to infinity at these locations. We deal with this problem by carrying out a \emph{regularization} procedure, which means using a coordinate transformation to reallocate the poles of the system.

There are different methods for regularizing the PCRTBP. Here, we choose the \emph{Levi-Civita transformation} \cite{Szebehely1967}, which is done locally and, therefore, needs to be performed for each primary separately. The regularized set of variables are named $(u,v,u^{ \prime},v^{\prime})$, where the prime notation stands for the derivative with respect to the new time variable $\tau$.
Using a complex notation, we can define $z = x + i y$ and $\omega = u + i v$. For regularization on the Earth's location, the Levi-Civita transformation is given by $z = \omega^2 -\mu$, whereas for the Moon's location $z = \omega^2 -\mu + 1$. Regarding the time variables, we have $dt=4(u^2+ v^2)d\tau$ in both cases.

The regularized equations of motion then becomes

\be
\begin{aligned}
u^{\prime\prime}-8(u^2+v^2)v^{\prime} &= \dfrac{\partial V_{E,M}}{\partial u},\\
v^{\prime\prime}+8(u^2+v^2)u^{\prime} &= \dfrac{\partial V_{E,M}}{\partial v},\\
\end{aligned}
\label{eq:motion_regularized}
\ee

\noindent where the new pseudo-potential for the regularization on the Earth's location, $V_E$, and on the Moon's location, $V_M$, are given by 

\begingroup
\setlength\abovedisplayskip{0pt}
\begin{multline}
V_{E}=4(1-\mu)+2(u^2+v^2)\Bigg\{{(u^2+v^2)}^2-2\mu(u^2-v^2)
\\
\left.+(\mu-C)+\dfrac{2\mu}{\sqrt{1+{(u^2+v^2)}^2-2(u^2-v^2)}}\right\},
\label{eq:V_Earth}
\end{multline}
\endgroup

\noindent and
 
\begingroup
\setlength\abovedisplayskip{0pt}
\begin{multline}
V_{M}=4\mu+2(u^2+v^2)\Bigg\{{(u^2+v^2)}^2+2(1-\mu)(u^2-v^2)
\\
\left.+(1-\mu-C)+\dfrac{2(1-\mu)}{\sqrt{1+{(u^2+v^2)}^2+2(u^2-v^2)}}\right\}.
\label{eq:V_Moon}
\end{multline}
\endgroup

As was done with Eqs.~\eqref{eq:motion_rotational}, we can rewrite the regularized equations of motion as a four-dimensional first-order differential equation:

\be
\dfrac{d}{d\tau}\left(u,v,u^{\prime},v^{\prime}\right)=\left(u^{\prime},v^{\prime},\dfrac{\partial V_{E,M}}{\partial u}+8(u^2+v^2)v^{\prime},\dfrac{\partial V_{E,M}}{\partial v}-8(u^2+v^2)u^{\prime}\right),
\label{eq:motion_regularized_first_order}
\ee

\noindent which has the same format of Eq.~\eqref{eq:ODE}.

In both transformations, about the Earth and the Moon, the origin of the reference frame is moved to the corresponding primary. By setting $(u,v)=(0,0)$ in Eqs.~\eqref{eq:V_Earth} and~\eqref{eq:V_Moon}, we can verify that $V$ is finite in these locations and, therefore, the pole is eliminated in both cases.

In practice, we define a radius around the Earth $\delta_{E}$, and a radius around the Moon $\delta_{M}$, and switch between Eqs.~\eqref{eq:motion_rotational_first_order} and~\eqref{eq:motion_regularized_first_order} every time an orbit enters or exits one of these regions. For comparison, we have $\delta_{E}=3.67\times10^{-2}$ and $\delta_{M}=1.00\times10^{-2}$, and the physical mean radii of the primaries are $r_E=1.66\times 10^{-2}$ and $r_M=4.52\times 10^{-3}$ in our dimensionless unit system \cite{Williams2019}. We also have to account for the time variable transformation in order to correctly trace an orbit in phase space. Since each time step $\Delta\tau$  in the regularized system is small, no higher than $10^{-3}$, we can use the \emph{trapezoidal rule} \cite{Press2007} to find the correct elapsed time $\Delta t$ in the original system,

\be
\Delta t=\int_0^{\tau}4\left[u^2(\tilde{\tau})+v^2(\tilde{\tau})\right]d\tilde{\tau}\approx2\Delta\tau\left[u^2(\tau)+v^2(\tau)+u^2(0)+v^2(0)\right],
\label{eq:trapezoidal_rule}
\ee

\noindent with an error which is proportional to ${\Delta\tau}^3$.

As a final remark on the regularization procedure, we can keep track of the numerical integration precision by calculating the integral of motion in the regularized variables, given by

\be
{u^{\prime}}^2+{v^{\prime}}^2=2V_{E,M}(u,v),
\label{eq:integral_regularized}
\ee

\noindent when the orbit is moving inside one of the regularization regions. It is worth noting, however, that we have to be careful with the errors introduced by the transformation itself since our investigation is performed in the non-regularized space.

\section{Phase space description}
\label{sec:phase_space}

\begin{figure}[h!]
\centering
\includegraphics[scale=0.85]{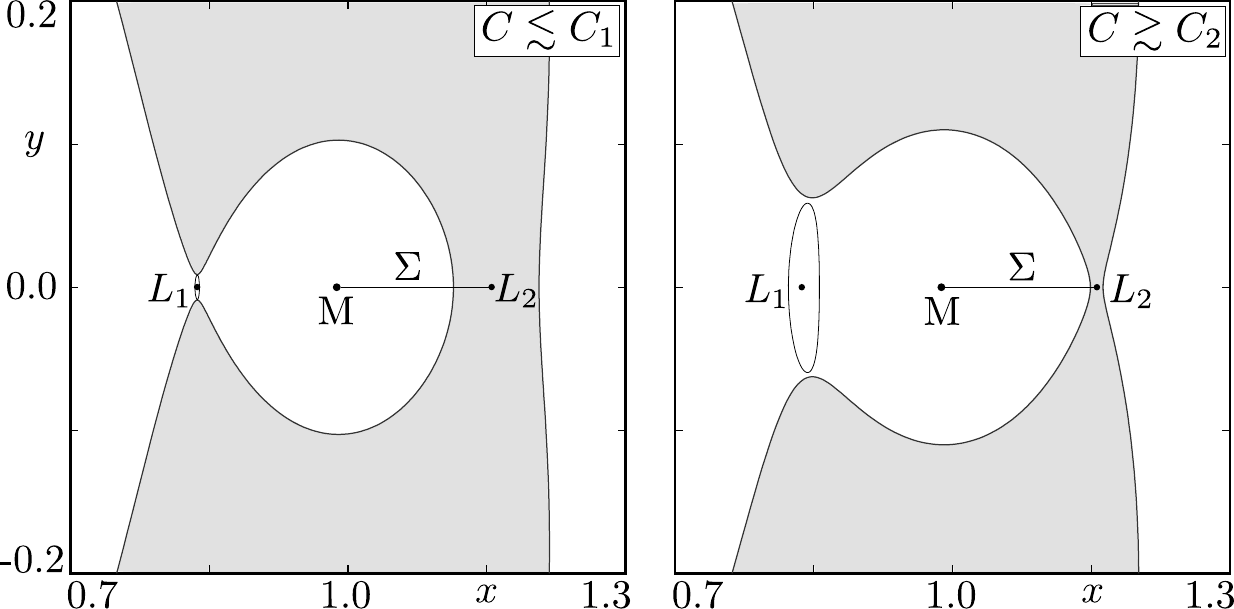}
\caption[The vicinity of the Moon]{Vicinity of the Moon for $C=3.188\lesssim C_1$ and $C=3.173\gtrsim C_2$. The surface of section $\Sigma$ is situated between the Moon and the Lagrangian point $L_2$, and it can be used to study the prograde orbits in the system, since it is defined for $\dot{y}>0$.}
\label{fig:ems_vicinity_model}
\end{figure}

With all the main features of the PCRTBP presented, we are now going to study the dynamical properties in the vicinity of the Moon. In order to do so, we define the following surface of section

\be
\Sigma=\{(x,y,\dot{x},\dot{y})\in\mathbb{R}^4~|~x_M< x<x_{L_2},~y=0,~\dot{y}>0\},
\ee

\noindent that is situated between the Moon and $L_2$ in the coordinate space $x$-$y$. We depict $\Sigma$ for $C\lesssim C_1$ and $C\gtrsim C_2$ in Fig.~\ref{fig:ems_vicinity_model}. The location in the $x$-axis of the Moon, $L_1$ and $L_2$ are given by $x_{M}=1-\mu=0.98785$, $x_{L_1}\approx0.8369$ and $x_{L_2}\approx1.1556$, respectively.

%\begin{figure}[h!]
%\centering
%\includegraphics[scale=0.85]{ems_vicinity_model.pdf}
%\caption[The vicinity of the Moon]{Vicinity of the Moon for $C=3.188\lesssim C_1$ and $C=3.173\gtrsim C_2$. The surface of section $\Sigma$ is situated between the Moon and the Lagrangian point $L_2$, and it can be used to study the prograde orbits in the system, since it is defined for $\dot{y}>0$.}
%\label{fig:ems_vicinity_model}
%\end{figure}

The equations of motion are numerically integrated with the explicit embedded Runge-Kutta Prince-Dormand 8(9) method \cite{Galassi2001}, as mentioned in Sec.~\ref{sec:dynamical_systems}. The errors associated with all the numerical procedures are controlled and kept below $10^{-10}$, and the Jacobi constant is verified to be preserved to, at least, the same order of magnitude.

\begin{figure}[h!]
\centering
\includegraphics[scale=0.868]{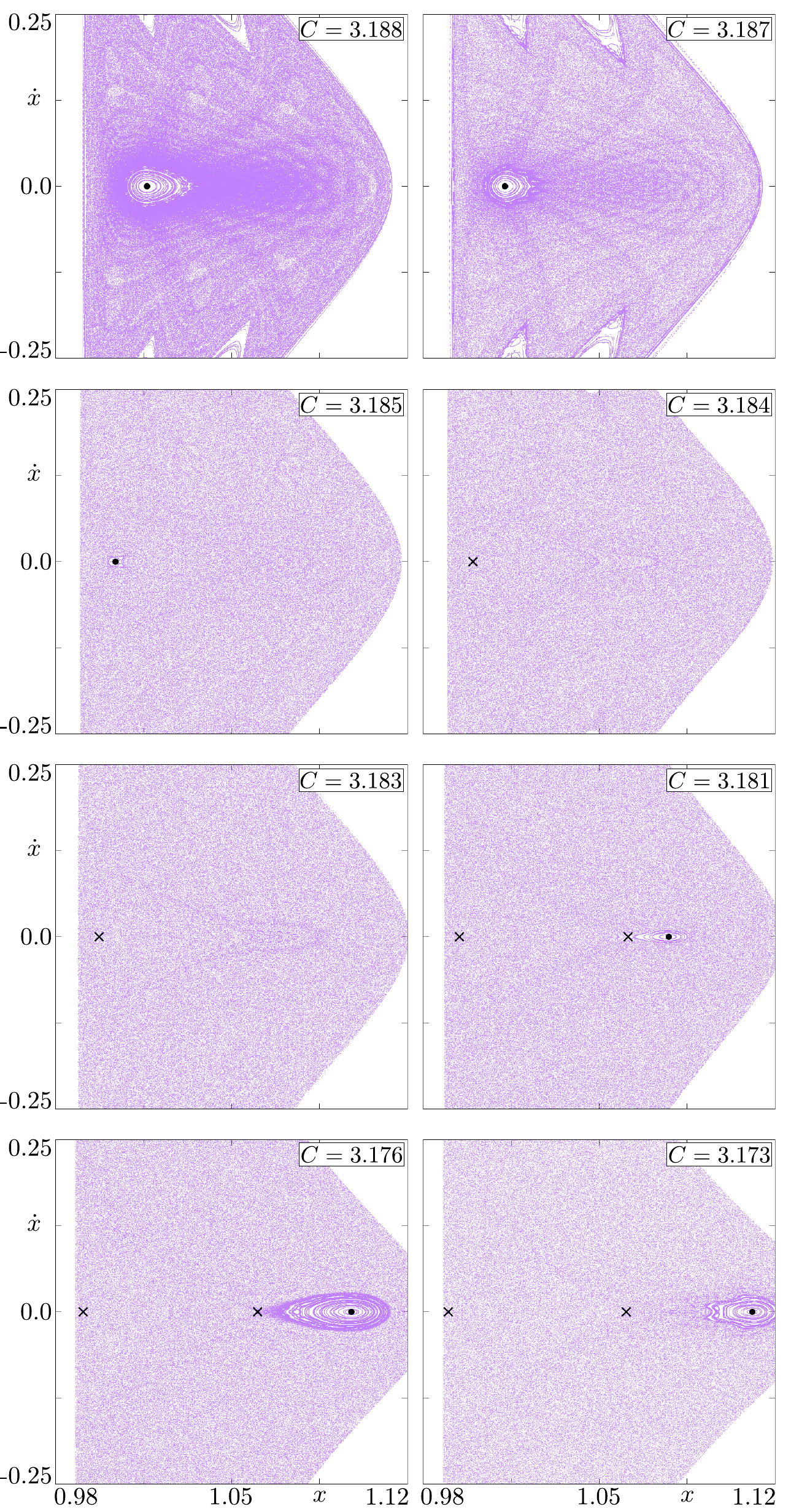}
\caption[Phase space of the Earth-Moon system]{Phase space $x$-$\dot{x}$ of the Earth-Moon system in the surface of section $\Sigma$ for different values of the Jacobi constant $C$. The system starts at a mixed scenario, goes through a global chaos scenario, and then back to a mixed scenario but with different stickiness properties. The black dots and crosses represent the main stable and unstable periodic orbits, respectively.}
\label{fig:ems_phase_space}
\end{figure}

The system's phase space $x$-$\dot{x}$ is presented in Fig.~\ref{fig:ems_phase_space}. The initial conditions are given by a $36\times36$ grid in $\Sigma$, and the orbits are integrated for $t=5\times10^3$ units of time both forward and backward. The first aspect we observe is the presence of three scenarios when we vary the Jacobi constant from $C=3.188$ to $C=3.173$. Scenario I. ($C=3.188$, $3.187$ and $3.185$) The system presents a mixed phase space and, as the Jacobi constant is lowered, the regular region starts to shrink. Scenario II. ($C=3.184$ and $3.183$) The invariant tori were destroyed and the chaotic sea dominates the phase space. Scenario III. ($C=3.181$, $3.176$ and $3.173$) The system becomes mixed again as a new stability region is created at another area of the phase space.

To determine for which Jacobi constant the transitions between the scenarios take place, and also to better understand what happens with the regular solutions, we study the stability of the main periodic orbits in each mixed case. The intersection of these orbits with $\Sigma$ is marked in Fig.~\ref{fig:ems_phase_space}, with dots for stable, and crosses for unstable, trajectories. 

\begin{figure}[h!]
\centering
\subfloat[Scenario I]{\includegraphics[scale=0.85]{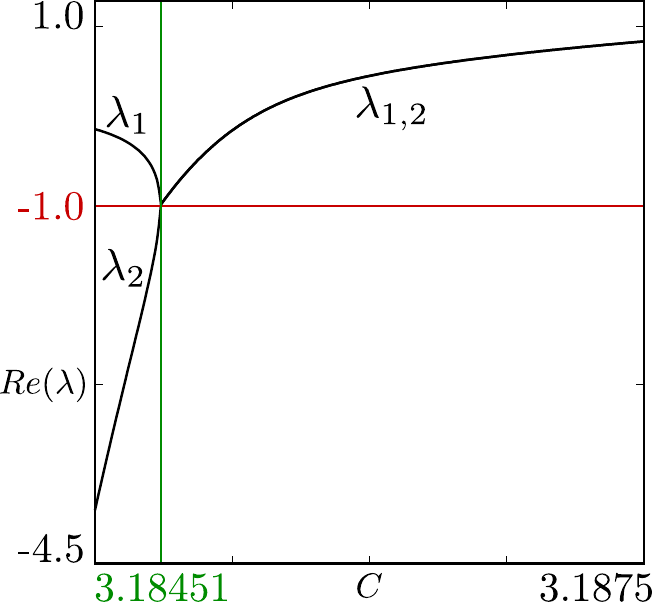}\label{subfig:inner_eigenvalues}}~~
\subfloat[Scenario III]{\includegraphics[scale=0.85]{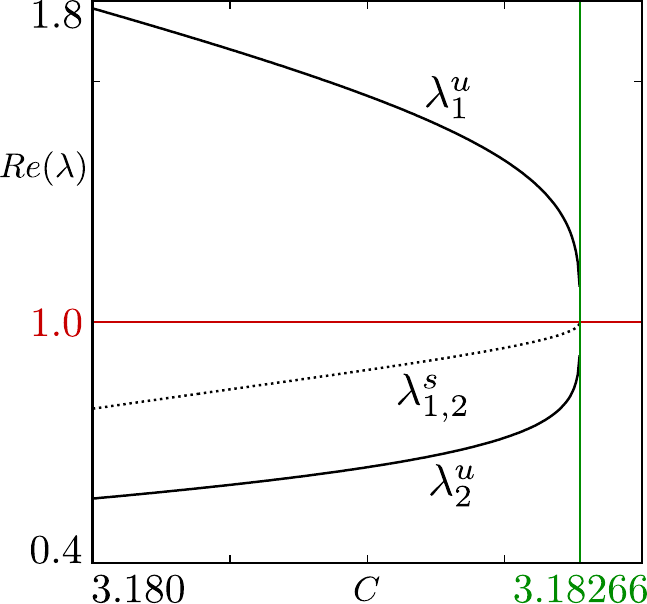}\label{subfig:outer_eigenvalues}}\\
\subfloat[Scenario III]{\includegraphics[scale=0.85]{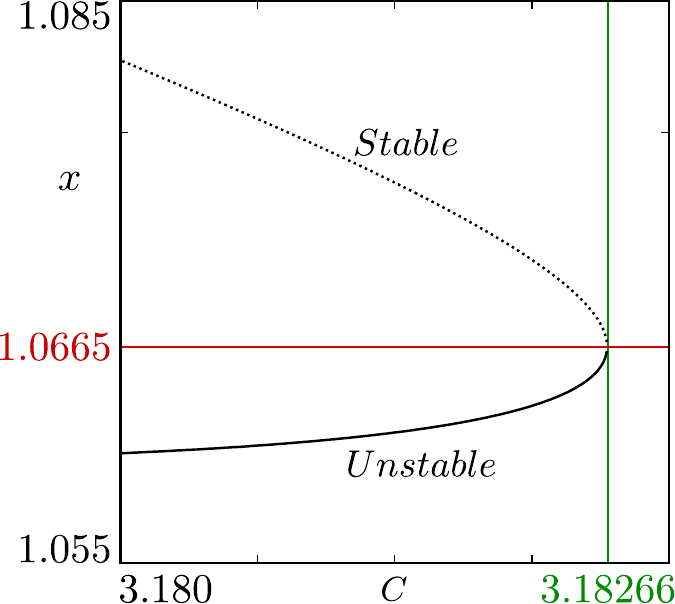}\label{subfig:outer_position}}
\caption[Bifurcation analysis for the direct periodic orbits]{Bifurcation analysis for the direct periodic orbits around the Moon. The real part of the two monodromy matrix eigenvalues that are related to stability is given as a function of the Jacobi constant for \protect\subref{subfig:inner_eigenvalues} Scenario I, and \protect\subref{subfig:outer_eigenvalues} Scenario III, showing two different types of bifurcation. In \protect\subref{subfig:outer_position}, both stable and unstable orbits of Scenario III are shown to collide if we increase $C$, as their location in the $x$-axis approach the same value.}
\label{fig:ems_bifurcation}
\end{figure}

In Scenario I, we have a periodic orbit of period 1, or fixed point, at the center of the regular region, as depicted in Fig.~\ref{fig:ems_phase_space}. The real part of the relevant monodromy matrix eigenvalues is shown as a function of $C$ in Fig.~\ref{subfig:inner_eigenvalues}. We can observe that, as we lower the Jacobi constant, the orbit goes from stable to unstable at approximately $C_{bif}^1=3.18451$, marking the transition between Scenarios I and II. Such bifurcation is called \emph{direct} or \emph{inverse}, depending on the stability of another periodic orbit formed outside of $\Sigma$ \cite{Contopoulos2004}.

In Scenario III, there is an additional pair of period-1 orbits, one at the center of the regular region and one to the left of it, as depicted in Fig.~\ref{fig:ems_phase_space}. We present the same analysis as before in Fig.~\ref{subfig:outer_eigenvalues} for both orbits. Initially, for $C=3.180$, we have one stable and one unstable orbit. As we increase the value of $C$, all eigenvalues tend to the same value and the orbits eventually disappear at $C_{bif}^2=3.18266$. In Fig.~\ref{subfig:outer_position}, we plot the $x$-axis position of both orbits and we observe that they indeed collide. We have, here, a saddle-node bifurcation \cite{Contopoulos2004}.

\begin{figure}[h!]
\centering
\subfloat[$C=3.187$]{\includegraphics[scale=0.85]{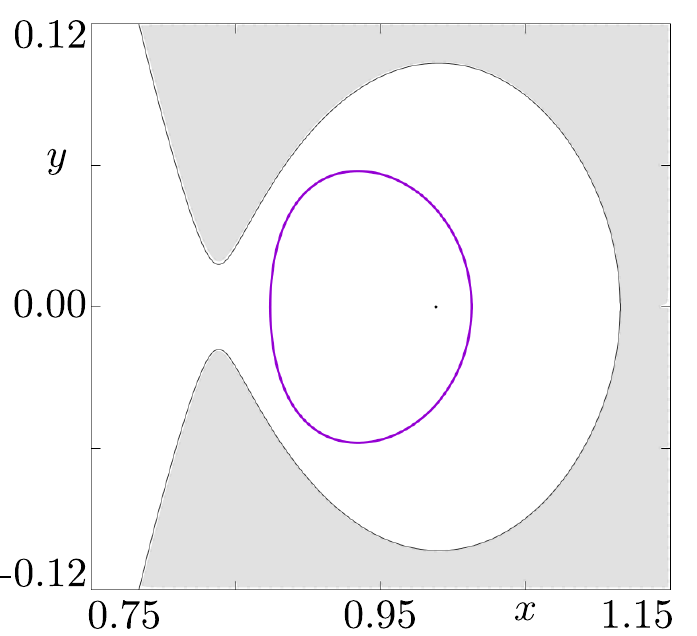}\label{subfig:lunar_orbit_I}} \ \  \ \ \ 
\subfloat[$C=3.176$]{\includegraphics[scale=0.85]{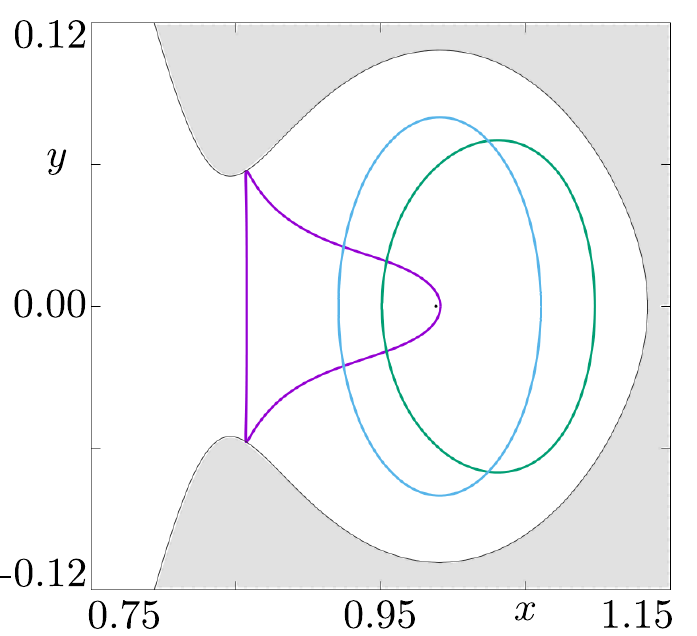}\label{subfig:lunar_orbit_III}}
\caption[Direct periodic orbits around the Moon]{Direct periodic orbits around the Moon. As we move from $C=3.187$ (Scenario I) to $C=3.176$ (Scenario III), one periodic orbit goes from stable to unstable (purple), while other two appear, one stable (green) and one unstable (blue). The intersection of these orbits with $\Sigma$ is marked in Fig.~\ref{fig:ems_phase_space}.}
\label{fig:ems_lunar_orbits}
\end{figure}

All periodic orbits that are studied in Fig.~\ref{fig:ems_bifurcation} belong to the class of \emph{direct periodic orbits} around the Moon, also referred to as the \emph{g class} \cite{Szebehely1967}.  In Fig.~\ref{fig:ems_lunar_orbits}, we trace these orbits for $C=3.187$ and $C=3.176$. Both the periodic orbit in Scenario I (purple) and the stable periodic orbit in Scenario III (green) are called \emph{Low Prograde Orbits}, while the unstable periodic orbit in Scenario III (blue) is called \emph{Distant Prograde Orbit} \cite{Restrepo2018}.

The second aspect we observe from Fig.~\ref{fig:ems_phase_space} is the different properties of stickiness between the mixed phase space scenarios. In Scenario III (e.g., $C=3.176$), the stickiness effect is highly concentrated around the regular region, as is commonly the case. However, in Scenario I (e.g., $C=3.187$), such effect also extends deep into the chaotic sea. This is an indication that the stickiness is probably being caused by invariant manifolds \cite{Contopoulos2010}, which are associated with UPOs located around the stable portion of the phase space. We will address this further in the next section.

An overview of the three different dynamical scenarios that are present in the system is given in Tab.~\ref{tab:scenarios}. The type ``Order'' is used to indicate the presence of regular solutions in the phase space. It is important to note that, as discussed before, the topology of the Hill region remains the same for all these Scenarios.

\renewcommand{\arraystretch}{1.2}
\begin{table}[h!]
\caption{Description of the dynamical scenarios that arise in the Earth-Moon system for $C_1>C>C_2$.}
\label{tab:scenarios}  
\centering
\begin{tabular}{cccc}
\hline\noalign{\smallskip}
Scenario & Range & Type & Stickiness  \\
\noalign{\smallskip}\hline\noalign{\smallskip}
I & $~~C_1>C>C_{bif}^1$ & Order & Non-localized \\
II & $C_{bif}^1>C>C_{bif}^2$ & Chaos & Absent \\
III & $C_{bif}^2>C>C_2~~$ & Order & Localized \\
\noalign{\smallskip}\hline
\end{tabular}
\end{table}

\section{Geometrical structures}
\label{sec:geometrical}

As we discussed before, $L_1$ is the only Lagrangian point inside the Hill region for our chosen range of Jacobi constant values. Furthermore, this point, and consequently the family of Lyapunov orbits around it, are localized between the primaries, separating the Earth's and the Moon's realms.

\begin{figure}[h!]
\centering
\subfloat{\includegraphics[scale=0.85]{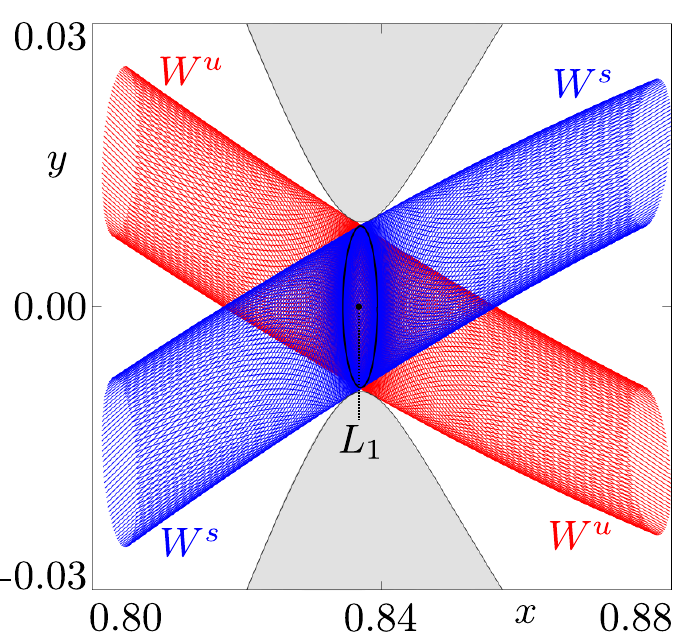}
\label{subfig:3_188_zoom}} \ \ \ \ \ 
\subfloat{\includegraphics[scale=0.85]{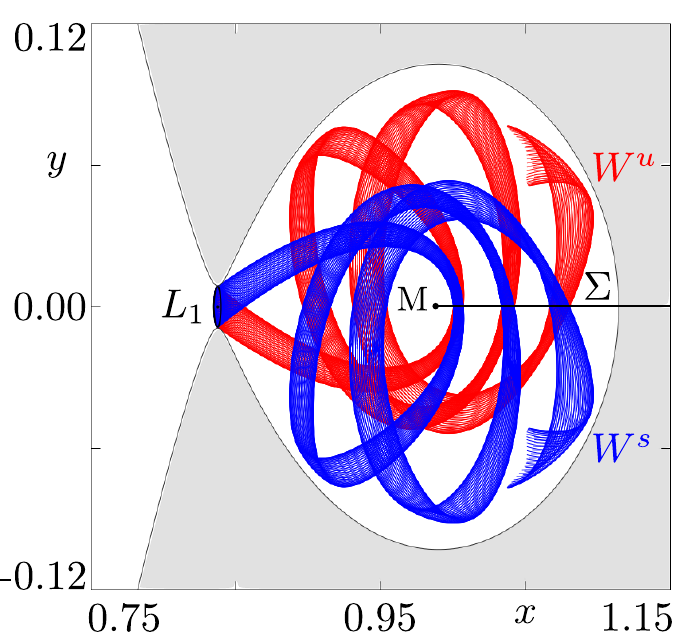}
\label{subfig:3_188_full}}
\caption[Invariant manifolds associated with a Lyapunov orbit]{A part of the stable (blue) and unstable (red) manifolds associated with the Lyapunov orbit for $C=3.188$ projected onto the coordinate space $x$-$y$. Is is interesting to note that each branch starts inside one of the realms.}
\label{fig:ems_manifolds_example}
\end{figure}

In Fig.~\ref{fig:ems_manifolds_example}, we show the invariant manifolds associated with the Lyapunov orbit for $C=3.188$ projected onto the coordinate space $x$-$y$. These structures were calculated following the method presented in Sec.~\ref{sec:2d_manifolds}. In the left panel, we can observe the cylindrical shape of the manifolds close to the Lyapunov orbit. In the right panel, we note the perpendicular crossings of the right branch of these invariant manifolds with our surface of section as it evolves inside the lunar realm.

\begin{figure}[h!]
\centering
  \includegraphics[scale=0.868]{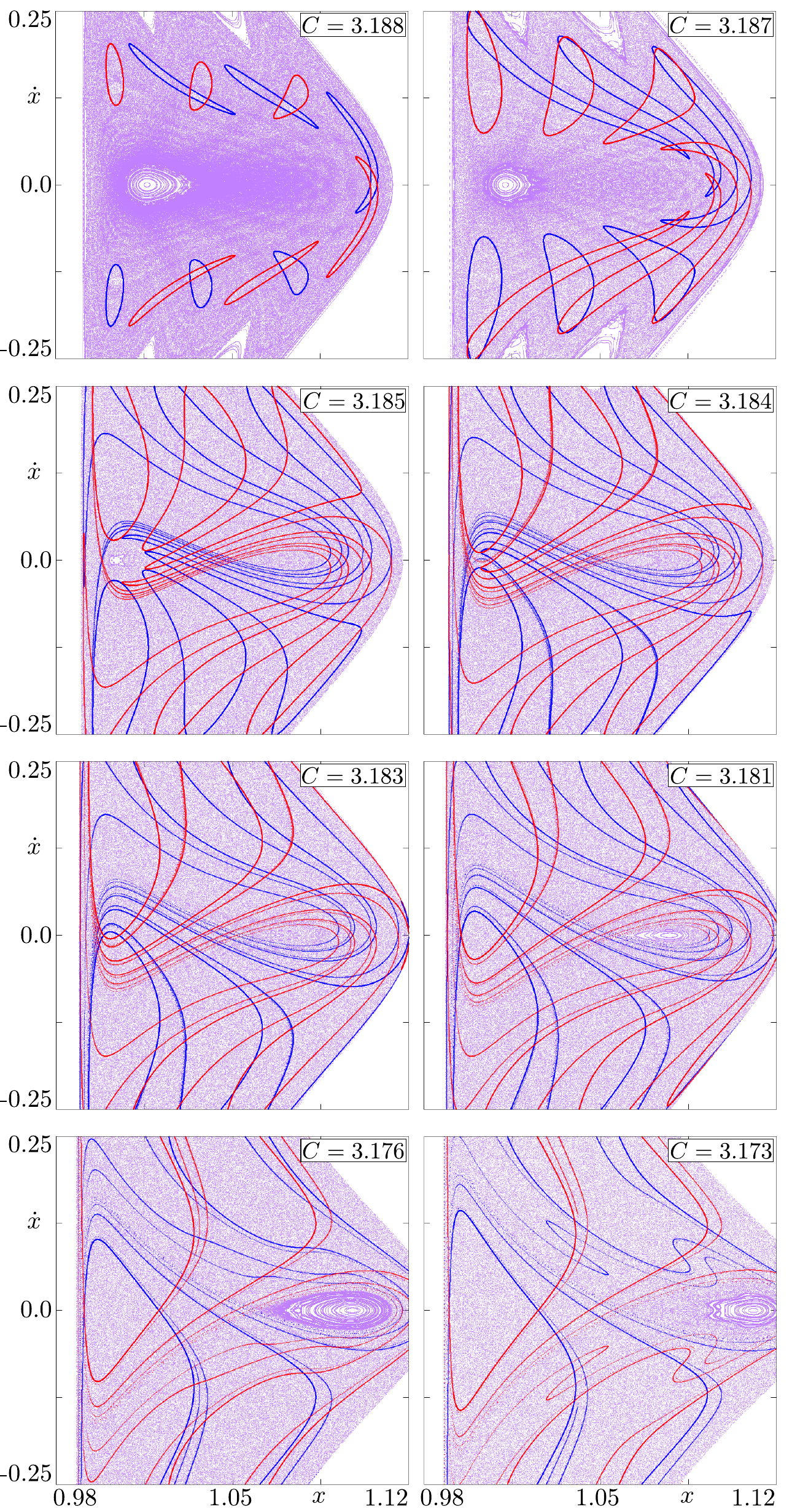}
\caption[Lyapunov orbit manifolds in the phase space]{Phase space $x$-$\dot{x}$ of the Earth-Moon system along with the first few crossings of the Lyapunov orbit manifolds with our surface of section. The geometrical structures and the dynamical configuration evolve together as the Jacobi constant goes from $C=3.188$ to $C=3.173$.}
\label{fig:ems_manifolds_phase_space}
\end{figure}

Let us name $\Gamma(L)$ the crossings between the invariant manifolds associated with a Lyapunov orbit and $\Sigma$. In Fig.~\ref{fig:ems_manifolds_phase_space}, we show the phase space $x$-$\dot{x}$ on our surface of section for the same selected $C$ as in Fig.~\ref{fig:ems_phase_space}, along with the first few components of $\Gamma(L)$. The first aspect we observe is that, as $C$ is lowered, the area of the phase space that is covered by the invariant manifolds increases, which is a reflection of the Lyapunov orbit increasing in length.

The fine relationship between the invariant manifolds associated with the Lyapunov orbit and the configuration of the phase space is the most notable result shown by Fig.~\ref{fig:ems_manifolds_phase_space}. Initially, the crossings happen far from the regular region. As $C$ gets lower, $\Gamma(L)$ starts to invade the stability region, which gets smaller accordingly. Eventually, all KAM curves are destroyed and the system is dominated by chaotic orbits. Finally, a new regular region is created in an area of the phase space that is not covered by the invariant manifolds. At the end of the chosen range, the new stability region starts to shrink as the manifolds oscillate around it.

Another interesting aspect observable from Fig.~\ref{fig:ems_manifolds_phase_space} is that the stickiness effect is somewhat restricted by $\Gamma(L)$. In Scenario I,  the manifolds do not occupy a large area of the phase space and stickiness extends deep into the chaotic sea. In Scenario III, there is a small portion of the phase space available and we can see such dynamical effect confined close to the regular region.

The stickiness phenomenon is likely caused by invariant manifolds associated with UPOs that are located around the regular regions. In order to verify this, we choose a Jacobi constant in each mixed phase space scenario and we calculate the UPO formed by the destruction of the last KAM curve. We then trace the invariant manifolds associated with such orbits and compare their spatial disposition to the stickiness effect.

\begin{figure}[h!]
\centering
\subfloat[$C=3.187$]{\includegraphics[scale=0.85]{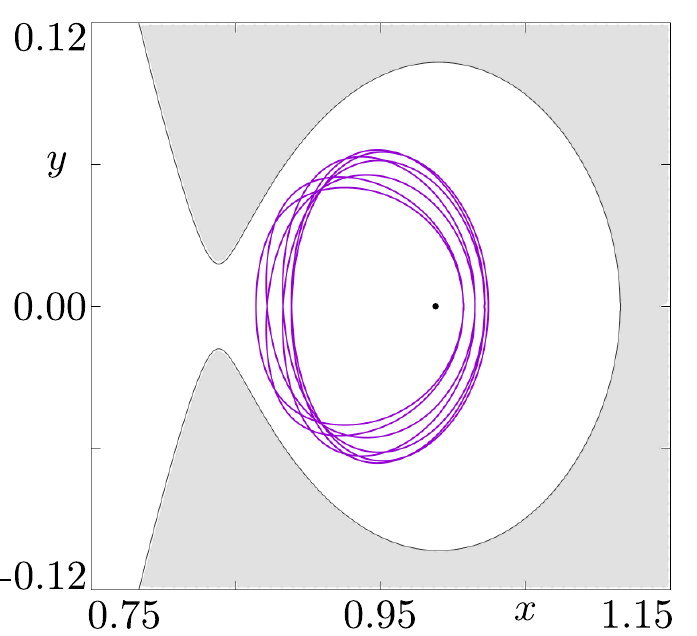}\label{subfig:lunar_satellite_I}} \ \  \ \ \ 
\subfloat[$C=3.176$]{\includegraphics[scale=0.85]{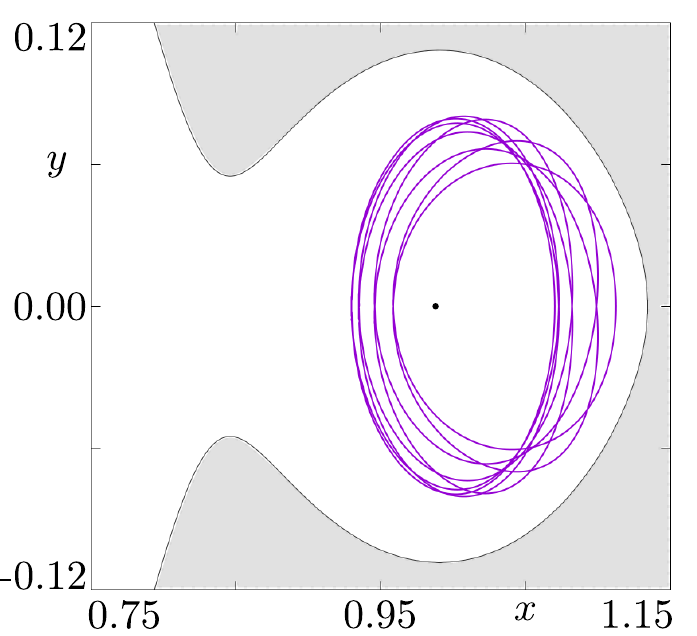}\label{subfig:lunar_satellite_III}}
\caption[Higher-order periodic orbits around the Moon]{Unstable periodic orbits around the Moon, which were formed from the destruction of the last invariant torus in the regular region, as seen from the coordinate space $x$-$y$. The UPO in Scenario I, $C=3.187$, crosses our surface of section seven times before closing in on itself. For Scenario III, $C=3.173$, we have an UPO of period 8.}
\label{fig:ems_higher_period_upo}
\end{figure}

\begin{figure}[h!]
\centering
\subfloat[$C = 3.187$]{\includegraphics[scale=0.85]{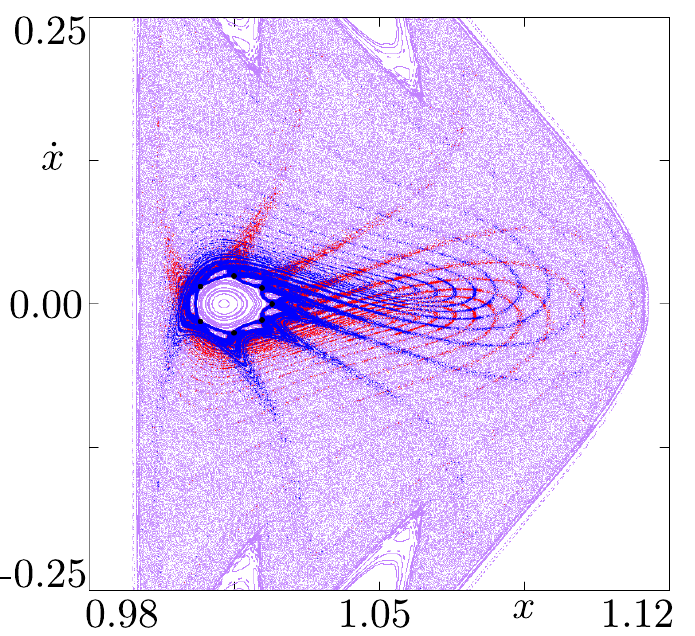}\label{subfig:manifolds_inner}}
\subfloat[$C = 3.187$]{\includegraphics[scale=0.85]{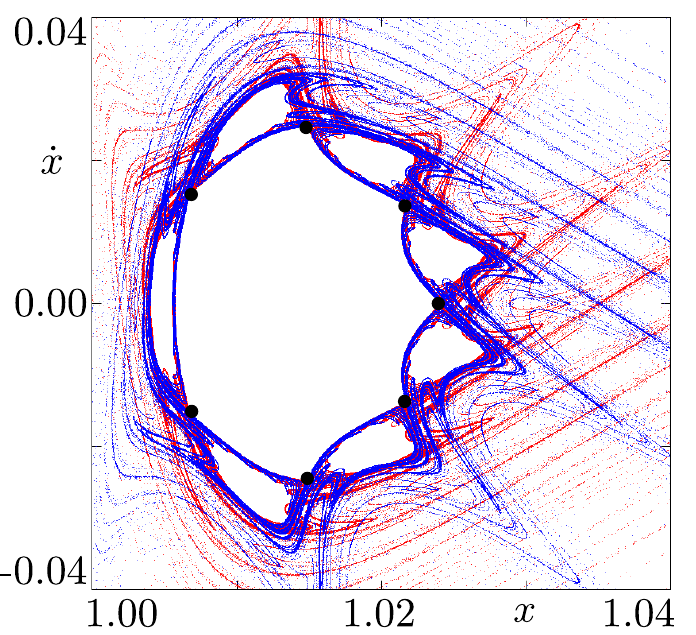}\label{subfig:manifolds_inner_zoom}}
\\
\subfloat[$C = 3.176$]{\includegraphics[scale=0.85]{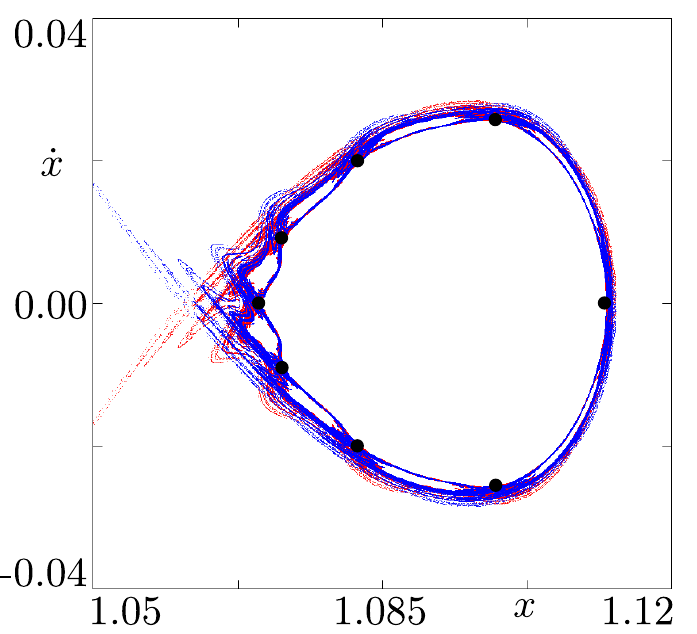}\label{subfig:manifolds_outer}}
\subfloat[$C = 3.181$]{\includegraphics[scale=0.85]{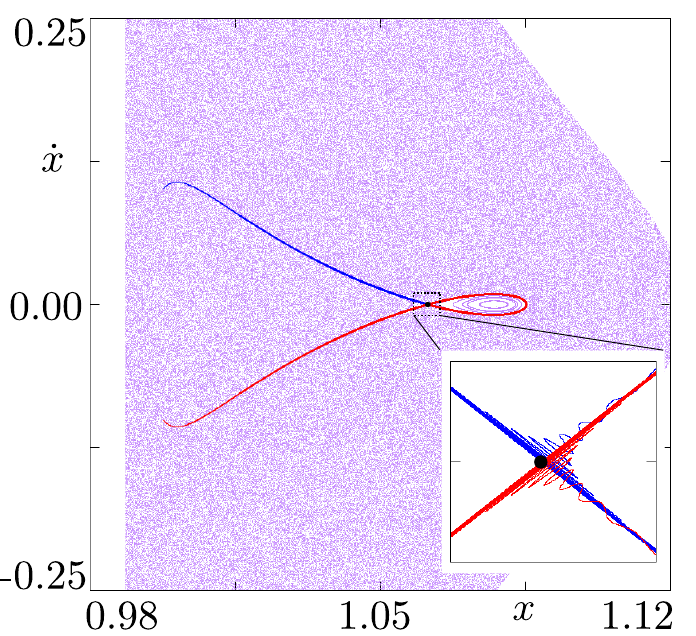}\label{subfig:saddle_manifolds}}
\caption[Invariant manifolds associated with specific UPOs]{Selected unstable periodic orbits and their respective invariant manifolds in the phase space $x$-$\dot{x}$. In \protect\subref{subfig:manifolds_inner} and \protect\subref{subfig:manifolds_inner_zoom} we present $\Gamma(P^7_{I})$, in \protect\subref{subfig:manifolds_outer} we trace $\Gamma(P^8_{III})$, and in \protect\subref{subfig:saddle_manifolds} we have $\Gamma(P^1_{III})$. The periodic orbits are marked with black points.}
\label{fig:ems_additional_manifolds}
\end{figure}

In Fig.~\ref{fig:ems_higher_period_upo}, we present both aforementioned unstable periodic orbits in coordinate space $x$-$y$. For Scenario I, we choose $C=3.187$ and we calculate an orbit of period 7 around the stability region. For Scenario III, we have $C=3.176$ and the UPO has period 8. We call these orbits $P^7_{I}$ and $P^8_{III}$, respectively, and we show the crossings of their invariant manifolds with our surface of section in Fig.~\ref{fig:ems_additional_manifolds}.

$\Gamma(P^7_{I})$ are presented in Fig.~\ref{subfig:manifolds_inner}, alongside the chaotic portion of the phase space, and in Fig.~\ref{subfig:manifolds_inner_zoom} as a close-up. We observe that these structures closely follow the pattern of the stickiness effect, extending deep into the chaotic sea. The same happens with $\Gamma(P^8_{III})$, shown in Fig.~\ref{subfig:manifolds_outer}. In this case, the manifolds stay close to the regular region, as we would expect. Hence, the stickiness phenomenon here can be interpreted as a consequence of the dynamics induced by these groups of manifolds.

In Fig.~\ref{subfig:saddle_manifolds}, we choose $C = 3.181$ and we trace the invariant manifolds associated with the saddle that was formed from the second bifurcation at $C^2_{bif}$, which we call $P^1_{III}$. We can see that $\Gamma(P^1_{III})$ is less complex than the aforementioned groups of manifolds, with just a small oscillation near the saddle. However, it is interesting to note that there exists a ghost effect in the phase space, right before the second bifurcation takes place, where orbits start to accumulate by following the same pattern as given by these manifolds. This phenomenon can be seen in Fig.~\ref{fig:ems_phase_space} for $C=3.183$.

\begin{figure}[h!]
\centering
\subfloat[$C = 3.187$]{\includegraphics[scale=0.85]{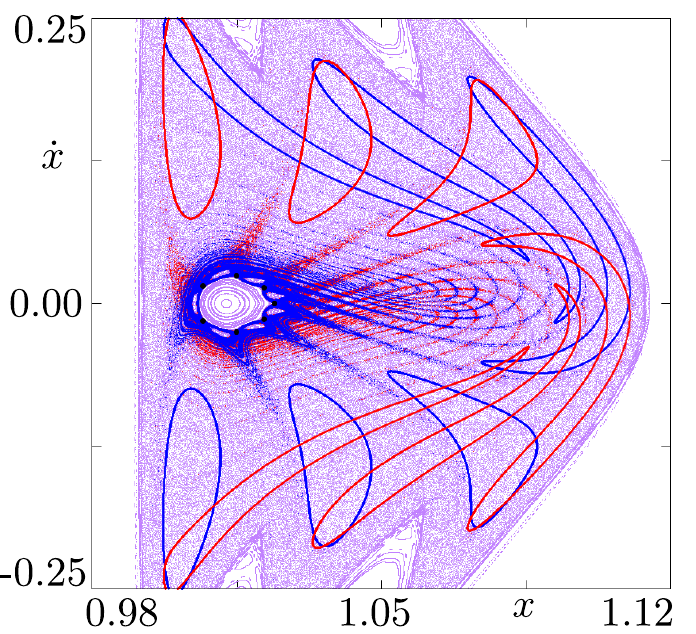}\label{subfig:all_manifolds_inner}}
\subfloat[$C = 3.176$]{\includegraphics[scale=0.85]{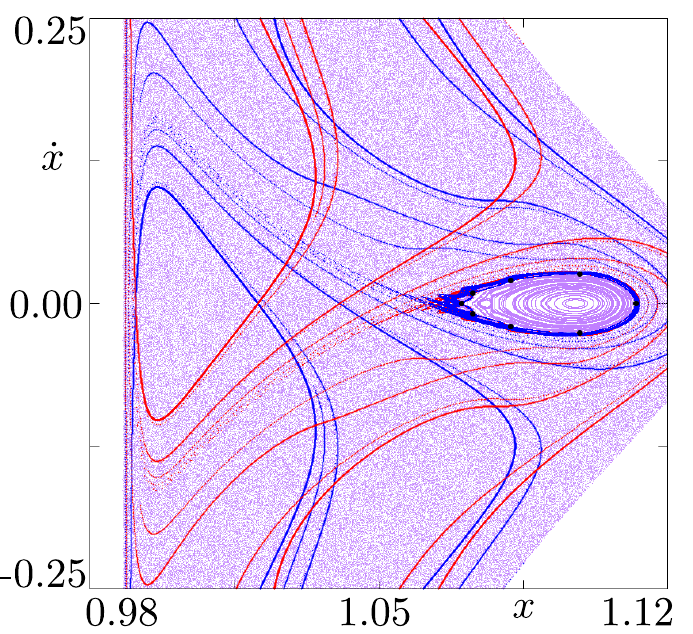}\label{subfig:all_manifolds_outer}}
\caption[Overview of the main geometrical structures in phase space]{Overview of the main geometrical structures in the phase space for $C = 3.187$ (Scenario I) and $C = 3.176$ (Scenario III). There is a close relationship between dynamics and geometry in the system.}
\label{fig:ems_manifolds_all}
\end{figure}

Differently from $\Gamma(L)$, the invariant manifolds depicted in Fig.~\ref{fig:ems_additional_manifolds} do not cross the surface $\Sigma$ transversally and, therefore, these structures resemble the manifolds traced in two-dimensional maps.
In Fig.~\ref{fig:ems_manifolds_all}, we give an overview of the main invariant manifolds that are present in the phase space $x$-$\dot{x}$ for $C=3.187$ and $C=3.176$.
All these groups of geometrical structures are clearly closely related to the different dynamical properties of the system.

\section{Transport properties}
\label{sec:transport}

One more aspect we observe in the system's phase space, Fig.~\ref{fig:ems_phase_space}, is the presence of areas with a low density of orbits, specially visible for $C=3.188$. By comparing Fig.~\ref{fig:ems_phase_space} to Fig.~\ref{fig:ems_manifolds_phase_space}, we can see that these regions coincide with the areas enclosed by $\Gamma(L)$. This is a consequence of the fact that the invariant manifolds associated with a Lyapunov orbit are responsible for the transport between the realms \cite{Koon2008}. The trajectories that are inside the first components of $\Gamma^s(L)$ leave the Moon's realm faster, thus making these areas less populated than other regions of the phase space.

\begin{figure}[h!]
\centering
\subfloat[Forward]{\includegraphics[scale=0.85]{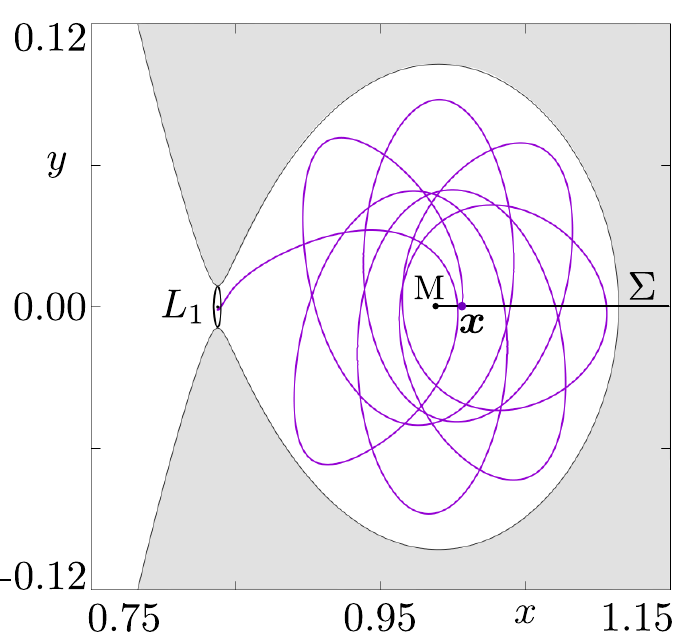}\label{subfig:orbit_fw}} \ \  \ \ \ 
\subfloat[Backward]{\includegraphics[scale=0.85]{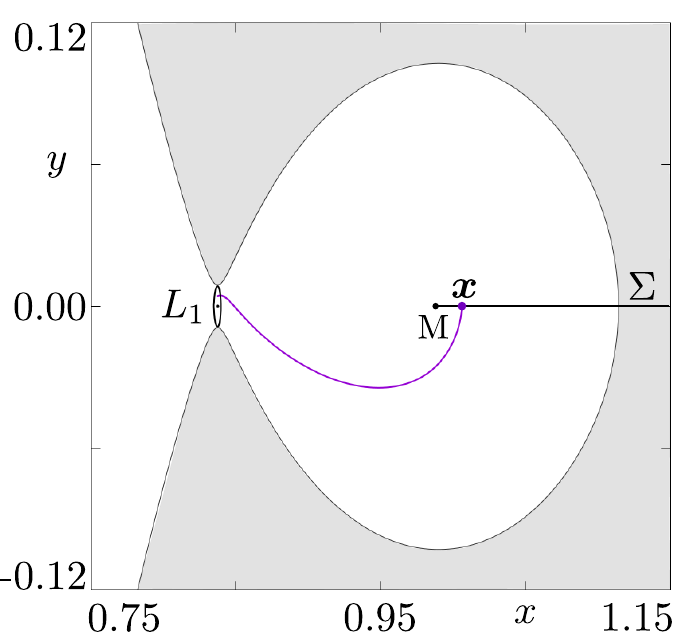}\label{subfig:orbit_bw}}
\caption[Forward and backward exit time]{Trajectory that starts in ${\bs{x}}\in\Sigma$ exiting the lunar realm \protect\subref{subfig:orbit_fw} forward and \protect\subref{subfig:orbit_bw} backward in time, which happens at a time $t=t_f$ and $t=t_b$, respectively. The multiplicative transit time is then given by $|t_ft_b|$, where $t_f>0$ and $t_b<0$.}
\label{fig:ems_transit_time_example}
\end{figure}

Let $t_f$ and $t_b$ be the time it takes for an orbit that starts in $\Sigma$ to exit the Moon's realm forward and backward in time, respectively, as shown in Fig.~\ref{fig:ems_transit_time_example}. In order to investigate the influence of the different groups of manifolds studied so far in the transport properties of the system, we introduce the \emph{multiplicative transit time}, which is defined as the absolute value of the \emph{product} of $t_f$ and $t_b$. By defining a transit time in this manner, we ensure that the orbits which stay the longest, or the shortest, amount of time inside the lunar realm are highlighted.

\begin{figure}[h!]
\centering
\includegraphics[scale=0.868]{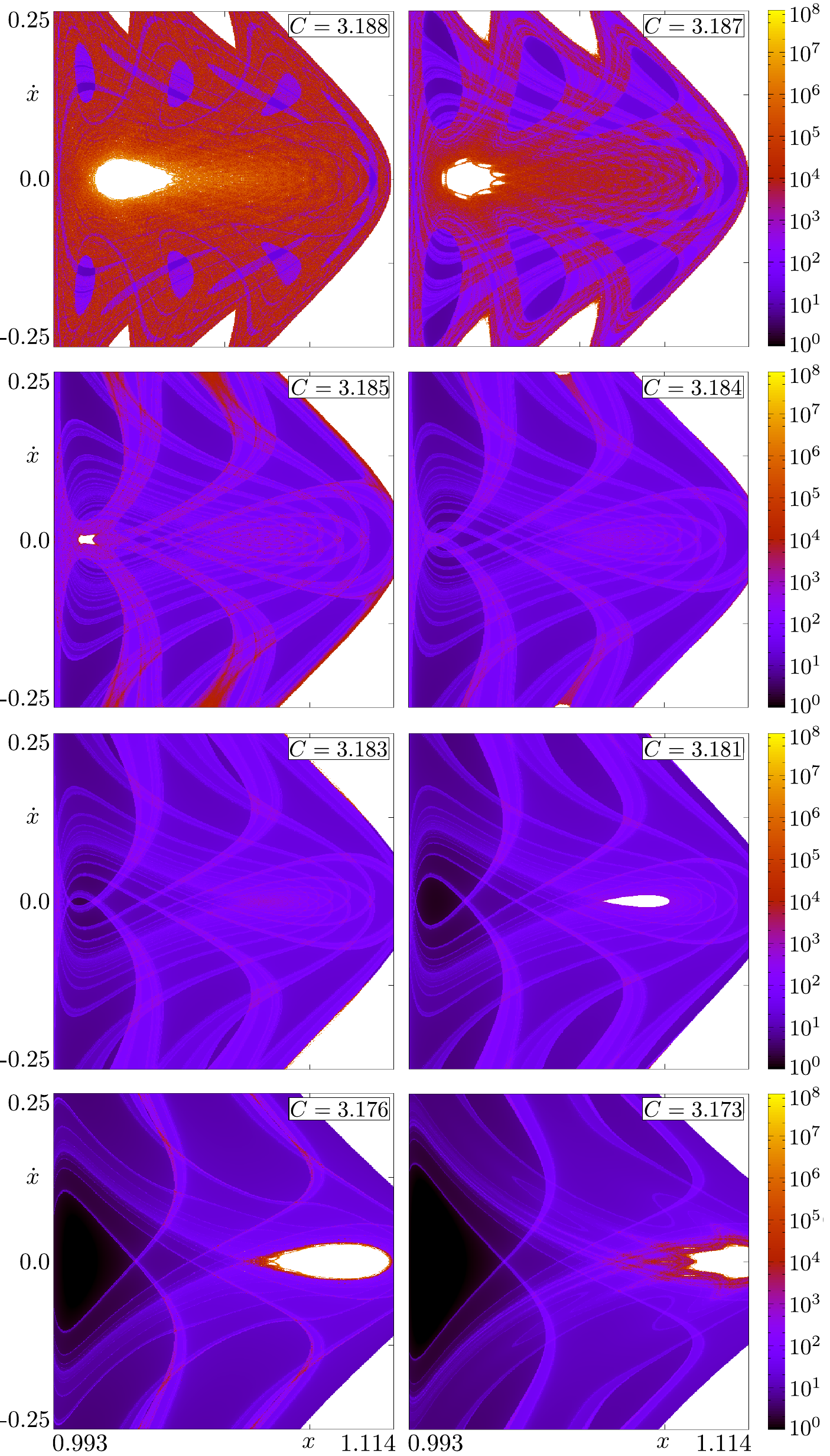}
\caption[Multiplicative transit time profile]{Profile of the multiplicative transit time in logarithmic scale assessed in the phase space $x$-$\dot{x}$. All orbits are chosen in $\Sigma$ and only the ones that exit the lunar realm are considered for analysis.}
\label{fig:ems_transit_time}
\end{figure}

In Fig.~\ref{fig:ems_transit_time}, we plot the profile of the multiplicative transit time in a logarithmic scale for the same values of Jacobi constant as before and assessed in the phase space $x$-$\dot{x}$. For this analysis, we chose a grid of $512 \times 1024$ initial conditions in $\Sigma$, integrated each of them for a maximum of $t = 5 \times 10^3$ both forward and backward in time, and considered only the orbits that did exit the lunar realm. 

By comparison to Figs.~\ref{fig:ems_manifolds_phase_space} and~\ref{fig:ems_additional_manifolds}, it is clear that the orbits with the lowest multiplicative transit time are the ones enclosed by $\Gamma(L)$, specially the orbits inside of the intersection between $\Gamma^s(L)$ and $\Gamma^u(L)$, for these are the ones that most rapidly enter and exit the lunar realm. It is also clear that, for $C=3.187$ and $C=3.176$, the orbits with the highest multiplicative transit time are the ones close to  $\Gamma(P^7_{I})$ and $\Gamma(P^8_{III})$.

What we observe, then, is a fine interplay between the roles of the different groups of invariant manifolds in the transport properties of the system. On the one hand, the manifolds associated with the Lyapunov orbit control the transfer of orbits from and to the lunar realm. On the other hand, the manifolds associated with higher-order UPOs dynamically trap the orbits inside the lunar realm.

One important aspect seen from the multiplicative transit time profiles is that all the orbits in the chaotic sea, except for a set of measure zero, do eventually exit the lunar realm. Given that the invariant manifolds associated with a Lyapunov orbit are responsible for the transport between the realms, this implies that $W(L)$ densely fills the chaotic portion of the phase space. It is important, then, to understand how these structures evolve over time.

\begin{figure}[h!]
\centering
\subfloat[$C = 3.188$]{\includegraphics[scale=0.85]{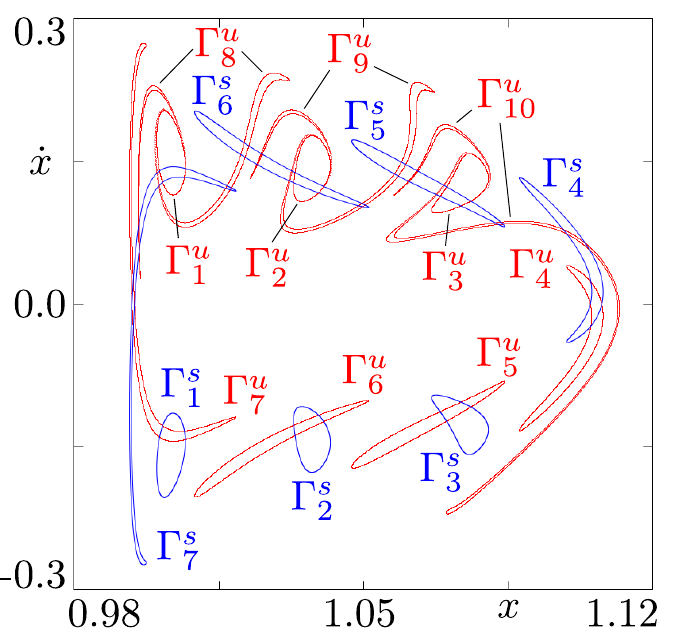}\label{subfig:break_3_188}} \ \  \ \ \ 
\subfloat[$C = 3.175$]{\includegraphics[scale=0.85]{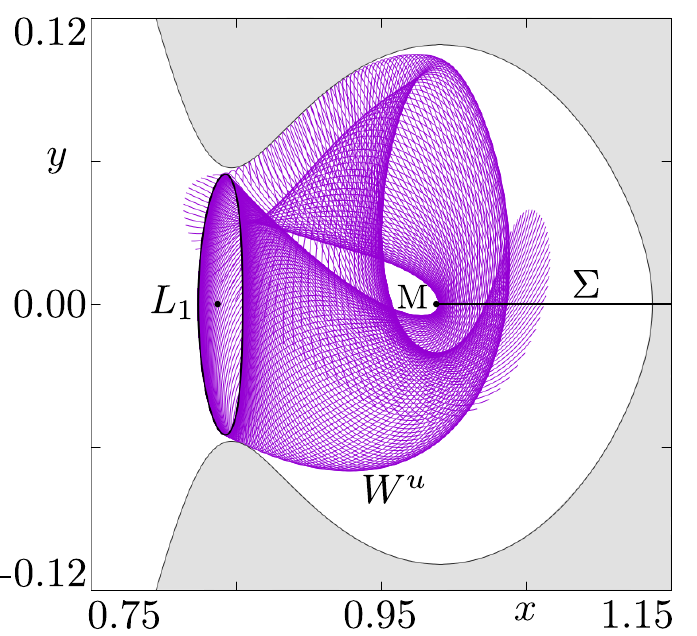}\label{subfig:break_3_175}}
\caption[Intersect and break process]{Intersect and break process on the Lyapunov orbit manifolds seen \protect\subref{subfig:break_3_188} in the phase space $x$-$\dot{x}$ for $C = 3.188$ and \protect\subref{subfig:break_3_175} in the coordinate space $x$-$y$ for $C = 3.175$.}
\label{fig:ems_intersect_and_break}
\end{figure}

The first few components of $\Gamma(L)$ are homeomorphic to circles, however, this property is eventually lost \cite{Gidea2007}. Let us focus on the right branch of the \emph{unstable} invariant manifold. After there is an intersection between a component of $\Gamma^u(L)$ and a component of $\Gamma^s(L)$, the manifold eventually breaks as the next components of $\Gamma^u(L)$ are divided into pieces. We name such process \emph{intersect and break}, and we exemplify it in Fig.~\ref{fig:ems_intersect_and_break}. 

%\begin{figure}[h!]
%\centering
%\subfloat[$C = 3.188$]{\includegraphics[scale=0.85]{ems_manifolds_break_phase_space_higher_resolution.pdf}\label{subfig:break_3_188}} \ \  \ \ \ 
%\subfloat[$C = 3.175$]{\includegraphics[scale=0.85]{ems_manifolds_break_coordinate_space.pdf}\label{subfig:break_3_175}}
%\caption[Intersect and break process]{Intersect and break process on the Lyapunov orbit manifolds seen \protect\subref{subfig:break_3_188} in the phase space $x$-$\dot{x}$ for $C = 3.188$ and \protect\subref{subfig:break_3_175} in the coordinate space $x$-$y$ for $C = 3.175$.}
%\label{fig:ems_intersect_and_break}
%\end{figure}

In Fig.~\ref{subfig:break_3_188}, we analyze the invariant manifolds as seen in the phase space $x$-$\dot{x}$ for $C=3.188$. In this case, the first component of the unstable manifold $\Gamma^u_1$ intersects the seventh component of the stable manifold $\Gamma^s_7$. The part of $\Gamma^u_1$ that is enclosed by $\Gamma^s_7$ follows the system's dynamics and transfers to the Earth's realm after mapping $\Sigma$ six more times. The other part of $\Gamma^u_1$ remains in the Moon's realm, but divided in two pieces. Then, what we see is that $\Gamma^u_8$ is formed by two segments that are not closed, but rather asymptotically approach $\Gamma^u_1$. This structure can be seen in Fig.~\ref{fig:ems_transit_time} for the same Jacobi constant.

The parts of $\Gamma^u$ that got disconnected also intersect $\Gamma^s$ and eventually break again. This process happens indefinitely, as the invariant manifolds densely fill the phase space. As a consequence, we can see the auto-similar structures formed in Fig.~\ref{fig:ems_transit_time}, specially visible for $C=3.187$.

In Fig.~\ref{subfig:break_3_175}, we look at the intersect and break process, but now from the coordinate space $x$-$y$, and for $C=3.175$. In this case, the process can already be seen after the first crossing of the unstable manifold $W^u$ with the surface of section. After this happens, a part of the manifold goes through the Lyapunov orbit and into the Earth's realm, while another part revolves around the Moon and crosses $\Sigma$ again.

It is important to emphasize here that the invariant manifolds $W$ are always continuous in the phase space, and it is their representation on the Poincaré map $\Gamma$ that breaks as $W$ form lobes that extend into the other realm.

\begin{figure}[h!]
\centering
\subfloat{\includegraphics[scale=0.85]{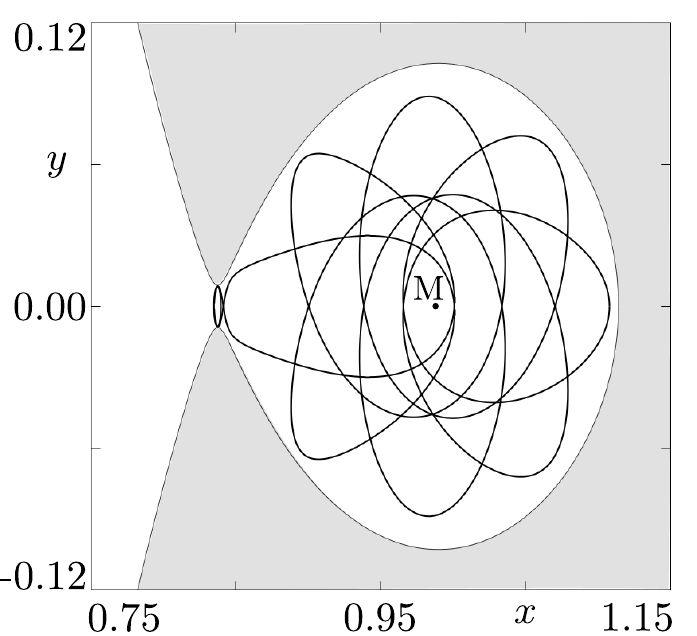}\label{subfig:homolcinic_orbit}} \ \  \ \ \ 
\subfloat{\includegraphics[scale=0.85]{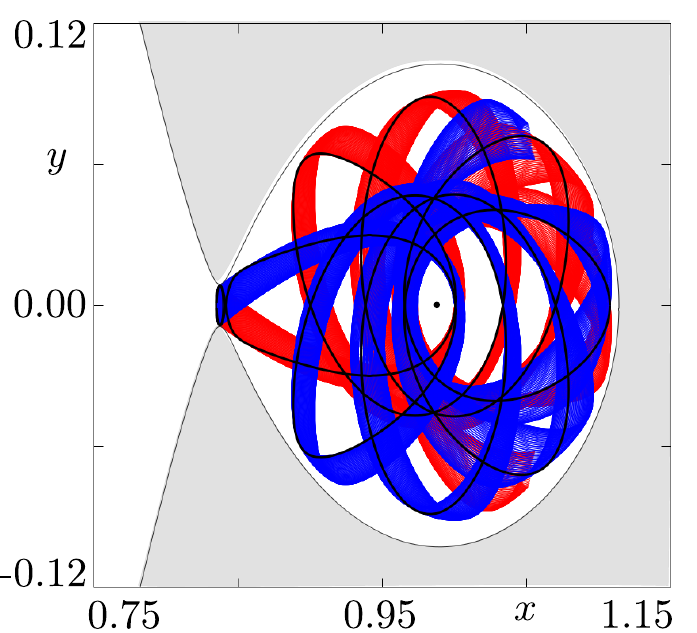}\label{subfig:homoclinic_orbit_manifolds}}
\caption[Homoclinic orbit in the Earth-Moon system]{Example of a homoclinic orbit in the Earth-Moon system. The trajectory starts and finishes at the Lyapunov orbit around $L_1$.}
\label{fig:ems_homoclinic_orbit}
\end{figure}

When an intersection between two invariant manifolds happens, there are also the orbits that belong to both sets at the same time. In Fig.~\ref{fig:ems_homoclinic_orbit}, we show one of these orbits from the intersection between $\Gamma^s_4$ and $\Gamma^u_4$ in Fig.~\ref{subfig:break_3_188}. Since both manifolds are associated with the same UPO, such orbit is \emph{homoclinic}. On the left panel, we can observe the traced orbit moving from and back to the Lyapunov orbit. On the right panel, we overlap the homoclinic orbit with a part of the respective invariant manifolds.

A somewhat similar process happens between the Lyapunov orbit manifolds and $\Gamma(P^7_{I})$ or $\Gamma(P^8_{III})$. Even though $\Gamma(L)$ seems to restrict the other manifolds in Fig.~\ref{fig:ems_manifolds_all}, they are allowed to and they do intersect. This is the mechanism which removes orbits trapped inside the lunar realm and transfers them to the Earth's realm, or the other way around. In this case, we have \emph{heteroclinic} orbits, since it involves two different UPOs.

The intersect and break process that was described here is a consequence of the fact that these invariant manifolds are two-dimensional surfaces. For one-dimensional manifolds, the intersection scheme is much simpler. In both cases, however, the homoclinic and heteroclinic orbits are of special importance since they represent natural connections within the system. 

One last feature we examine is the role of invariant manifolds on the system's transport properties when collision with the primaries is considered. In Fig.~\ref{subfig:break_3_175}, for example, we can observe that part of the unstable manifold passes very close to the Moon, implying that these structures influence the manner in which collision happens in the system. In order to account for this effect, we define a radius around the Moon given by $r_M=4.52\times 10^{-3}$, which corresponds to the satellite's physical mean radius converted to our dimensionless units \cite{Williams2019}, and we stop the numerical integration if the orbit reaches this region.

\begin{figure}[h!]
\centering
\includegraphics[scale=0.868]{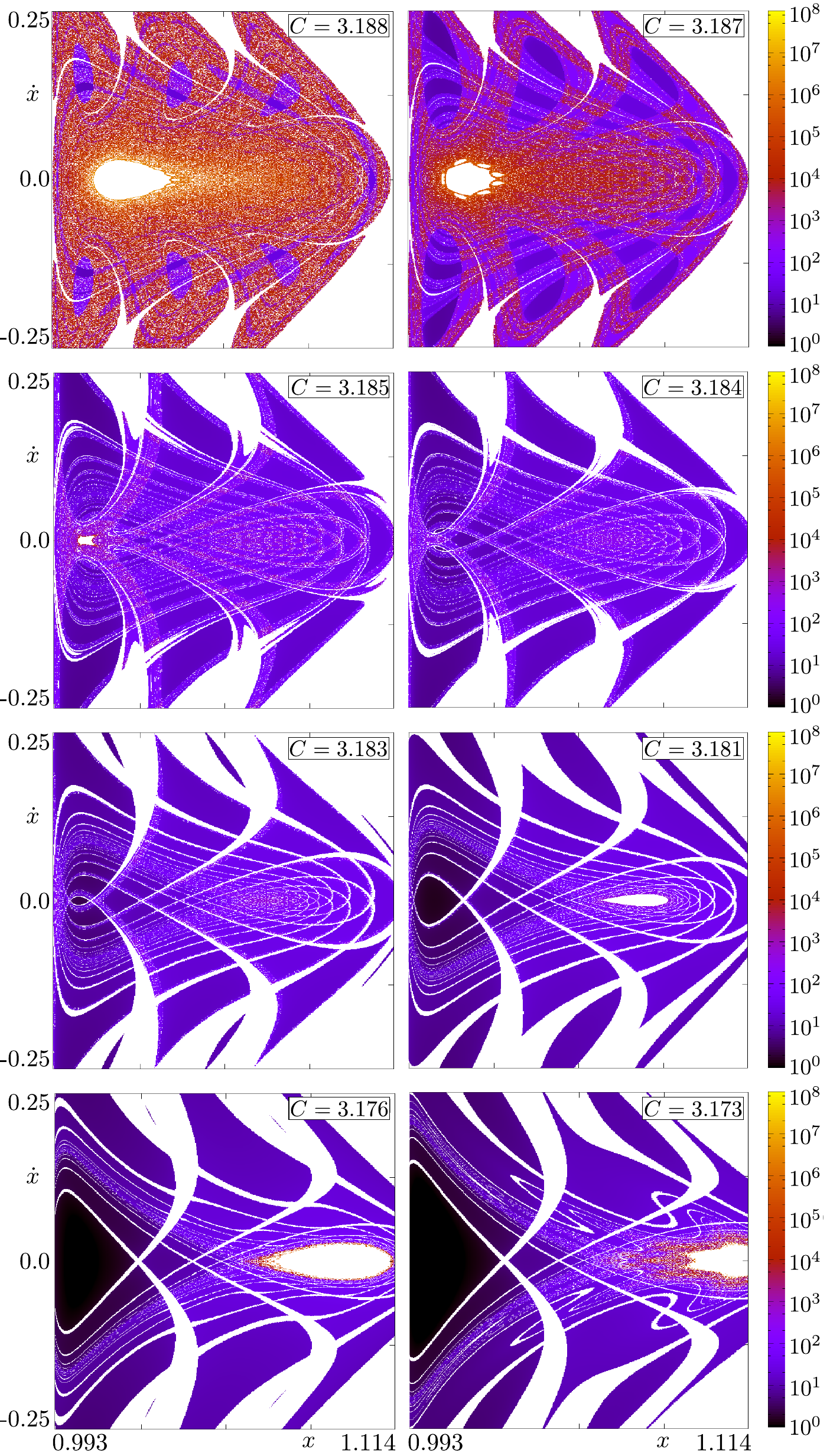}
\caption[Multiplicative transit time profile considering collision]{Multiplicative transit time profiles assessed in the phase space $x$-$\dot{x}$ on a logarithmic scale and considering collision. As before, all orbits are chosen in $\Sigma$ and only the ones that exit the lunar realm are considered for analysis.}
\label{fig:ems_transit_time_with_collision}
\end{figure}

In Fig.~\ref{fig:ems_transit_time_with_collision}, we present the multiplicative transit time profiles considering collision. In practice, we cannot associate a multiplicative transit time to a Moon-collisional orbit, since it ends before leaving the lunar realm, a phenomenon also known as \emph{leaking} \cite{Assis2014}. Therefore, the set of collisional orbits appear in white in the profiles. By comparing Fig.~\ref{fig:ems_transit_time_with_collision} to Fig.~\ref{fig:ems_transit_time}, we can divide this set in two groups.

The first group forms a riddled structure all over the chaotic sea. As the Jacobi constant gets smaller, this structure almost disappears along with the regular region. Later, it grows back again, but it is mostly localized around the new regular region. Such scheme suggests a close relationship between this group of collisional orbits and the invariant manifolds associated with higher-order UPOs. The second group is composed of collisional areas that grow larger as we decrease the Jacobi constant. By comparing the location of such areas to the spatial disposition of the invariant manifolds associated with the Lyapunov orbits, we find these to be closely related as well.

\section{Escape behavior}
\label{sec:escape}

For orbits that \emph{begin} in the vicinity of the Moon, transferring to the Earth's realm may be interpreted as an \emph{escape} from the lunar realm. In this sense, we can consider our system to be \emph{open} and, consequently, consider the dynamics of the escape orbits prior to exiting the system to be a \emph{transient} motion. We are now going to investigate how the dynamical phenomena studied so far, such as stickiness and transport, are experienced by particular ensembles of orbits. In other words, we will investigate how the transient dynamics in the lunar realm is affected by the choice of initial conditions.

To help with this investigation, we develop a method that follows the path of the escape orbits in the phase space. Let $\bs{\varphi}(\bs{x},t)$ be a solution of our dynamical system with \emph{escape time} $T^e$, and let us cover our phase space with a grid composed of $N$ two-dimensional boxes. We define the \emph{escape measure} on a box $B_i$ as

\begin{equation}
    \mu_i^e=\dfrac{\eta(B_i,\bs{\varphi}(\bs{x},t),T^e)}{T^e},
    \label{eq:escape_measure_def}
\end{equation}

\noindent where $\eta(B_i,\bs{\varphi}(\bs{x},t),T^e)$ is the time spent by the solution $\bs{\varphi}(\bs{x},t)$ inside $B_i$ during the time interval $0\leq t <T^e$. By considering the whole grid, we have $\sum^N_{i=1}\mu^e_i=1$.

The escape measure is a finite-time version of the \emph{natural measure} \cite{Ott2002}, and it shows what parts of the phase space are most or least visited by an escape orbit, therefore highlighting the aforementioned dynamical effects. It is probable, though, that a given escape orbit will visit only a small portion of the phase space before exiting the system. Hence, instead of looking at the motion of an individual trajectory, we are going to consider an ensemble $U$ composed of $M$ initial conditions that are very close together.  With this, we define the \emph{mean escape measure} 

\begin{equation}
    \nu_i={\langle\mu_i^e\rangle}_{U}=\dfrac{1}{M}\sum^{M}_{j=1}\mu_{i,j}^e,
    \label{eq:mean_escape_measure_def}
\end{equation}

\noindent where $\mu_{i,j}^e=\eta(B_i,\bs{\varphi}(\bs{x}_j,t),T^e_j)/T^e_j$ is the escape measure on box $B_i$ for the initial condition $\bs{x}_j\in U$, and $T^e_j$ is the escape time of solution $\bs{\varphi}(\bs{x}_j,t)$. Again, we have that $\sum_{i=1}^{N}\nu_i=1$.

\begin{figure}[h!]
\centering
\includegraphics[scale=0.9]{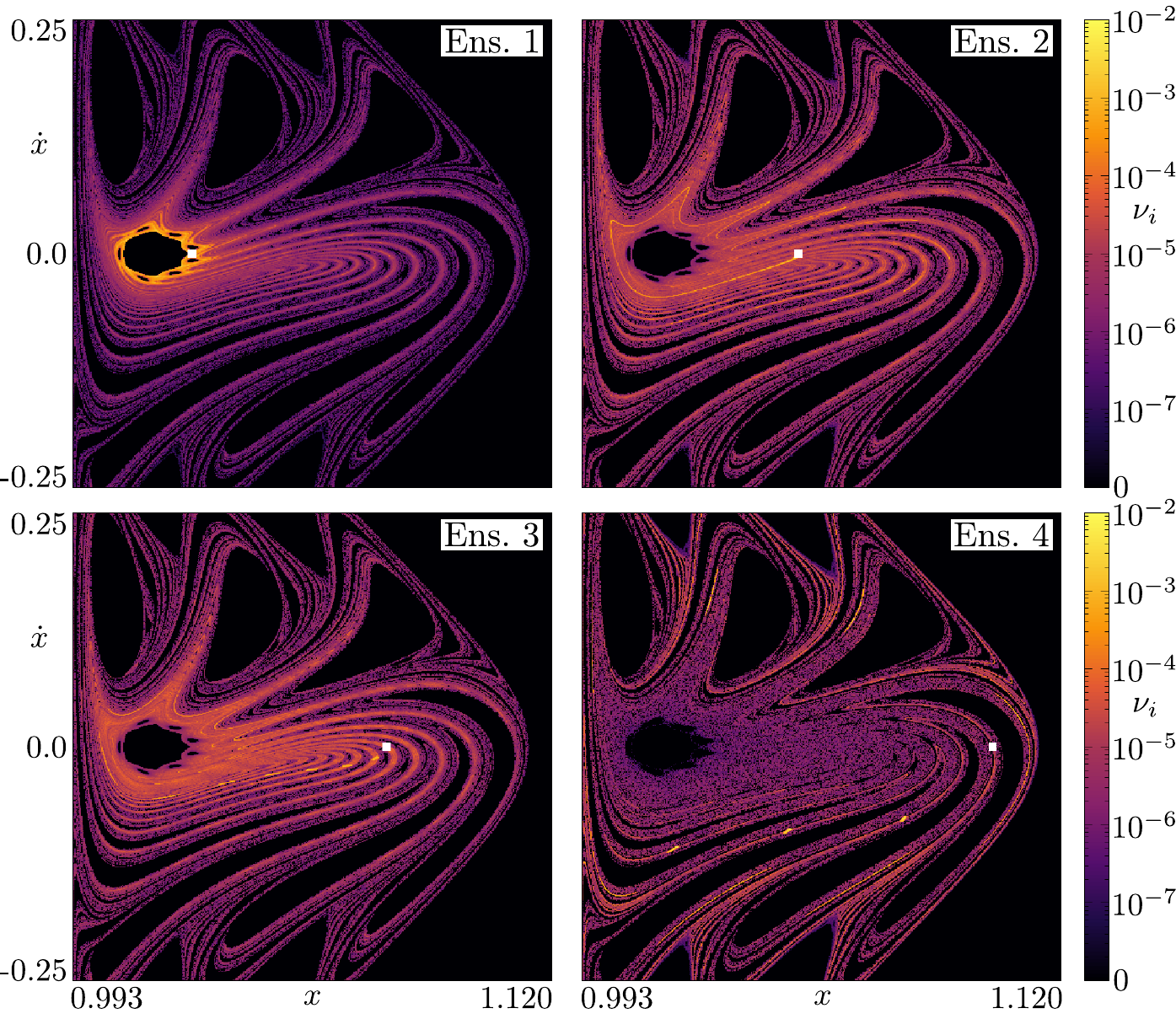}
\caption[Mean escape measure profile]{Profile of the mean escape measure on a logarithmic scale, assessed in the phase space $x$-$\dot{x}$ for $C=3.187$. The white squares mark the location of the ensembles of initial conditions.}
\label{fig:ems_mean_escape_measure_profile}
\end{figure}

In Fig.~\ref{fig:ems_mean_escape_measure_profile}, we present the profile of the mean escape measure on a logarithmic scale for $C=3.187$, calculated on a grid of $512\times512$ boxes for four different ensembles of initial conditions, which are represented by white squares. Each ensemble has size $5\times10^{-4}$ by $4\times10^{-3}$ in the phase space $x$-$\dot{x}$, and is composed of $M=10^4$ points uniformly distributed in a grid. All orbits are chosen in the chaotic sea and are integrated for at most $T_{max}=5\times10^3$ units of time. We choose $T_{max}$ as to make sure that, at least, 85\% of the orbits from each ensemble escape the system. For the ones that do not, we consider $T^e=T_{max}$ in Eq.~\eqref{eq:mean_escape_measure_def}.

%The ensembles are chosen uniformly on the $\dot{x}=0$ line, with Ensemble 1 centered at $(1.025,~0.0)$ and Ensemble $4$ at $(1.102,~0.0)$, approximately.
The first ensemble is located next to the regular region, and we can observe a high visitation frequency around this area. Ensembles 2 and 3, on the other hand, generate orbits that prefer the mid-section of the phase space. Finally, the orbits that begin in Ensemble 4 seem to concentrate more on the outer parts of the phase space, with a low visitation frequency on the area close to the stability region. Hence, what we observe is the existence of three very different types of transient motion, with the stickiness effect being experienced only by the orbits that start in the first ensemble. Such ensemble was actually chosen on top of the period-7 UPO on Scenario I, $P^7_{I}$, and we can also observe that the mean escape measure profile delineates the unstable manifolds associated with this orbit.

\begin{figure}[h!]
\centering
\includegraphics[scale=0.9]{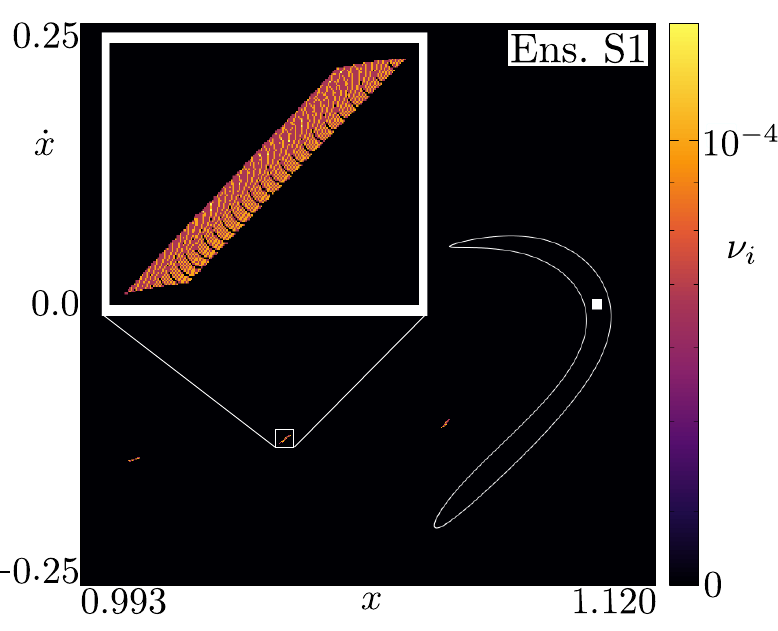}
\includegraphics[scale=0.9]{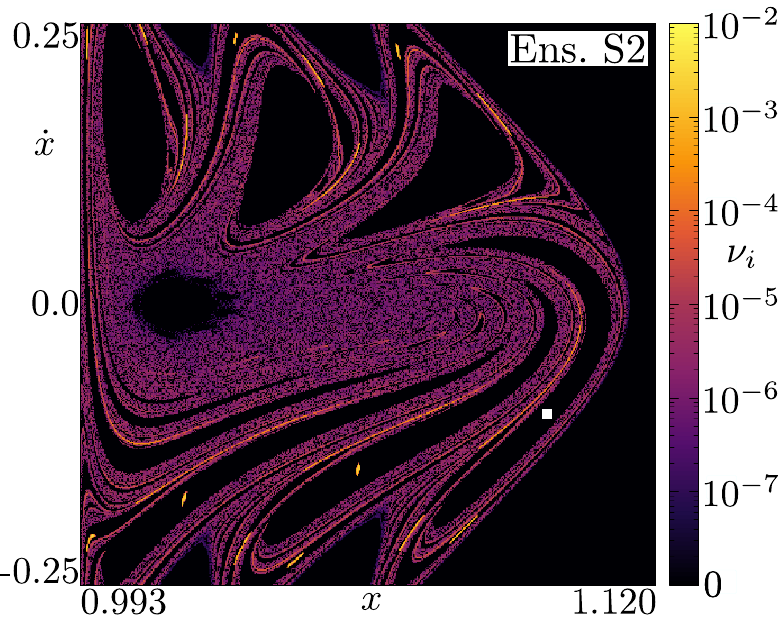}
\caption[Escape channels on the mean escape measure profile]{Mean escape measure profile on a logarithmic scale, assessed in the phase space $x$-$\dot{x}$ for Ensembles S1 and S2. These sets of initial conditions were chosen inside the ``black lakes'' and demonstrate the presence of escape channels in the system.}
\label{fig:ems_mean_escape_measure_profile_special_cases}
\end{figure}

One more feature we observe in Fig.~\ref{fig:ems_mean_escape_measure_profile} is the presence of ``black lakes'', areas on the phase space that are not visited and that do not correspond to a regular or a forbidden region. If we compare this figure to Fig.~\ref{fig:ems_manifolds_phase_space} for $C=3.187$, it is clear that these lakes correspond to the areas enclosed by the unstable manifolds associated with the Lyapunov orbit. Fig.~\ref{fig:ems_mean_escape_measure_profile_special_cases} presents the mean escape measure profile for two ensembles which were chosen inside one of these regions.

All the orbits that begin in Ensemble S1 move closely together and leave the system after crossing $\Sigma$ three times, as is shown in the left panel. Conversely, the orbits that begin in Ensemble S2 visit the whole phase space before exiting the lunar realm, as is shown in the right panel. Such distinct transient behaviors happen due to the fact that Ensemble S1 is actually located inside the intersection between $\Gamma^s_4$ and $\Gamma^u_4$ (see Fig.~\ref{fig:ems_manifolds_phase_space} and also the left panel of Fig.~\ref{fig:ems_intersect_and_break} for reference). Therefore, the orbits that begin in S1 rapidly follow the stable manifold through the Lyapunov orbit and onto the Earth's realm. On the other hand, Ensemble S2 is solely inside the unstable manifold, which leads the orbits around the lunar realm.

Figures~\ref{fig:ems_mean_escape_measure_profile} and~\ref{fig:ems_mean_escape_measure_profile_special_cases} have shown that the transient behavior of the system is strongly dependent on the choice of initial conditions and that the dynamical effects, such as stickiness and transport, may vary from case to case. By defining suitable quantities, it is possible for us to quantify such behaviors and attribute a degree of complexity to the transient motion \cite{deOliveira2020measure}.

\chapter{Conclusion and perspectives}
\label{ch:conclusion}

% \subsubsection{Conclusion}

In this manuscript, we investigated the relationship between dynamical properties and geometrical structures in a near-integrable Hamiltonian system, the Earth-Moon system, which was modeled by the planar circular restricted three-body problem.

As we varied the Jacobi constant, the system underwent three dynamical scenarios with very different phase space structures, without changing the topology of the Hill region. Scenarios I and III presented mixed phase spaces with distinct stickiness behavior, while Scenario II was dominated by global chaos. By carrying out a linear stability analysis, we showed that the transitions between these scenarios were due to specific bifurcations that occurred with the direct periodic orbits around the Moon.

We illustrated how the invariant manifolds of the main unstable periodic orbits in the system related to the aforementioned dynamical scenarios. The invariant manifolds associated with the members of the Lyapunov orbit family determined the shape and size of the regular regions, while the manifolds of the resonant unstable periodic orbits determined the behavior of the stickiness effect in the mixed phase space scenarios. Even though these two groups of manifolds were two-dimensional structures, they crossed our surface of section in different manners, hence displaying different geometrical properties in the Poincaré map.

By defining and calculating the multiplicative transit time, we were able to depict the influence of the invariant manifolds on the transport properties of the system. What was observed was a fine interplay between the Lyapunov orbit manifolds, which controlled the transition of orbits between the Earth and the Moon's realms, and the manifolds of the resonant unstable periodic obits, responsible for temporarily trapping the transit orbits. We also observed a close relationship between these geometrical structures and the sets of collisional orbits when we considered a collision radius around the Moon. 

Lastly, we analyzed the behavior of orbits that originated from different ensembles of initial conditions in the phase space, before leaving the lunar realm. Their motion was quantified by the mean escape measure, that determined which regions of the phase space were more or less visited by these trajectories. We found that dynamical effects such as stickiness were differently experienced by the ensembles, and that these trajectories could be influenced by the fast escape routes in the system that were formed by the Lyapunov orbit manifolds.

As a final remark, by studying the dynamical influence of invariant manifolds in the area-preserving Hénon map, and comparing it to the Earth-Moon system, we were able to observe similarities between the one-dimensional and two-dimensional cases. In both situations, we had the destruction of invariant tori with the advancement of manifolds in the phase space, and also concluded that these structures may have different roles in the transport phenomena. However, the intersection between higher-dimensional manifolds was shown to involve a more complex scenario, with these structures eventually breaking in the Poincaré map representation as a result of intersecting each other. Furthermore, the fast escape route seen from the mean escape measure profile for Ensemble~S1 is only possible because the Lyapunov orbit manifold disconnects from the surface of section and reintersects it transversally.

In short, our work provided an extensive view on the bounded planar Earth-Moon system, and reinforced the necessity of understanding both the dynamical and geometrical aspects of a Hamiltonian system.

There are a number of different ways for us to continue our research. For example, we can increase the range of the Jacobi constant as to incorporate heteroclinic intersections between different families of Lyapunov orbits. We can also consider different Hamiltonian systems or different versions of the three-body problem. In special, we intend to include small dissipative terms, which would reproduce physical effects seen in asteroid dynamics \cite{Murray1994,Bottke2006}. With this, it is possible to study the robustness of invariant manifolds and of their influence in the system's dynamics.

\chapter{Intellectual Production}
\label{ch:intellectual}

In this chapter, we lay out the intellectual production resulted from our research. 

\section{Scientific papers}

In total, we wrote three preprints, all of which were uploaded to the arXiv online repository (\url{https://arxiv.org/}). From these preprints, two were published and one is currently under review in a peer-reviewed scientific journal.

The first article was published in \emph{The European Physical Journal Special Topics}, and was named ``Dynamical trapping in the area-preserving Hénon map'' \cite{deOliveira2020dynamical}. In this paper, we carry out the investigation presented in Sec.~\ref{sec:1d_manifolds} where we have shown how the one-dimensional invariant manifolds are related to dynamical trapping in an area-preserving version of the Hénon map. We also analyzed how the distribution of homoclinic and heteroclinic intersections varies as the system goes from presenting stickiness to the total absence of this phenomenon.

The second article was published in \emph{Celestial Mechanics and Dynamical Astronomy}, entitled ``Order-chaos-order and invariant manifolds in the bounded planar Earth-Moon system'' \cite{deOliveira2020}. This paper exposes the main results presented in Ch.~\ref{ch:earth_moon_system} regarding the relationship between dynamical properties and geometrical structures in the bounded version of the Earth-Moon system, as modeled by the planar circular restricted three-body problem. 

The third preprint was entitled ``Measure, dimension, and complexity of escape in open Hamiltonian systems'' \cite{deOliveira2020measure}. In this work, we developed the concepts defined in Sec.~\ref{sec:escape}, and investigated the statistical properties of the mean escape measure profile in the Earth-Moon system, and also in a two-dimensional map used for modeling the dynamics of the magnetic field lines of a plasma which is confined to a single-null divertor tokamak. Furthermore, we defined and calculated two quantities capable of quantifying the distinct transient behaviors that emerge in these systems.

\section{Open source code}

The computer program that we developed for our numerical simulations was written in C/C\texttt{+}\texttt{+} programming language \cite{Press2007}, with the numerical integrator and linear algebra package being implemented from the GNU Scientific Library \cite{Galassi2001}. Its code is fully available at the Oscillations Control Group Wiki page (\url{http://yorke.if.usp.br/OscilControlWiki/index.php/Main_Page}) and also at Github (\url{https://github.com/vitor-de-oliveira/PCRTBP-explorer}).

\appendix

\chapter{Further information}
\label{app:further_3BP}

In this Appendix, we provide more information regarding the basic concepts of the planar circular restricted three-body problem, which were addressed in Sec.~\ref{sec:model}.

\section{PCRTBP in Hamiltonian form}
\label{app:PCRTBP_hamiltonian}

In order to write the planar circular restricted three-body problem in a Hamiltonian form, we use the following transformation from the rotational coordinate system \cite{Belbruno2004}

\be
\begin{aligned}
q_1&=x,\\
q_2&=y,\\
p_1&=\dot{x}-y,\\
p_2&=\dot{y}+x,
\end{aligned}
\ee

\noindent where the Hamiltonian function is $\mathcal{H}(p_1,p_2,q_1,q_2)=-C(x,y,\dot{x},\dot{y})/2$, which is given by

\begin{equation}
 \mathcal{H}(p_1,p_2,q_1,q_2)=\dfrac{(p_1+q_2)^2}{2}+\dfrac{(p_2-q_1)^2}{2}-\Omega(q_1,q_2).
\end{equation}

The equations of motion for the problem in the new variables $(q_1,q_2,p_1,p_2)$ are then given by the \emph{Hamilton equations}

\begin{equation}
 \begin{aligned}
  \dot{q}_k &= \dfrac{\partial \mathcal{H}}{\partial p_k}, \\
  \dot{p}_k &= -\dfrac{\partial \mathcal{H}}{\partial q_k}.
 \end{aligned}
\end{equation}

The PCRTBP can also be written in a more standardized form for near-integrable Hamiltonian systems by use of the \emph{Delauney variables}. In this case, the Hamiltonian function is given by a non-perturbed term, which corresponds to the energy of the two-body system, and a perturbed term that is proportional to the mass parameter \cite{Ferraz2007,Celletti2010}.

\section{Lagrangian points location and stability}
\label{app:stability}

The Lagrangian equilibrium points can be found by setting $\dot{x}=\ddot{x}=\dot{y}=\ddot{y}=0$ in Eqs.~\eqref{eq:motion_rotational}, which is the same as to calculate the critical points of the pseudo-potential $\Omega$:
 
\be
\dfrac{\partial\Omega}{\partial x}=\dfrac{\partial\Omega}{\partial y}=0.
\label{eq:equilibrium_condition_appendix}
\ee

From Eq.~\eqref{eq:omega}, we have

\be
\dfrac{\partial\Omega}{\partial x}=x\left[1-\frac{1-\mu}{r_{31}^3}-\frac{\mu}{r_{32}^3}\right]-\mu(1-\mu)\left[\frac{1}{r_{31}^3}-\frac{1}{r_{32}^3}\right],
\label{eq:omega_x}
\ee

\noindent and

\be
\dfrac{\partial\Omega}{\partial y}=y\left[1-\frac{1-\mu}{r_{31}^3}-\frac{\mu}{r_{32}^3}\right].
\label{eq:omega_y}
\ee

Let us first consider the case $y\neq0$. By combining Eqs.~\eqref{eq:equilibrium_condition_appendix} and~\eqref{eq:omega_y}, we have

\be
1-\frac{1-\mu}{r_{31}^3}-\frac{\mu}{r_{32}^3}=0.
\label{eq:condition_1}
\ee 

Substituting the condition \eqref{eq:condition_1} in Eq.~\eqref{eq:omega_x} and equating to zero gives

\be
\mu(1-\mu)\left[\frac{1}{r_{31}^3}-\frac{1}{r_{32}^3}\right]=0 \ \ \Rightarrow \ \ r_{31}=r_{32}=r,
\label{eq:condition_2}
\ee

\noindent which means that the distance from the test particle is the same for both primaries.

Putting \eqref{eq:condition_2} back into Eq.~\eqref{eq:condition_1} gives $r=1$. Since the units of measurement are normalized, this result indicates that the distance between the third particle and the primaries are equal to the distance between the primaries, which is only possible if the particles are located at the vertices of an equilateral triangle.

Finally, using the expressions for $r_{31}$ and $r_{32}$ given in Eqs.~\eqref{eq:r_1_2_rotational}, we have the following location, in the coordinate space $x$-$y$, for the two \emph{triangular equilibrium points}:

\be
L_4 = \left(\frac{1}{2}-\mu,\frac{\sqrt{3}}{2}\right) \ \ \text{and} \ \ L_5 = \left(\frac{1}{2}-\mu,-\frac{\sqrt{3}}{2}\right).
\label{eq:l4_l5}
\ee

Since we are able to analytically calculate the position of the equilibrium points for $y\neq0$, we know that there are exactly two points that satisfy this criterion. We proceed to show that there are exactly three equilibrium points that satisfy $y=0$.

Let us consider the function \cite{Arnold2007}

\be
{\Omega|}_{y=0}=\frac{x^2}{2}+\frac{1-\mu}{|x+\mu|}+\frac{\mu}{|x-(1-\mu)|},
\label{eq:omega_y_zero}
\ee

\noindent which has second derivative, with respect to $x$, given by

\be
\frac{d^2}{dx^2}({\Omega|}_{y=0})=1+\frac{2-2\mu}{|x+\mu|^3}+\frac{2\mu}{|x-(1-\mu)|^3}.
\label{eq:omega_y_zero_second_derivative}
\ee

According to Eq.~\eqref{eq:omega_y_zero}, ${\Omega|}_{y=0}>0$ for all interval and diverges for $x\to\pm\infty$, $x\to-\mu$, and $x\to1-\mu$. Given that Eq.~\eqref{eq:omega_y_zero_second_derivative} is also positive for the entire real line, there exists exactly three minima of the function \eqref{eq:omega_y_zero}, which correspond to three \emph{collinear equilibrium points} in the $x$-axis, each belonging to one of the following intervals: $(-\infty,-\mu)$, $(-\mu,1-\mu)$, and $(1-\mu,\infty)$.

There is no analytical solution for the exact location of the collinear Lagrangian points. One possible way to obtain an approximate solution is by use of a series expansion along with Lagrange's inversion method \cite{Murray1999}. However, with little computational effort, we can numerically calculate an approximate solution in the following manner.

Let us consider $x\in(-\mu,1-\mu)$, and use condition \eqref{eq:equilibrium_condition_appendix} to rewrite Eq.~\eqref{eq:omega_x} as

\be
{(x+\mu)}^2{(1-\mu-x)}^2x-(1-\mu){(1-\mu-x)}^2+\mu{(x+\mu)}^2=0.
\label{eq:polynomial_L1}
\ee

By rearranging the terms in Eq.~\eqref{eq:polynomial_L1}, we end up with a fifth-order polynomial equation in $x$. We then calculate the location of $L_1$ in the $x$-axis by using \emph{Newton's method} starting at $x_0=(1-\mu)-{(\mu/3)}^{1/3}$. The same can be done for the other two collinear points \cite{Szebehely1967}.

The stability of the Lagrangian points is given by a linear analysis around them \cite{Murray1999}. From Eqs.~\eqref{eq:motion_rotational}, we have that the \emph{Jacobian matrix} of the system in the basis $(x,y,\dot{x},\dot{y})$ is given by

\be
{\bs{DJ}}=\left(
\renewcommand{\arraystretch}{1.6}
\begin{array}{cccc}
0 & 0 & 1 & 0 \\
0 & 0 & 0 & 1 \\
\dfrac{\partial^2\Omega}{\partial x^2} & \dfrac{\partial^2\Omega}{\partial xy} & 0 & 2 \\
\dfrac{\partial^2\Omega}{\partial yx} & \dfrac{\partial^2\Omega}{\partial y^2} & -2 & 0 \\
\end{array} \right),
\label{eq:jacobian_matrix}
\ee

\noindent with

\be
\dfrac{\partial^2\Omega}{\partial x^2}=1-\frac{1-\mu}{r_{31}^3}\left[{1-3\frac{{(x+\mu)}^2}{r_{31}^2}}\right]-\frac{\mu}{r_{32}^3}\left[{1-3\frac{{(x-(1-\mu))}^2}{r_{32}^2}}\right],
\label{eq:omega_xx}
\ee

\be
\dfrac{\partial^2\Omega}{\partial y^2}=1-\frac{1-\mu}{r_{31}^3}\left[{1-3\frac{{y}^2}{r_{31}^2}}\right]-\frac{\mu}{r_{32}^3}\left[{1-3\frac{{y}^2}{r_{32}^2}}\right],
\label{eq:omega_yy}
\ee

\noindent and

\be
\dfrac{\partial^2\Omega}{\partial xy}=\dfrac{\partial^2\Omega}{\partial yx}=3y\left[\frac{(1-\mu)(x+\mu)}{r_{31}^5}+\frac{\mu(x-(1-\mu))}{r_{32}^5}\right].
\label{eq:omega_xy}
\ee

Equation~\eqref{eq:jacobian_matrix} has \emph{four} eigenvalues, which are given by

\be
\lambda_{1,2}=\pm\frac{\sqrt{2}}{2}\left\{\Omega_{xx}+\Omega_{yy}-4-{\left[{(4-\Omega_{xx}-\Omega_{yy})}^2-4(\Omega_{xx}\Omega_{yy}-\Omega_{xy}^2)\right]}^{\frac{1}{2}}\right\}^{\frac{1}{2}},
\label{eq:lambda_12}
\ee

\noindent and

\be
\lambda_{3,4}=\pm\frac{\sqrt{2}}{2}\left\{\Omega_{xx}+\Omega_{yy}-4+{\left[{(4-\Omega_{xx}-\Omega_{yy})}^2-4(\Omega_{xx}\Omega_{yy}-\Omega_{xy}^2)\right]}^{\frac{1}{2}}\right\}^{\frac{1}{2}}.
\label{eq:lambda_34}
\ee

For the collinear equilibrium points, we have \cite{Szebehely1967}

\be
\dfrac{\partial^2\Omega}{\partial x^2}>0, \ \ \dfrac{\partial^2\Omega}{\partial y^2}<0, \ \ \text{and} \ \ \dfrac{\partial^2\Omega}{\partial xy}=0.
\label{eq:conditions_collinear_points}
\ee

Putting conditions \eqref{eq:conditions_collinear_points} into Eqs.~\eqref{eq:lambda_12} and~\eqref{eq:lambda_34}, we find that the eigenvalues $\lambda_{1,2}$ are purely imaginary, whilst $\lambda_{3,4}$ are real. Hence, the collinear equilibrium points are \emph{unstable}.

For the triangular equilibrium points, Eq.~\eqref{eq:l4_l5}, we have

\be
\dfrac{\partial^2\Omega}{\partial x^2}=\frac{3}{4}, \ \ \dfrac{\partial^2\Omega}{\partial y^2}=\frac{9}{4}, \ \ \text{and} \ \ \dfrac{\partial^2\Omega}{\partial xy}=\pm\frac{3\sqrt{3}}{4}(1-\mu).
\label{eq:conditions_triangular_points}
\ee

Substituting (\ref{eq:conditions_triangular_points}) into Eqs.~\eqref{eq:lambda_12} and~\eqref{eq:lambda_34}, we arrive at the condition that all the eigenvalues are purely imaginary if, and only if,

\be
1-27(1-\mu)\mu\geq0 \ \ \Rightarrow \ \ \mu\leq0.0385.
\label{eq:stability_condition_l4_l5}
\ee

Hence, if \eqref{eq:stability_condition_l4_l5} is satisfied, which is the case for the Earth-Moon system, $L_4$ and $L_5$ are \emph{stable}.

\section{Derivation of the Jacobi constant}
\label{app:jacobi}

The Jacobi constant can be derived in the following manner. First, we multiply the top expression in Eqs.~\eqref{eq:motion_rotational} by $\dot{x}$, and the bottom one by $\dot{y}$:

\be
\begin{aligned}
\dot{x}\ddot{x}-2\dot{x}\dot{y} &= \dot{x}\dfrac{\partial\Omega}{\partial x},\\
\dot{y}\ddot{y}+2\dot{y}\dot{x} &= \dot{y}\dfrac{\partial\Omega}{\partial y}.\\
\end{aligned}
\label{eq:jacobi_calculations}
\ee 

We then add both expressions in Eqs.~\eqref{eq:jacobi_calculations}, which gives

\be
\dot{x}\ddot{x}+\dot{y}\ddot{y}=\dot{x}\dfrac{\partial\Omega}{\partial x}+\dot{y}\dfrac{\partial\Omega}{\partial y}=\frac{d\Omega}{dt}.
\label{eq:jacobi_calculations_2}
\ee

Next, we integrate Eq.~\eqref{eq:jacobi_calculations_2} with respect to time. Integrating by parts the left-hand side gives

\be
\nonumber
\int{(\dot{x}\ddot{x}+\dot{y}\ddot{y}) \ dt}=\dot{x}^2 - \int{\dot{x}\ddot{x} \ dt} + \dot{y}^2 - \int{\dot{y}\ddot{y} \ dt} \Rightarrow \int{(\dot{x}\ddot{x}+\dot{y}\ddot{y}) \ dt} = \frac{\dot{x}^2}{2} + \frac{\dot{y}^2}{2},
\ee 

\noindent while directly integrating the right-hand side gives $\Omega+K$, with $K$ constant.

Finally, we choose $K=-C/2$, and we end up with

\be
\frac{\dot{x}^2}{2} + \frac{\dot{y}^2}{2}=\Omega+\frac{-C}{2} \ \ \Rightarrow \ \ C=2\Omega(x,y)-\dot{x}^2-\dot{y}^2.
\label{eq:jacobi_constant_appendix}
\ee

The Jacobi constant can also be written in the inertial reference frame by applying the transformation \eqref{eq:rotation} to Eq.~\eqref{eq:jacobi_constant_appendix}. If we proceed with this transformation, but choose $K=C$, instead of $K=-C/2$, we have \cite{Belbruno2004}

\be
C=\dfrac{1}{2}(\dot{\xi}^2+\dot{\eta}^2)-\Psi(\xi,\eta,t)-(\xi\dot{\eta}-\eta\dot{\xi}).
\label{eq:jacobi_constant_appendix_inertial}
\ee

The first three terms in Eq.~\eqref{eq:jacobi_constant_appendix_inertial} represent the total energy of the third particle, while the last two terms compose the massless angular momentum of $P_3$ \cite{Battin1987}. Therefore, even though these quantities are not conserved for $\mu\neq0$, the Jacobi constant can be seen as a combination of the two.

\singlespacing
\bibliographystyle{ieeetr}
\bibliography{citation_phd}

\end{document}